\documentclass[12pt]{report}         %% LaTeX2e document.
\usepackage{utthesis2}              %% Preamble.
\usepackage{amssymb}
\usepackage{bm}
\usepackage{amsfonts}
\newcommand{\beq}{\begin{equation}}
\newcommand{\eeq}{\end{equation}}
\newcommand{\beqa}{\begin{eqnarray}}
\newcommand{\eeqa}{\end{eqnarray}}
\newcommand{\w}{\wedge}

\newcommand{\ts}{\textstyle}
\newcommand{\dl}{\bm{\delta}}
\newcommand{\nn}{\nonumber}
\newcommand{\vep}{\varepsilon}
\newcommand{\cR}{\stackrel{\circ}{R}}
\newcommand{\bR}{\stackrel{\bullet}{R}}

\singlespace                       %% Uncomment one of these if you want
\renewcommand{\thesisauthor}{Andrew Culp Randono}    %% Your% official UT name.

\renewcommand{\thesismonth}{August}     %% Your month of graduation.

\renewcommand{\thesisyear}{2007}      %% Your year of graduation.

\renewcommand{\thesistitle}{In Search of Quantum de Sitter Space: Generalizing the Kodama State}     %% The title of
\renewcommand{\thesisauthorpreviousdegrees}{B.S.}
\renewcommand{\thesissupervisor}{Richard A. Matzner}
\renewcommand{\thesiscommitteemembera}{Philip J. Morrison}
\renewcommand{\thesiscommitteememberb}{Sonia Paban}
\renewcommand{\thesiscommitteememberc}{Lawrence C. Shepley}
\renewcommand{\thesiscommitteememberd}{Karen K. Uhlenbeck}
\renewcommand{\thesisauthoraddress}{
9 Marland Rd.\\
        Colorado Springs, CO 80906
      }
\renewcommand{\thesisdedication}{I dedicate this dissertation to my parents}
\begin{document}
\bibliographystyle{utphys}

\thesiscopyrightpage                 %% Generate the copyright page.

\thesiscertificationpage             %% Generate the PhD. certification page.

\thesistitlepage                     %% Generate the title page.

%\thesissignaturepage                %% Generate the Master's signature page.

\thesisdedicationpage                %% Generate the dedication page.

\begin{thesisacknowledgments}        %% Use this to write your%% acknowledgments; it can be anything
Let me begin by thanking those people who encouraged me to pursue quantum gravity at UT despite the odds, especially
Abhay Ashtekar, Carlo Rovelli, Richard Matzner, Lee Smolin, and my undergraduate advisor Allen Everett to whom my first
paper on the main subject of this dissertation is
dedicated. I would like to thank the relativity crew at UT, Richard Matzner, and Larry Shepley, for
providing a great working environment. Thanks to Spencer Stirling and Jerry Schirmer for learning quantum gravity with me in
the beginning. Let me thank Lee Smolin for being my long distance advisor, for many
discussions on quantum gravity in general, for reading too many papers I've sent you, and for hosting me at the Perimeter
Institute in the summer of 2006.

In regards to the dissertation I would like to thank Ted Jacobson, Don Witt, Chopin Soo, and Abhay Ashtekar for brief but
helpful comments and exchanges. Many thanks to Stephon Alexander for helping me understand the action of CPT on the
Kodama state and the deeper implications of quantum gravity on cosmology and the standard model. Thanks to Laurent
Freidel for discussions on torsion and fermion interactions, and for his comments on the generalized Kodama states
that helped me progress.
I'd like to thank Ed Witten for his comments on my papers and for a discussion about the nature of the Kodama state
that inspired me to think deeper about the matter. I'd also like to thank Lee Smolin again for writing a review of
quantum gravity and de Sitter space that inspired me, and others, to start thinking about the problem.

More personally, I should thank the J-team for making sure I didn't work too hard in Austin, and for being the (self-proclaimed)
Jabronis. I'll miss you guys. Thank you, Antonia, for your love. And finally, thanks to my parents for their endless
support.
   
\end{thesisacknowledgments}          %% allowed in LaTeX2e par-mode.

\begin{thesisabstract}               %% Use this to write your thesis
The Kodama state is unique in being an exact solution to all the constraints of quantum gravity that also has a well
defined semi-classical interpretation as the quantum version of a classical spacetime, namely de Sitter or anti-de
sitter space. Despite this, the state fails to pass some of the key tests of a physically realistic quantum state. In
an attempt to resolve this problem, we track down the root of the problem to a choice for a particular parameter: the
Immirzi parameter. The Kodama state takes this parameter to be complex, whereas modern formulations of canonical
quantum gravity require that the parameter is real. We generalize the Kodama state to real values of the Immirzi
parameter, and find that the generalization opens up a large Hilbert space of states, one of which can be directly
interpreted as particular slicing of de Sitter space. We then show that these states resolve, or are expected to resolve
many of the problems associated with the original version of the Kodama state. In order to resolve the interpretation of
the multitude of states, we develop a new model of covariant classical and quantum gravity where the full Lorentz group
is retained as a local symmetry group, and the canonical evolution generated by the constraints has a close relation to
a larger group: the de Sitter group. This formalism gives strong evidence that the multitude of generalized Kodama
states can be unified into a single quantum state that is quantum de Sitter space.
      
\end{thesisabstract}                 %% allowed in LaTeX2e par-mode.

\tableofcontents                     %% Generate table of contents.
% \listoftables                      %% Uncomment this to generate list
                                     %% of tables.
% \listoffigures                     %% Uncomment this to generate list
                                     %% of figures.

\chapter{Prologue}                        %% Begin your thesis text here; follow the report style and group your text
                                                      %% in chapters, sections, etc.

\section{Introduction}
Perturbative techniques in quantum field theory and their extension to quantum gravity are unparalleled in
computational efficacy. In addition, because one can always retreat to the physical picture of particles as small field
perturbations propagating on a classical background, perturbation theory maximizes the ease of transition from quantum to
classical mechanics, and many processes can be viewed as quantum analogues of familiar classical events.
However, the transparent
physical picture disappears in systems where the distinction between background and perturbation to said background
is blurred. Such systems include strongly interacting systems, such as QCD, or systems where there is no preferred
background structure, such as general relativity. In contrast, non-perturbative and background independent approaches
to quantum gravity do not
distinguish background from perturbation, and are, therefore, appropriate for modeling the quantum mechanical ground
state of the universe itself, which it is hoped will serve as the vacuum on which perturbation theory can be based.
However, this is often at the expense of losing the smooth transition from a quantum description to its classical or
semi-classical counterpart as evidenced, for example, by the notorious problem of finding the low energy limit of Loop
Quantum Gravity. The sticking point is that pure quantum spacetime may be sufficiently divorced from our classical
understanding of fields on a smooth Riemannian manifold that matching quantum or semi-classical states with classical
analogues may be extremely difficult.

In this respect the Kodama state is unique. Not only is the state an exact solution to all the constraints of canonical
quantum gravity, a rarity in itself, but it also has a well defined physical interpretation as the quantum analogue of a 
familiar classical spacetime, namely de Sitter
or anti-de Sitter space depending on the sign of the cosmological constant\cite{Kodama:original, Kodama:original2,
Smolin:kodamareview}. Thus, the state is a candidate for the fulfillment of one of the
distinctive advantages of non-perturbative approaches over perturbative techniques: the former has the potential to
predict the purely
quantum mechanical ground state on which perturbation theory can be based.

In addition, the Kodama state has many beautiful mathematical properties relating the seemingly disparate fields of
abstract knot theory and quantum field theory on a
space of connections\cite{Witten:knots}. In particular, the exact form of the state is known in both the connection
representation
where it is the exponent of the Chern-Simons action, and in the q-deformed spin network representation where it is a
superposition of all framed spin networks with amplitudes given by the Kauffman bracket\footnote{The loop transform is
well understood and rigorous at the level of mathematical physics for Euclidean signature spacetime. For Lorentz
signature
spacetime, the loop transform is believed to be the Kauffman bracket, but the proof requires integrating along a real
contour in the complex plane, and it is not as rigorous as in the real case (see e.g. \cite{Pullin:Book}). The de Sitter
state that we will
present shares loop transform properties in common with the Euclidean signature Kodama state, so it is well defined in
the loop basis.}\cite{Kauffman:TemperleyLieb}.
This connection played a pivotal
role in the development of the loop approach to quantum gravity. One offshoot of the connection between the state and
knot theory is that the relation with quantum groups allows for a reinterpretation of the role cosmological constant as
the modulator of the deformation parameter of the quantum deformed group.

Ultimately, however, observation and experiment are the arbiters of the relevance of a physical theory. Cosmological
evidence suggests that we live in an increasingly vacuum dominated universe, which is asymptotically approaching
de Sitter space in the future as matter fields are diluted by the expansion of the universe, and possibly in the past as
well as evidenced by the success of inflation models. Thus, the state with positive $\lambda$ is particularly relevant to
modern cosmology. Since the state has the status of a ``wave
function of the universe", it opens up the possibility of making unique quantum gravitational predictions of a
cosmological
nature---a route that appears to be the most promising direction towards quantum gravity 
phenomenology\cite{Glikman:QGPhenomenology}.

Despite all the promise the state offers, it is plagued with problems. In particular, it does not or is not known to
satisfy some key tests of a physically realistic quantum state. Most notably, there is good evidence that the wave
function cannot be normalized under any inner product, it is not invariant under the discrete CPT symmetry, and
perturbations to the state may have negative energy sectors. Most of these problems have been argued to apply to the
Kodama state by analogy with a corresponding Chern-Simons state in Yang-Mills, rather than through explicit
demonstration\cite{Witten:note}.
Since the Kodama state is subtly, but significantly different from the
Yang-Mills solution, the analogy is not precise. However, the issue of normalizability has been
investigated\cite{Smolin:linearkodama}, and the non-normalizability of linearized perturbations to the state has been
confirmed.

With these problems in mind, in this dissertation we set out to generalize the state in order to overcome these
difficulties. In order to do this, we will exploit the subtle differences between the Kodama state in quantum gravity and
the Chern-Simons state in Yang-Mills theory. We will first track the problems of the original Kodama state and the
analogous Chern-Simons state down to the complexification necessary in its construction. In contrast to the Yang-Mills
state, we will show that this complexification is not necessary in the construction of the Kodama state, but
corresponds to a particular choice for a parameter in the quantum theory called the Immirzi parameter, $\beta$. When
this parameter is taken to be $\beta=-i$, the original Kodama state can be constructed, but modern formulation of Loop
Quantum Gravity take the parameter to be real precisely in order to avoid complications from complexification. 
At present, the parameter is believed to have physical ramifications, in that it modulates the size
of quanta of space at the Planck length, and its value is believed to be fixed by a matching of the semi-classical
derivation of black hole entropy with the full quantum gravitational derivation. Thus, choosing $\beta$ to be real
is more than simply a way to avoid the problems associated with the Kodama state---it allows the Kodama state to fit
into the framework of modern canonical quantum gravity.

Once the state, or as we will see {\it states}, are constructed we will show that they solve or are expected to solve
many of the problems of the original Kodama state. In addition, the generalization will introduce some interesting
paradoxes. In particular, we will see that it appears to open up a large Hilbert space of states that solve the
Hamiltonian constraint. The interpretation of this multitude of states is a delicate matter---the states could be
perturbations to de Sitter space, or they could all be related by a sector of the gauge degrees of freedom that is cut
off by the standard gauge fixing procedure of canonical quantum gravity. In an attempt to answer this question, we
will develop a new model of covariant classical and quantum gravity where this partial gauge fixing is avoided. Once
this is developed, we will find strong evidence that that the multitude of generalized Kodama states can be unified into
a single state.

\section{Outline}
The logical structure of the dissertation proceeds as follows. The first six chapters will be devoted to developing the
necessary background material. Our reviews are intended to be pedagogical rather than exhaustive. In chapter \ref{ECReview} we will present a general overview of the Einstein-Cartan approach to classical gravity and the
extension that is the usual starting point of Loop Quantum Gravity. We will also discuss the coupling of fermions to
Einstein-Cartan gravity and its extension, and the ramifications of spin-torsion coupling in the effective field
theory. This last part is outside of the logical structure of the dissertation, but it does serve to illuminate the role
played by the Immirzi parameter in the classical theory. In chapter \ref{dSReview} we will review classical de
Sitter spacetime, and most importantly, the standard slicings of de Sitter space and the initial data formulation. In
chapter \ref{MMReview} we will review the Macdowell-Mansouri formulation of gravity, which will be useful in
understanding the connection between the Kodama state and local de Sitter gauge symmetry. In this chapter we will also
present a group theoretical reason for the positivity of the bare cosmological constant. In chapter \ref{HamiltonianGR}
we will review the Hamiltonian formulation of the various versions of Einstein-Cartan gravity.
Those familiar with the $3+1$ formulation of gravity may still wish to read this section since we will use notation
that is likely to be unfamiliar. In chapter \ref{LQGIntro} we will summarize the basics of Loop Quantum Gravity at a,
more or less, introductory level. Special emphasis will be placed on the features that are relevant to our discussion
of the Kodama state. In chapter \ref{KState0} we will review the construction and interpretation of the original form
of the Kodama
state and conclude with a discussion of the problems associated with the state. In chapter \ref{KState1} we will
track down the source of the problems of with the Kodama state and suggest a resolution. Then we will proceed to
construct the generalized Kodama states and the unique state in the sector that solves all of the quantum constraints. In
chapter \ref{KState2} we will discuss the physical interpretation of the state from an initial data perspective and
from a more covariant WKB analysis. The latter will provide evidence that the full sector of states can be unified
into a single umbrella state. We will discuss how the state solves many of the problems associated with the original,
and how to construct the physical inner product of the state which will yield an intriguing connection with the
Macdowell-Mansouri formulation of gravity. Finally, in an attempt to resolve the interpretation of the multitude of
states, we will develop a new framework for covariant classical and quantum gravity in chapter \ref{GaugeFree}.
There we will see strong evidence that the full sector of generalized state is unified in a covariant setting, and the
single state is the umbrella state found in the previous chapter. This will highlight the true nature of the Kodama
state and its relation to de Sitter symmetry.

\section{Conventions}
Throughout this dissertation we will work with a Lorentzian metric with signature $\eta=diag(-1,1,1,1)$ (unless otherwise
indicated) on a four dimensional manifold $M$. A generic three-dimensional Cauchy slice of $M$ is denoted by $\Sigma$.
Indices in the tangent and cotangent spaces of the base manifold, spacetime indices are represented by Greek letters
$\{\mu,\nu,\alpha,\beta,...\}$, and spatial indices are represented by lower case Roman indices in the beginning of the
alphabet $\{a,b,c,...\}$. Upper case Roman
indices $\{I,J,K,L,...\}$ represent spacetime indices in an orthonormal frame or in the adjoint $Spin(3,1)$
representation space and they range from $0$
to $3$. Lower case Roman indices $\{i,j,k,...\}$ are three dimensional indices in the adjoint representation of $SU(2)$, and
range from $1$ to $3$. The metric volume form $\epsilon_{IJKL}$ is defined such that 
$\epsilon_{0123}=-\epsilon^{0123}=+1$. 

We will often have use for a Clifford algebra notation. Here vectors are valued
in the grade-1 elements of the Clifford algebra, $V=\frac{1}{2}\gamma_{I}V^{I}$, where the factor of $\frac{1}{2}$
occurs for convenience so that $V_{I}W^{I}=Tr(VW)$. Bivectors and elements of the $so(3,1)$ Lie algebra then take the
form $\frac{1}{4}\gamma_{[I}\gamma_{J]}\alpha_{IJ}$. The volume element of the Clifford algebra is
$\star=-i\gamma_{5}=\frac{1}{4!}\epsilon_{IJKL}\gamma^{I}\gamma^{J}\gamma^{K}\gamma^{L}$. When
writing integrals over forms that are valued in the
Clifford algebra or the Lie algebra of a gauge group, the trace is assumed in the integral and will generally be in the
fundamental or adjoint representation as indicated. We will often use the short hand notation where the explicit wedge
products are dropped. 

The cosmological constant is taken to be positive when the scalar curvature is positive in vacuum, and negative when
the scalar curvature is negative. The numerical coefficients are fixed so that $R=4\lambda$ in vacuum. 

The Immirzi parameter is defined so that the Ashtekar-Barbero connection is $A^{ij}=\Gamma^{ij}-\beta
{\epsilon^{ij}}_{k}K^{k}$
and the left handed Ashtekar connection results from $\beta=-i$.

\chapter{Classical Einstein-Cartan Gravity\label{ECReview}}
The typical starting point of Loop Quantum Gravity is a modification of the Einstein-Cartan action. The
Einstein-Cartan action differs from the Einstein-Hilbert action in that it allows for non-zero torsion in the presence
of matter (see e.g. \cite{GockelerSchucker, Ortin:book}). At the level of the
Einstein equations in vacuum the two actions yield the same equations of motion. Thus, from a classical perspective,
the vacuum dynamics of the gravitational field in Einstein-Cartan theory
and Einstein-Hilbert theory are indistinguishable. In the presence of matter, however, the Einstein-Cartan action
allows for spin-torsion coupling, which is excluded by hand in the Einstein-Hilbert approach. The torsion in the
former is non-propagating, and it generally is negligible unless spin-currents are extremely large. Current
experimental data cannot distinguish between the two theories. On the other hand, from a theoretical perspective, it
is extremely natural, and very compelling, to allow for spin-torsion coupling. As we will see, this provides the first
steps towards understanding gravity in the framework of the ordinary gauge theories of the standard model, placing
them on similar, if not the same footing. From a quantum perspective, allowing for torsion versus excluding it by hand
changes things considerably. Even if we restrict attention to the vacuum sector, since quantum dynamics depends on
contributions from the the entire space, including off-shell contributions where the equations of motion do not hold,
the Einstein-Cartan and Einstein-Hilbert approaches will likely yield different quantum theories. This has been
verified in three-dimensional gravity, where the quantization is fairly well understood. In $3+1$ gravity, torsion
likely does play a fundamental, though poorly understood, role in the standard formulation of Loop Quantum Gravity.
Thus, in choosing to begin with the Einstein-Cartan action, we are taking a theoretical leap of faith. However, again
from theoretical
arguments, this leap is not without ample justification\footnote{Some of the justification comes from hindsight---the
original Wheeler-Dewitt formulation of canonical quantum gravity, which begins with the Einstein-Hilbert action, is
fraught with problems, many of which have not been solved since it was introduced some forty years ago. On the other
hand, much more is understood from approaches that apply the basics of the Wheeler-Dewitt formalism to the 
Einstein-Cartan formalism.}. 

\section{Preliminaries}
The Einstein-Cartan approach begins with a basis that trivializes the metric components. At each point of the
manifold, choose a basis of one-forms $e^{I}$
such that the metric components are put in canonical diagonal form:
\beq
g=\eta_{IJ}\,e^{I}\otimes e^{J}\,,
\eeq
where $\eta_{IJ}$ is given by
\beq
\eta_{IJ}=\eta^{IJ}=
\left[\begin{array}{cccc} -1&0&0&0 \\
0&1&0&0\\
0&0&1&0\\
0&0&0&1
\end{array}\right]\,.
\eeq
Such a basis goes by various names, which we may have occasion to use: orthonormal basis, tetrad, veirbein, or frame
field. This choice of basis can always be made, though the basis will not be a coordinate basis in general. To map
between a coordinate basis and the orthonormal frame, it is convenient to view the components of the frame as a
$4\times 4$ matrix so that $e^{I}=e^{I}_{\mu}dx^{\mu}$. This matrix must be everywhere invertible if we are to be able
to define the inverse metric at each point. We will write the inverse simply as $e^{\mu}_{I}$. This matrix serves as
the components of an orthonormal vector basis, $\bar{e}_{I}=e^{\mu}_{I}\,\frac{\partial}{\partial x^{\mu}}$.
Invertibility then simply means $e^{I}(\bar{e}_{J})=\delta^{I}_{J}$.
The equivalence principle underpins the philosophy behind the frame
fields. At each point we can always boost to a frame such that in a small enough neighborhood of the point, within the
finite experimental error of the measuring devices of the observer, the local physics will local like special
relativity in Minkowski space. The frame field simply defines a different instance of local Minkowski space at each
point. From a computational perspective, all of the information about the metric (aside from the signature) is now
contained in the frame, $e^{I}$, itself. Thus, $e^{I}$ will now be one of the dynamical fields that occurs in the
action. Since the equivalence principle is almost certain to break down at very small length scales, choosing the
metric or the tetrad to be dynamical is like choosing an answer to {\it which came first, the chicken or the egg?} As we
will see later, there are good theoretical reasons for taking the tetrad to be the more fundamental object. 

There is considerable freedom in choosing the tetrad at each point. Since only the canonical form of the metric must
be preserved, we can rotate or boost any set of frame fields with a different rotation or boost parameter at each
point and the resulting frame will still be a good frame field. Thus, the local gauge freedom is (locally) the
Lorentz group $SO(3,1)$. Since we are still free to redefine points on the manifold, the diffeomorphism invariance of
the base manifold is still intact. Geometrically, we are simply defining a fibre bundle with each fibre being
identical to flat Minkowski space. The ``metric", $\eta_{IJ}$, lives in the fiber, and the frame field (together with
its inverse) allows one to go back and forth between the fiber and the base manifold by projecting vectors and forms
in the fibre to the tangent space $T^{*}M$ of the base manifold. On this bundle we add a connection
$\omega^{IJ}={\omega^{IJ}}_{\mu}dx^{\mu}$ which allows one to parallel transport objects in the fiber at one point to
objects in the fiber at another. The connection is defined to be compatible with the metric, $\eta^{IJ}$, in the fiber
so that
\beqa
D_{\omega}\eta^{IJ}&=&d\eta^{IJ}+{\omega^{I}}_{K}\eta^{KJ}+{\omega^{J}}_{K}\eta^{IK}\nn\\
&=&\omega^{IJ}+\omega^{JI}\nn\\
&=& 0\,.
\eeqa
The last line tells us that the components of the connection must be anti-symmetric. Since the Lie algebra of
$SO(3,1)$ is spanned by the anti-symmetric $4\times 4$ matrices, in a local basis of a trivializing neighborhood of the
fiber bundle the connection
takes values in the Lie algebra $so(3,1)$. To obtain the ordinary Levi-Civita connection in this basis we have to
impose one more condition: compatability with the tetrad. This is equivalent to the vanishing of the torsion:
\beq
De^{I}=T^{I}=e^{I}_{\alpha} \,{\ts\frac{1}{2}}\,{T^{\alpha}}_{\mu\nu}\,dx^{\mu}\w dx^{\nu}=0\,.
\eeq
It can be shown that the connection that satisfies these two conditions, which we will denote by $\Gamma^{IJ}$ or
$\omega^{IJ}[e]$,
is the ordinary Levi-Civita connection expressed in the orthonormal basis. It can be solved explicitly in terms of the
tetrad and its inverse by
\beq
{\Gamma^{IJ}}_{\mu}={\omega^{IJ}}_{\mu}[e]=2e^{\nu[I}\partial_{[\mu}e^{J]}_{\nu]}
+e_{\mu K}e^{\nu I}e^{\sigma J}\partial_{[\sigma}e_{\nu]}^{K}\ .\label{Levi-Civita}
\eeq
The connection $\omega^{IJ}$ is called the spin connection since it can be used to define the covariant derivative of
a spinor. To see this, it is useful to work with the Clifford algebra defined by
\beq
\gamma^{I}\gamma^{J}+\gamma^{J}\gamma^{I}=2\eta^{IJ}\,.
\eeq
In the standard matrix representation, the matrices $\gamma^{I}$ are $4\times 4$ complex matrices. We can define a
sixteen dimensional vector space, which we will often have occasion to use, out of products of the Clifford matrices.
The vector space has the convenient basis 
\beqa
&{\ts\frac{1}{4}}\,\mathbb{I}& \\
&{\ts\frac{1}{2}}\gamma^{I} & \\
&{\ts \frac{1}{2}}(\gamma^{[I}\gamma^{J]})&\\
& {\ts \frac{1}{2}}\gamma^{I} \star&\\
& {\ts\frac{1}{4}} \star &
\eeqa
where $\star=-i\gamma_{5}=\gamma^{0}\gamma^{1}\gamma^{2}\gamma^{3}$. As a vector space over the complex numbers, under
the action of matrix multiplication, the above algebra forms a basis for the Lie algebra $gl(4,\mathbb{C})$. This is
simply because any complex $4\times 4$ is an element of $gl(4,\mathbb{C})$, and any complex matrix can be written as a
complex linear combination of the above. The vector space also has a natural inner product. Using the short hand notation
$\Gamma^{\hat{I}}$ where $\hat{I}$ ranges from $1$ to $16$ for the basis given above, the inner product is
\beqa
\langle A , B\rangle &=&Tr(AB)\nn\\
&=&A_{\hat{I}}B_{\hat{J}}\,Tr(\Gamma^{\hat{I}}\Gamma^{\hat{J}})\nn\\
&=&A_{\hat{I}}B_{\hat{J}}\,\eta^{\hat{I}\hat{J}}
\eeqa
It can be shown using the familiar trace properties of the Dirac matrices, that $\eta^{\hat{I}\hat{J}}$ is a diagonal
$16\times 16$ matrix with all $1$'s and $-1$'s in the diagonals. The subalgebra formed by the bivector elements of the
Clifford algebra is isomorphic to $so(3,1)$. This then gives the spinor representation of the Lie algebra of the group
$Spin(3,1)$, which is the double cover of $SO(3,1)$ and defines the action of the Lorentz group on spinors. In this
representation, the spin connection becomes
\beq
\omega=\omega^{IJ}\,{\ts \frac{1}{4}}\gamma_{I}\gamma_{J}
\eeq
and its action on a Dirac spinor $\psi$ is simply
\beq
D\psi=d\psi+\omega\,\psi\,.
\eeq

\section{The Einstein-Cartan action}
We now construct the Einstein-Cartan action. The ingredients we will use to build the action are the frame fields,
$e^{I}$, and the spin connection, $\omega^{IJ}$, and possibly non-dynamical geometric objects like the metric,
$\eta^{IJ}$, and the alternating symbol, $\epsilon_{IJKL}$. Since the Einstein-Hilbert and Einstein-Cartan
approaches differ classically only by their treatment of torsion, we require that the action reduces to the
Einstein-Hilbert action when the torsion is non-zero. Defining the curvature of the spin-connection
${R^{I}}_{J}=d{\omega^{I}}_{J}+{\omega^{I}}_{K}\wedge {\omega^{K}}_{J}$ the action we will consider is 
\beq
S_{EC}=\frac{1}{4k}\int_{M}\epsilon_{IJKL}\,e^{I}\w e^{J}\w R^{KL}-
{\ts \frac{\lambda}{6}}\,\epsilon_{IJKL}\,e^{I}\w e^{J}\w e^{K}\w e^{L}\,.
\eeq
The second term in the action involves the cosmological constant, $\lambda$. Defining the Ricci scalar $R(\omega)=R^{IJ}(\bar{e}_{I},\bar{e}_{J})$,
and the determinant $e=det(e^{I}_{\mu})$, in a coordinate basis, this action takes the form
\beq
S_{EC}=\frac{1}{2k}\int_{M} (R(\omega)-2\lambda) \,e\,d^{4}x\,.
\eeq
where we have used the identity $e=\sqrt{|det\,g|}$. When the torsion is zero, the Ricci-scalar for $\omega$ reduces to
the ordinary Ricci
scalar for the Levi-Civita connection. Thus, the action does reduce to the Einstein-Hilbert action when the torsion is
constrained to be zero. It will be useful to write this in an index-free Clifford notation. Defining
$e=\frac{1}{2}\gamma_{I}e^{I}$ and $R=R_{IJ}\frac{1}{4}\gamma^{I}\gamma^{J}=d\omega+\omega\wedge \omega$, the action
becomes 
\beq
S_{EC}=\frac{1}{k}\int_{M}Tr\left(\star\,e\w e\w R-{\ts\frac{\lambda}{6}}\star\,e\w e\w e\w e\right)\,.
\eeq
Often we will use a short hand notation where we drop the explicit $Tr(\bullet)$ and the explicit wedge products. With
these conventions, the action looks like
\beq
S_{EC}=\frac{1}{k}\int_{M}\star\,e\,e\,R-{\ts\frac{\lambda}{6}}\star\,e\,e\,e\,e\,.
\eeq 

We will now look at the equations of motion obtained by a variational principle. Since the true dynamical variables
are $e^{I}$ and $\omega^{IJ}$, we will consider the equations of motion obtained by the fixed points of
$S_{EC}=S_{EC}[e,\omega]$ in the function space spanned by $\omega$ and $e$. By construction,
$\omega^{IJ}+\delta\omega^{IJ}$ must also be an $\eta$-compatible spin connection. Thus, we will only consider
variations which preserve the anti-symmetry of $\omega^{IJ}$. Setting the variation of $S_{EC}$ with respect to
$\omega^{IJ}$ equal to zero for arbitrary $\delta\omega^{IJ}$, we have 
\beq
\epsilon_{IJKL}D_{\omega}(e^{K}\wedge e^{L})=0\,. \label{ECw}
\eeq
Varying $e^{I}$ and setting the variation of the action to zero, we have
\beq
\epsilon_{IJKL}e^{J}\w R^{KL}-{\ts\frac{\lambda}{3}}\epsilon_{IJKL}e^{J}\w e^{K}\w e^{L}=0\,. \label{ECe}
\eeq
In the index free notation, the same equations of motion are
\beq
e\,\star R-\star R\,e=-\frac{2\lambda}{3}\star e\, e\,e
\eeq
and
\beq
D(\star e\,e)=0\ .
\eeq
Using the invertibility of the tetrad, the the first equation, (\ref{ECw}), can be solved to give
\beq
D_{\omega}e^{I}=T^{I}=0\,.
\eeq

Thus, vanishing torsion in vacuum is a {\it dynamical} equation of motion in the Einstein-Cartan theory. Using this, the
remaining equation, (\ref{ECe}), written in a coordinate basis is equivalent to
\beq
R_{\mu\nu}-{\ts\frac{1}{2}}g_{\mu\nu}R=-\lambda g_{\mu\nu}\,,
\eeq
where $R_{\mu\nu}$ and $R$ are the Ricci tensor and scalar of the Levi-Civita connection. This is the ordinary vacuum
field equation for general relativity. The Ricci scalar and the Ricci tensor are completely determined by the metric
and the cosmological constant in vacuum. To see this, solve the above to give
\beqa
R_{\mu\nu}&=&\lambda g_{\mu\nu}\nn\\
R&=&4\lambda\,.
\eeqa
All the degrees of freedom in vacuum, then, come from the trace free part of the Riemmann tensor, also called the Weyl
tensor, $C_{\mu\nu\alpha\beta}$.
In tetrad language, the general solution to (\ref{ECe}) is given by
\beq
R^{IJ}={\ts\frac{\lambda}{3}}\,e^{I}\w e^{J}+C^{IJ} 
\eeq
where $C^{IJ}=C^{[IJ]}$ is defined to be completely trace free: $C^{IJ}(\bar{e}_{I},\ )=0$. We will often have use for
this expression in future chapters. We note that this is true independent of (\ref{ECw}). When this equation of motion
is solved, we have an additional constraint (employing the Bianchi identity, $DR^{IJ}=0$),
\beq
DC^{IJ}=0\,.
\eeq

\section{Spin-Torsion coupling in Einstein-Cartan theory}
We will now couple Dirac spinors to Einstein-Cartan theory and show that the torsion couples to spin-currents. To
allow for torsion, first
we have to learn how to write the Dirac Lagrangian in tetrad language. The Dirac Lagrangian in an arbitrary curved
spacetime is given by
\beq
S_{D}=\kappa\int_{M}{\ts\frac{i}{2}}\left(\bar{\psi}\gamma^{\mu}\nabla_{\mu}\psi-
\overline{\nabla_{\mu}\psi}\gamma^{\mu}\psi\right)\sqrt{-g}\,d^{4}x
\eeq
where $\nabla_{\mu}=\partial_{\mu}+\Gamma_{\mu}$ is a spinor representation of the Levi-Civita connection. To allow for
torsion, we replace $\Gamma$ with an arbitrary spin connection $\omega$ and couple the theory to the Einstein-Cartan
action. Making this substitution and expressing the above in tetrad language, we have\cite{GockelerSchucker}
\beqa
S_{D}&=&\kappa\int_{M}{\ts\frac{i}{2}}\,\left(\bar{\psi}\gamma^{I}e^{\mu}_{I}D_{\mu}\psi-
\overline{D_{\mu}\psi}\,e^{\mu}_{I}\gamma^{I}\psi\right)\,e\,d^{4}x \nn\\
&=&\frac{\kappa}{6}\int_{M}{\ts\frac{i}{2}}\,\epsilon_{IJKL}\,e^{I}\w e^{J}\w e^{K}\w 
\left(\bar{\psi}\gamma^{I}D\psi-\overline{D\psi}\gamma^{I}\psi\right)\,.
\eeqa
In the index-free Clifford notation the above is given by
\beq
S_{D}=\kappa\int_{M}{\ts\frac{i}{2}}
\left(\bar{\psi}\star e\,e\,e\,D\psi+\overline{D\psi}\star e\,e\,e\,\psi
\right)
\eeq
The total action is then $S=S_{EC}+S_{D}$. Varying this action with respect to $\omega$ we have 
\beqa
\delta_{\omega}S=\frac{1}{k}\int_{M}-\star D(e\,e)\,\delta\omega+\kappa\int_{M}
-{\ts\frac{i}{2}}\,\{\psi\bar{\psi},\star e\,e\,e\}\,\delta\omega
\eeqa
where in the above (recall that the trace is assumed) we are viewing $\psi\bar{\psi}$ as a complex $16\times 16$
matrix. The Fierz identity allows us to break up this matrix into the standard Clifford basis given above, and relates
the coefficients in the linear combination to spin-currents. For our case, the Fierz identity is
\begin{equation}
-4\psi\bar{\psi}=(\bar{\psi}\psi)\mathbf{1}-(\bar{\psi}\star\psi)\star
+(\bar{\psi}\gamma_{I}\psi)\gamma^{I}+(\bar{\psi}\gamma_{I}\star \psi)\gamma^{I}\star
-\frac{1}{2}(\bar{\psi}\gamma_{[I}\gamma_{J]}\psi)\gamma^{[I}\gamma^{J]}\,.
\end{equation}
Using this, and the trace properties of the Dirac matrices, only the axial-vector term in the Fierz decomposition enters
into the trace leaving us with
\beq
\delta_{\omega}S=\frac{1}{k}\int_{M}-\star D(e\,e)\,\delta\omega+\kappa\int_{M}
-{\ts\frac{i}{8}}\,\bar{\psi}\gamma_{I}\star\psi\,\{\gamma^{I},\,e\,e\,e\}\,\delta\omega\,.
\eeq
Using the identity $\{\gamma^{I},\,e\,e\,e\}=4\,e^{I}\,e\,e$ and setting the variation to zero for an arbitrary
$\delta\omega$, we have
\beq
\star\,D(e\,e)=k\kappa\,{\ts\frac{1}{2}}\,A_{I}e^{I}\,e\,e \label{ECTorsion}
\eeq
where $A^{I}=\bar{\psi}\gamma_{5}\gamma^{I}\psi$ is the axial-current. In index notation, this becomes
\beq
T^{I}\w e^{J}-e^{I}\w T^{J}=k\kappa\,{\ts\frac{1}{4}}\,{\epsilon^{IJ}}_{KL}\,A\wedge e^{K}\wedge e^{L}\,.
\eeq
This can be solved to yield a final expression for the torsion:
\beq
{T^{I}}_{JK}={T^{I}}_{[JK]}=-k\kappa\,{\ts\frac{1}{2}}A_{M}{\epsilon^{MI}}_{JK}\,.
\eeq

It is useful to consider the effective field theory that one obtains from plugging this result back into the action.
The effective field theory yields what one would expect to see experimentally at low
energies. Since the spin-torsion coupling is very small, it is likely that any experimental observation will be in a
very low energy regime, thereby justifying the use of the effective field theory for phenomenological models. To
obtain this, we first recall that any connection can be split into a metric compatible connection and a tensorial piece
which contains all the information about the torsion:
\beq
{\omega^{IJ}}_{\mu}={\Gamma^{IJ}}_{\mu}+{C^{IJ}}_{\mu}\,.
\eeq
Here, $\Gamma$ is the metric compatible, torsion free Levi-Civita connection and ${C^{IJ}}_{\mu}$, called the
contorsion tensor, must satisfy ${C^{I}}_{K[\mu}e^{K}_{\nu]}={T^{I}}_{\mu\nu}$. We can solve for the contorsion
explictly in terms of the torsion yielding
\beq
C_{IJK}=\frac{1}{2}(T_{IJK}+T_{JKI}+T_{KJI})\,.
\eeq
In our case, the contorsion is 
\beq
C_{IJK}=-k\kappa {\ts\frac{1}{4}}A^{M}\,\epsilon_{MIJK}\,.
\eeq
In terms of the contorsion, the relevant piece of the Einstein-Cartan action takes the form
\beq
\frac{1}{4k}\int_{M}\epsilon_{IJKL}e^{I}\, e^{J}\, R(\omega)^{KL}
=\frac{1}{4k}\int_{M}\epsilon_{IJKL}e^{I}\, e^{J}\, R(\Gamma)^{KL}
+\epsilon_{IJKL}e^{I}\, e^{J}\, {C^{K}}_{M}\, C^{ML}\,.
\eeq
Inserting our expression for the contorsion into the last term and solving, the interaction term becomes
\beq
S^{(1)}_{int}=\frac{1}{4k}\int_{M}\epsilon_{IJKL}e^{I}\, e^{J}\, {C^{K}}_{M}\, C^{ML}
=\frac{3k\kappa^{2}}{16}\int_{M}A\cdot A\ e\,d^{4}x
\eeq
The remaining interaction comes from part of the Dirac Lagrangian involving the contorsion tensor given by
\beqa
S^{(2)}_{int}&=&\kappa \int_{M}
\bar{\psi}\star e\,e\,e\,C\psi-\bar{\psi}C\,\star\,e\,e\,e\,\psi\nn\\
&=& -\frac{3k\kappa^{2}}{8}\int_{M}A\cdot A\ e\,d^{4}x\,.
\eeqa
Putting everything together we have
\beqa
S_{eff}&=&\frac{1}{2k}\int_{M}\left(R(\Gamma)-2\lambda
+{\ts\frac{i}{2}}\,k\kappa\left(\bar{\psi}\gamma^{\mu}\nabla_{\mu}\psi-
\overline{\nabla_{\mu}\psi}\gamma^{\mu}\psi\right)\right)\,e\,d^{4}x \nn\\
&\  & -\frac{3k\kappa^{2}}{16}\int_{M}
\bar{\psi}\gamma_{5}\gamma^{I}\psi\ \bar{\psi}\gamma_{5}\gamma_{I}\psi\ e\,d^{4}x\,.
\eeqa
Thus, we see that the low-energy signature of Einstein-Cartan theory is an axial-axial spin-current interaction in the
effective field theory that is a relic of the spin-torsion interaction.

\section{The Holst modification of the Einstein-Cartan action}
The starting point of Loop Quantum Gravity is a modification of the Einstein-Cartan action, called the Holst
action\cite{Holst}.
The reason for this modification will become more clear in the canonical analysis we will present later on. For now,
it will be sufficient to justify the Holst modification by showing that there are non-trivial terms that can be added
to the gravitational action that do not effect the equations of motion. Consider the action
\beqa
S_{H}&=&\frac{1}{k}\int_{M}\star e\,e\,R+{\ts \frac{1}{\beta}}e\,e\,R
 -{\ts\frac{\lambda}{6}}\star e\,e\,e\,e \nn\\
 &=& \frac{1}{4k}\int_{M}\epsilon_{IJKL}\, e^{I}\w e^{J}
 \left(R^{KL}-{\ts\frac{\lambda}{6}}e^{K}\w e^{L}\right)-{\ts\frac{2}{\beta}}e_{I}\w e_{J}\w R^{IJ}\,.
\eeqa
The action is identical to
the Einstein-Cartan action with the addition of a parity violating term. Here, the parameter $\beta$, 
called the Immirzi parameter, is a real or complex constant. Presently, the meaning and interpretation of this parameter
is at best poorly understood. It is clear that in some sense the parameter is a measure of a degree of parity
violation built into the action. The term can be rewritten as follows:
\beq
\frac{1}{k\beta}\int_{M}e\,e\,R=-\frac{1}{2k\beta}\int_{M}T\,T
+\frac{1}{2k\beta}\int_{\partial M}e\,De\,.
\eeq
The boundary term is the topological Nieh-Yan class. We see from this that the Immirzi parameter will measure the
width of torsional fluctuations in the path integral\cite{Starodubtsev:MMgravity}. The parameter has a dramatic effect in the quantum
theory---in
particular, we will see in the quantum theory that area will be quantized in multiples of $\beta l_{Pl}^{2}$. Thus, at
the quantum level,
the parameter fine tunes the discretization scale. At the classical level, perhaps surprisingly, the parameter has no
effect whatsoever (at least in vacuum). To see this we consider the equations of motion of the Holst action. It will be
useful to introduce the operator $P_{\star}=\star+\frac{1}{\beta}$. This operator is invertible whenever 
$\beta\neq\pm i$, and the inverse is given by
 $P_{\star}^{-1}=-\frac{\beta^{2}}{1+\beta^{2}}\left(\star-\frac{1}\beta\right)$.
The equations of motion obtained by varying the Holst action with respect to $\omega$ are
\beq
P_{\star}D(e\,e)=0\,.
\eeq
Inverting $P_{\star}$, the remaining equation is the familiar equation that can be solved to yield $T=0$. Thus,
despite $\beta$ being a measure of the width of torsional fluctuations on the full phase space, on-shell the torsion
is still zero. The remaining equation of motion found from varying the action with respect to $e$ is
\beqa
P_{\star}R\,e-e\,P_{\star}R &=& \star R\,e-e\,\star R+{\ts\frac{1}{\beta}}\,DT\nn\\
&=&{\ts\frac{2\lambda}{3}}\star e\,e\,e
\eeqa
Since the torsion vanishes on-shell, the above reduces to the equations of motion of Einstein-Cartan theory. In the
special case where $\beta=\pm i$, the matrix $P_{\star}$ becomes a projection operator, and the Holst action reduces to
Ashtekar's simplification (see e.g. \cite{Ashtekar:book}) of the Einstein-Cartan action given by
\beqa
S_{A}&=&\frac{2}{k}\int_{M}\star e\,e\,R_{(L/R)}
 -{\ts\frac{\lambda}{6}}\star e\,e\,e\,e \nn\\
\eeqa
where $R_{(L/R)}=\frac{1}{2}(1\mp\gamma_{5})R$. Here $P_{(L/R)}=\frac{1}{2}(1\mp\gamma_{5})$ is the left/right chiral
projection operator. The curvature $R_{(L/R)}$ is the curvature of $\omega_{(L/R)}=P_{(L/R)}\omega$. This connection
defines the parallel transport of left/right-chiral spinors. When reality constraints are imposed on the tetrad, it
can be shown that the equations of motion from this action are equivalent to those of the Einstein-Cartan action.
Again, the main advantages of this action are its implications in the canonical theory, which we will see later.

\subsection{Coupling spinors to the Holst action}
Coupling spinors to the Holst action is a delicate matter. In the previous discussion, we showed that the equations of
motion from the Holst action are equivalent to the Einstein-Cartan action {\it in vacuum}. This worked out because the
additional Holst-modified term in the Einstein equations is torsion dependent, and the remaining equations of motion
gave vanishing torsion. This begs the question, {\it does the parity violating Immirzi term affect the equations of
motion when spinors are added to gravity via the spin-torsion coupling?} For a short time it was hoped that the
Immirzi term would produce a different signature in the effective field theory\cite{Rovelli:Torsion}. This effect was
believed to yield
vector-axial and vector-vector coupling terms which were to be dependent on the Immirzi parameter. Although this is
partially true, it was eventually revealed that the effect also depends on some additional structure, implicitly built
into the action\cite{Freidel:Torsion}.
It turns out that in the presence of torsion there
is considerable freedom in defining the Dirac Lagrangian through non-minimal coupling terms. These are terms such as 
\beq
\alpha_{1}\int_{M}\bar{\psi}\,e\,e\,e \,D\psi -\overline{D\psi}\,e\,e\,e\,\psi
\eeq
and 
\beq
\alpha_{2}\int_{M}\bar{\psi}\star e\,e\,e \,D\psi -\overline{D\psi}\star e\,e\,e\,\psi\,.
\eeq
Both of these terms are total derivatives when the torsion is zero, but they lead to non-trivial current-current
interactions in the effective field theory via spin-torsion coupling. Although the new current-current interaction terms
in the effective field theory when these terms are coupled to the Holst action are dependent on the Immirzi parameter,
they are also dependent on the non-minimal coupling coefficients $\alpha_{1}$ and $\alpha_{2}$, such that the
interactions vanish
when these non-minimal couplings are turned off. Thus, these signature interactions have as much to do with the values of
the non-minimal coupling coefficients as to do with the value of the Immirzi parameter. A characteristic parity violating
vector--axial-vector
interaction does emerge for particular values of the coupling coefficients, but, again only when the non-minimal
coupling terms are turned on.

In light of the above discussion, rather than discuss the details of effective field theory with the non-minimal
coupling terms, we would like to show that one can choose an appropriate Dirac Lagrangian such that when added to the
Holst action for any value of the Immirzi parameter (real or imaginary), it produces the same equations of motion and
effective field theory as the Einstein-Cartan action\cite{Randono:Torsion, Mercuri:Torsion}. To this end, let us consider
the Dirac Lagrangian
\beqa
S_{D}&=&\kappa\int_{M}{\ts\frac{i}{2}}\left(\bar{\psi}\star e\,e\,e \,D\psi
+\overline{D\psi}\star e\,e\,e\,\psi\right)\nn\\
& &\qquad\qquad{\ts\frac{i\alpha}{2}}\left(\bar{\psi}\,e\,e\,e\,D\psi
-\overline{D\psi}\,e\,e\,e\,\psi\right)\nn\\
&=&\kappa\int_{M}\Big({\ts\frac{i}{2}}\left(\bar{\psi}\,\gamma^{\mu}\,D_{\mu}\psi
-\overline{D_{\mu}\psi}\,\gamma^{\mu}\,\psi\right)\nn\\
& &\qquad\qquad-{\ts\frac{i\alpha}{2}}\left(\bar{\psi}\star\gamma^{\mu}\,D_{\mu}\psi
+\overline{D_{\mu}\psi}\star\gamma^{\mu}\,\psi\right)\Big) \,e\,d^{4}x\ .
\eeqa
Varying the full action with respect to $\omega$ and repeating the same steps in the previous derivation, we arrive at
the equation of motion
\beq
P_{\star}\star D(e\,e)=k\kappa (\star-\alpha){\ts\frac{1}{2}}\,A_{I}e^{I}\,e\,e
\eeq
If we now set $\alpha=-\frac{1}{\beta}$, the operator on the right-hand side reduces to
$P_{\star}=\star+\frac{1}{\beta}$. Assuming $\beta \neq \pm i$, we can invert this operator on both sides to obtain
\beq
\star\,D(e\,e)=k\kappa\,{\ts\frac{1}{2}}\,A_{I}e^{I}\,e\,e\ , 
\eeq
which is the same result that we obtained for pure Einstein-Cartan gravity, (\ref{ECTorsion}), without the parity
violating Immirzi terms. In the special case when $\beta=\pm i$, since the expression is complex, we can solve the
real and imaginary parts separately. The end result for {\it any} value of $\beta$, real or complex, is
precisely the same result we obtained for the torsion of the ordinary Einstein-Cartan case:
\beq
{T^{I}}_{JK}={T^{I}}_{[JK]}=-k\kappa\,{\ts\frac{1}{2}}A_{M}{\epsilon^{MI}}_{JK}\,.
\eeq
Inserting this back into the action yields the effective theory which now contains all of the ordinary Einstein-Cartan
interactions, but it also contains the parity violating terms
\beqa
S^{(1)}_{odd}&=&\frac{1}{k}\int_{M} e\,e\,C\,C \nn\\
S^{(2)}_{odd}&=& \kappa\int_{M} -{\ts\frac{i}{2\beta}}\left(\bar{\psi}\,e\,e\,e\,C\psi
+\overline{\psi}C\,e\,e\,e\,\psi\right)\,.
\eeqa
where $C=\frac{1}{4}\gamma_{I}\gamma_{J}C^{[IJ]}$ is the contorsion tensor. When evaluated using the torsion given
above, both of these terms vanish\footnote{One easy way to see this is the following. The terms must be parity odd,
and they must contain two factors of $A^{I}$, and four factors of $e^{J}$. The only other tensor that enters the
calculation is $\epsilon_{IJKL}$, but this must not be present in the final expression since it must be parity odd.
There is no way to put these ingredients together to get a parity odd expression without contracting two $e$'s, which
gives zero. This can be confirmed through an explicit calculation.}. The remaining interactions are identical to the
Einstein-Cartan case. 
Thus, we have shown that there is a natural generalization to the Holst action coupled to Dirac spinors that
reproduces the equations of motion and the effective field theory of the Einstein-Cartan action coupled to Dirac
spinors, thereby reinforcing the notion that the Barbero-Immirzi-Holst transformation is a canonical transformation even
in the presence of matter.

\chapter{Review of de Sitter Space\label{dSReview}}

In this section we will present a brief review of de Sitter spacetime. Our aim is primarily to present in a
pedagogical way the properties of de Sitter space that we will use in proceeding chapters (for good reviews see
\cite{Hawking:LSS, Moschella:deSitter}). 

Simply put, de Sitter space is the simplest vacuum solution to Einstein's equations with a positive cosmological
constant. This does not mean, however, that the spacetime is trivial. In fact, the spacetime has many of the
interesting,
subtle, and potentially confusing properties that characterize general relativity. Most notably, any geodesic observer in
the spacetime will
have particle and event horizons. This also makes it one of the most interesting spacetimes from a quantum mechanical
perspective---it is simple enough that we can construct quantum mechanical models based on it, yet it is rich
enough, particularly in the horizon structure, to allow one to analyze generic properties of
quantum spacetime. de Sitter space is characterized by constant, {\it positive} curvature. Since general relativity is
background independent, the obvious question is {\it if the curvature is constant, what is it constant with respect
to?} The answer is that it is constant with respect to the only field defined on the manifold, the field that defines
the geometry of spacetime itself: the metric. Let us build the spacetime from the ground up starting from the Ricci
scalar. There is no ambiguity in parallel transporting a scalar---a scalar is constant if it is constant with respect to
itself. Thus we define the (Levi-Civita) Ricci scalar 
\beq
R=4\lambda
\eeq
where $\lambda$ is a positive quantity (taking it to be negative would define anti-de Sitter space) and the factor of
four is just a convention. We now need to define the Ricci tensor. The Ricci tensor must be constructed out of the
metric alone and it must have the appropriate symmetries, namely $R_{\mu\nu}=R_{\nu\mu}$. The natural choice is
$R_{\mu\nu}\sim g_{\mu\nu}$. Since the Ricci scalar must be $4\lambda$, we have
\beq
R_{\mu\nu}=\lambda g_{\mu\nu}.
\eeq
Now we need to construct the full curvature tensor. Again, the expression must be built out of the metric alone, it
must have the same symmetries as the Riemann curvature tensor, and its Ricci scalar and tensor must be given by the
expression above. The natural choice is
\beq
R_{\mu\nu\alpha\beta}=\frac{\lambda}{3}(g_{\mu\alpha}g_{\nu\beta}-g_{\mu\beta}g_{\nu\alpha}).
\eeq
This serves to define the local geometric structure of de Sitter space. All spacetimes with the above property are
unique up to topology. It is easy to see that this satisfies Einstein's equations. It can be shown that the
curvature of a generic solution to Einstein's equations is 
\beq
R_{\mu\nu\alpha\beta}=\frac{\lambda}{3}(g_{\mu\alpha}g_{\nu\beta}-g_{\mu\beta}g_{\nu\alpha})
+C_{\mu\nu\alpha\beta}
\eeq
where $C_{\mu\nu\alpha\beta}$ is the Weyl tensor, which is zero in de Sitter space. Most of our expressions will be in
the tetrad formulation, so we should express the above in tetrad language. Defining a local orthonormal frame $e^{I}$,
and an $so(3,1)$ connection $\omega^{IJ}$ with curvature $R^{IJ}=d\omega^{IJ}+{\omega^{I}}_{K}\wedge\omega^{KJ}$, de
Sitter space is locally defined by the condition
\beq
R^{IJ}={\ts\frac{\lambda}{3}}\,e^{I}\wedge e^{J}. \label{dSspace}
\eeq
It is important to recognize that generically, the connection $\omega^{IJ}$ and the tetrad $e^{I}$ may have non-zero
torsion. However, from the Bianchi identity, we have
\beq
D(e^{I}\wedge e^{J})={\ts\frac{3}{\lambda}}\,DR^{IJ}=0.
\eeq
We recognize this as the Einstein-Cartan equation that can be solved to give $T^{I}=De^{I}=0$. The remaining
Einstein-Cartan equation
\beq
\epsilon_{IJKL}\,e^{J}\wedge\left(R^{KL}-{\ts\frac{\lambda}{3}}\,e^{K}\wedge e^{L}\right)=0
\eeq
is also obviously solved by (\ref{dSspace}). Thus, (\ref{dSspace}) is alone sufficient to solve the Einstein-Cartan
equations and define the local structure of de Sitter space. 

Now, let's analyze the de Sitter space from an initial data formulation. Choosing a slicing of de Sitter space, the
dynamical variables pulled back tot he Cauchy surface will satisfy a set of equations characteristic of de Sitter
space.
Since we will often have need to work in the time gauge where $e^{0}_{a}=0$, and $E^{i}_{a}\equiv e^{i}_{a}$, it will
be useful to analyze the constant curvature condition in this gauge. To this end, using the identities 
$^{4}R^{ij}=^{3}R^{ij}+K^{i}\wedge K^{j}$ and ${R^{i}}_{0}=^{3}DK^{i}$, where
 $K^{i}\equiv {\omega^{i}}_{0}$, and it is understood that all variables are pulled back to the Cauchy surface, the de
Sitter condition becomes
 \beqa
 ^{3}R^{ij}+K^{i}\wedge K^{j}&=&{\ts\frac{\lambda}{3}}\,E^{i}\wedge E^{j}\nn\\
 ^{3}DK^{i}&=&0\,. \label{dScondition0.1}
 \eeqa

The easiest way to visualize a space of constant curvature is to embed it into a larger space of one-dimension higher.
Picture a two-sphere, which has constant positive curvature embedded in three dimensions. Indeed, for a Euclidean
signature metric, de Sitter space can be represented by the 4-sphere embedded in a five-dimensional space with a
Euclidean metric. The Euclidean de Sitter metric is then simply the Euclidean metric in the embedding space pulled
back to the sphere. The radius of the sphere is the radius of curvature of the space, which is directly related to the
cosmological constant. This provides an easy way to determine the isometry group of Euclidean de Sitter space. The
isometry group is the set of global transformations on the space that leave the metric structure unchanged, and it is
determined by the algebra of the Killing vectors under the Lie bracket. Any
five dimensional rotation about an axis that goes through the origin of the sphere will not change the embedding of the
sphere. Thus, the isometry group is isomorphic to the five-dimensional rotation group $SO(5)$. Since this is a ten
dimensional group, and the maximum number of Killing vectors of any four dimensional space is ten\footnote{In
Minkowski space, the Killing vectors consist of three rotations, three boosts, and 4 translations for a total of ten
vector fields. In de Sitter space, the identifications are very similar.},
Euclidean de Sitter space is maximally symmetric. For Lorentzian de Sitter space we need to embed a space of constant
curvature into a five dimensional space with Lorentzian signature:
\beq
ds^{2}=-dx_{0}^{2}+dx_{1}^{2}+dx_{2}^{2}+dx_{3}^{2}+dx_{4}^{2}
\eeq
 Since the signature is Lorentzian, the space of
constant positive curvature is a hyperboloid with open ends in the ``time" direction. Thus, de Sitter space is the
4-hyperboloid
\beq
-x_{0}^{2}+x_{1}^{2}+x_{2}^{2}+x_{3}^{2}+x_{4}^{2}= \frac{3}{\lambda}
\eeq
The algebra of the set of global transformations that leaves the above invariant is the set of Lorentzian ``rotations",
$SO(4,1)$. Thus, the algebra of the Killing vectors under the Lie bracket is isomorphic to $SO(4,1)$ and we will
simply refer to this as the de Sitter group $dS_{4}$. To discuss the intrinsic metric, we need to define a set
of coordinates on the hyperboloid. There are three standard coordinate charts of de Sitter space, each corresponding
to different ways of slicing the spacetime into space and time. These slicings are the analogues of the three
different types of conic sections: we can slice the hyperboloid horizontally, diagonally, or vertically. The
resultant, not necessarily complete, coordinate charts cover manifolds with topology $\mathbb{R}\times
\mathbb{S}^{3}$, $\mathbb{R}\times \mathbb{R}^{3}$, and $\mathbb{R}\times
\mathbb{H}^{3}$, respectively. The simplest is the horizontal slicing. In this slicing, at each instant in time, the
three-space is a three-sphere with constant spatial curvature. The radius of curvature of the space initially
contracts until it reaches a minimum value, $r_{0}=\sqrt{3/\lambda}$, at the throat (which we define to be at $t=0$)
then it expands again to infinity. Explicitly, if we define the coordinates on the hyperboloid by inverting the
relations
\beqa
x_{0}&=& r_{0}\,\sinh (t/r_{0}) \nn\\
 x_{1}&=& r_{0}\,\cosh(t/r_{0})\,\cos\chi\nn\\
x_{2}&=& r_{0}\,\cosh(t/r_{0})\,\sin\chi\,\cos\theta\nn\\
x_{3}&=& r_{0}\,\cosh(t/r_{0})\,\sin\chi\,\sin\theta\,\cos\phi\nn\\
x_{4}&=&  r_{0}\,\cosh(t/r_{0})\,\sin\chi\,\sin\theta\,\sin\phi
\eeqa
the metric pulled back to the hyperboloid becomes
\beq
ds^{2}=-dt^{2}+r_{0}^{2}\,\cosh^{2}(t/r_{0})\,d\Omega^{2}
\eeq
where $d\Omega^{2}$ is the standard constant curvature metric on $\mathbb{S}^{3}$ given by
\beq
d\Omega^{2}=d\chi^{2}+\sin^{2}\chi(d\theta^{2}+\sin^{2}\theta\,d\phi^{2})\,.
\eeq
The singularities in this metric are all trivial singularities from the polar coordinates, and this coordinate chart
covers the full manifold. We can define an orthonormal frame in the time-gauge based on the above metric by
$e^{0}_{0}=1$, $e^{i}_{0}=0$, $e^{0}_{a}=0$, and
\beq
E^{i}_{a}=
\left[\begin{array}{ccc} e^{t/r_{0}} & 0 & 0 \\
0 & e^{t/r_{0}}\sin\chi & 0 \\
0 & 0 & e^{t/r_{0}}\sin\chi\sin\theta 
\end{array}\right]\,.
\eeq
In this basis, the extrinsic curvature, and the three-dimensional Levi-Civita curvature are
\beqa
K^{i} &=& {\ts\frac{1}{r_{0}}}\tanh(t/r_{0})\,E^{i} \\
^{3}R^{ij}& =&\frac{1}{r_{0}^{2}\cosh^{2}(t/r_{0})}\,E^{i}\wedge E^{j}\,.
\eeqa
The second line clearly shows that the three curvature is always intrinsically constant, but the constant grows to a
maximum value of $\frac{\lambda}{3}$ and then shrinks again.

Alternatively, we could define a time variable by slicing the hyperboloid along $45$-degree angles. Defining the new
coordinates along this slicing by
\beqa
t &=& r_{0}\,\ln\left(\frac{x_{0}+x_{4}}{r_{0}}\right) \nn\\
x &=& \frac{r_{0}\,x_{1}}{x_{0}+x_{4}}\nn\\
y &=& \frac{r_{0}\,x_{2}}{x_{0}+x_{4}}\nn\\
z &=& \frac{r_{0}\,x_{3}}{x_{0}+x_{4}}\,.
\eeqa
In these coordinates, the metric becomes
\beq
ds^{2}=-dt^{2}+e^{2t/r_{0}}(dx^{2}+dy^{2}+dz^{z})\,.
\eeq
This coordinate chart does not cover the full manifold but only one half of the hyperboloid. Despite this, in some
sense this is the most physical description of de Sitter space. To see this, consider the Freidman-Robertson-Walker
metric for $k=0$. Choose a set of observers that are co-moving with respect to the fields that make up the constant
energy density. Now take the limit as this energy density (other than the vacuum density) goes to zero. We end up with
precisely this form of the metric. It follows that the lines of constant $x^{i}$ are timelike geodesics. Since the
proper distance between two neighboring observers on this congruence at one time increases as time passes, the space
appears to be inflating.

The fact that the above chart does not cover the whole manifold has a physical explanation. Although de Sitter space
is geodesically complete, there are points in de Sitter space that cannot be connected by any geodesic, and any
geodesic family of observers will have both particle and event horizons. The portion of the space not covered by the
above chart is the portion of de Sitter space that is outside of the past light cones of this family of observers, and,
therefore, out of causal contact. Indeed, one can verify from the conformal diagram that the diagonal boundary is, in
fact, $\mathcal{I}^{-}$ for this set of observers. As opposed to the horizontal slicing, $\mathcal{I}^{-}$ is null,
whereas $\mathcal{I}^{+}$ remains spacelike. This is what one would expect in an inflating universe. If one were to send
out a light signal at the present, points farther than $r_{0}$ on the line of incidence will be receding fast enough
that the light ray will never catch up to them. Thus, the point of intersection of the null cone with the spacelike
$\mathcal{I}^{+}$ defines the boundary between points an observer can influence, and points he will never be able to
influence. On the other hand, for every two observers, no matter how far apart, if we trace back far enough in time we
will come to a time when these points were close enough together to be in causal contact. Thus, an observer
will be able to see out to infinity so long as he has a powerful enough telescope to look arbitrarily far back in time.

From the form of the three-metric, it is clear that the spatial curvature is flat. This can be verified in the tetrad
language, where the solution becomes $K^{i}=\sqrt{\lambda/3}\,E^{i}$, and $\omega^{ij}=0$. From this, it is clear that
the spatial curvature is zero. Employing the standard Cartesian basis, we can set $E^{i}_{a}=\delta^{i}_{a}$, so that 
$K^{i}_{a}=\sqrt{\lambda/3}\,\delta^{i}_{a}$, and $\omega^{ij}=\Gamma^{ij}[E]=0$, which one can verify solves the
full set of equations. The, flat three-space representation of de Sitter space will be particularly useful to us later
on.

\chapter{The Macdowell-Mansouri Formulation of Gravity\label{MMReview}}

In many ways, gravity is tantalizingly similar to an ordinary gauge theory. Nowhere is the connection more 
obvious than the Macdowell-Mansouri approach to general relativity\cite{MMoriginal}. The Einstein-Cartan makes approach
makes 
significant steps towards describing gravity in terms of an ordinary gauge theory. It has most of the right
ingredients---in particular, the use of locally inertial frames allows one to introduce the underlying group
$SO(3,1)$ as the gauge freedom in choosing an arbitrary frame at each point. Then one introduces the spin-connection,
which is none-other than an ordinary connection over a principle g-bundle, just as in ordinary gauge theories. Relevant
fields now live in the fiber, but can be projected down to the tangent space of the base manifold using frame-field. The
equations of motion allow one to write the connection in terms of the metric and recover ordinary general relativity.
The odd-ball fields in this approach are the frame fields, which have no analogue in ordinary Yang-Mills type gauge
theories. This is where the Macdowell-Mansouri approach shines---it combines the frame fields and the spin connection
into a single connection based on a new group: the de Sitter group. The Macdowell-Mansouri action constructed solely
out of the curvature of this connection looks eerily similar to the Yang-Mills action. However, there are important
differences, which we will discuss. For our purposes, the Macdowell-Mansouri action will provide considerable
insight into the nature of the Kodama state. 

\section{Why the de Sitter group?\label{deSitterChapter}}
The first question we must address before building a theory of gravity based on the de Sitter group is {\it why the de
Sitter group?} This question can be partially answered with the benefit of hindsight---the Macdowell-Mansouri action
seems to work and it gives us new insight into the nature of gravity. However, there are some fine distinctions that
cannot be answered with this hindsight. In particular, the Macdowell-Mansouri approach does not seem to give us any
insight into the magnitude of the cosmological constant, and it does not even tell us what sign the cosmological
constant should take. Furthermore, one could even raise the objection that the approach is not even truly based on the de
Sitter group since the action itself is not de Sitter invariant---the action could just as well have been built out of
the $SO(3,1)$ spin connection and frame-fields without losing anything. In this section we would like to give an
alternative reason for choosing to base our theory on the de Sitter group as opposed to the Lorentz group or the
anti-de Sitter group. This argument will closely follow \cite{Randono:dSgroup}.

The obvious, and somewhat na\"{i}ve, answer to the question posed above is that observational evidence suggests that the
cosmological constant is, indeed, non-zero and positive, albeit very small. That said, there are many effects that all
contribute to the value of the cosmological constant, including the vacuum modes of each and every dynamical field,
renormalization, quantum anomalies, exotic matter, etc... Thus, the hitherto unknown aggregate of all these effects
must yield the observed value of the constant. Our argument applies to the ``bare" cosmological constant by which we
mean the contribution to the cosmological constant from pure, group theoretic reasons or other mathematical constructs,
prior
to the contribution from other sources. This will become more clear as the argument proceeds. There are many theoretical
reasons, primarily from supersymmetry and string theory as to why the bare cosmological constant should be negative.
We will give a theoretical argument suggesting the bare cosmological constant should be non-zero and {\it positive}.

In contrast to all of the other forces in nature gravity is unique in that it is {\it universal}. That is gravity,
as we know it, appears to couple to all known types of matter and energy, including all standard model particles, 
extremely weakly interacting particles such as neutrinos, and dark matter and energy (whatever it may be). It is
generally
believed that gravity falls under the category of a gauge theory (albeit a strange one) based on the Lorentz group or
a larger symmetry group containing a Lorentz subgroup. Since the groups comprising the standard model all come from a
particular fermionic inner product which must be preserved, it is natural to ask, {\it which inner products are
preserved by the gravitational gauge group?} Conversely, we can turn this question around and ask: {\it Given the
universal fermionic inner products of the standard model, what is the universal gauge group that preserves these inner
products?} Such a universal gauge group is likely to be the gauge group underpinning the gravitational interaction.
In the simplest extensions of the standard model to include neutrino mass, the universal fermionic inner products
involved are the Dirac and Majorana inner products (see e.g. \cite{Zee:QFT, Ramond}). We will show that the largest gauge
group
that preserves both these inner products is locally isomorphic to the four-dimensional de Sitter group. Thus, demanding
that the gravitational interaction is universal and preserves both these inner products not only implies the existence
of a (bare) cosmological constant, but it also implies that the cosmological constant must be {\it positive}. We
then go on to show that, by a simple extension of the Macdowell-Mansouri mechanism, gravity can be formulated in a way
that is strictly de Sitter invariant. The calculations in this section are slightly more elegant when the metric
signature is $(+,-,-,-)$, and since this is the standard convention in particle physics, we will work with this
signature for this section only. At the end we will give relevant results in the $(-,+,+,+)$ signature.

\subsection{The Lie algebra of the Dirac inner product}
The ordinary Dirac inner product which enters into the mass term of the Dirac Lagrangian is given by 
\beq
\langle \phi,\psi\rangle=\phi^{\dagger}\gamma^{0}\psi
\eeq
where $\gamma^{0}$ is the time component of a 4-dimensional complex representation of the Clifford algebra defined by
$\gamma^{I}\gamma^{J}+\gamma^{J}\gamma^{J}=2\eta^{IJ}$ where $\eta^{IJ}=diag(1,-1,-1,-1)$. The group that
preserves this inner product
is the 4--dimensional conformal group which is locally isomorphic to $U(2,2)\times U(1)$. This can be easily seen by
working in the Dirac representation where $\gamma^{0}$ has the form $diag(1,1,-1,-1)$.
Thus, the group preserving the Dirac inner product in this representation must satisfy
$g^{\dagger}\gamma^{0}g=\gamma^{0}$,
which is by definition $U(2,2)\times U(1)$.

It will be useful to work in a Clifford representation to define a basis for Lie algebras. Since the Dirac matrices
are linearly independent, they define a basis for the complex Lie algebra $gl(4,\mathbb{C})$ \cite{GockelerSchucker}. A
convenient choice for the basis is:
\beqa
&1& \nn\\
&\frac{1}{2}\gamma^{I} & \nn\\
& \frac{1}{4}\gamma^{[I}\gamma^{J]}&\nn\\
& \frac{1}{2}\gamma^{I} \star&\nn\\
& \star &
\eeqa
where $\star=-i\gamma_{5}=i\gamma^{0}\gamma^{1}\gamma^{2}\gamma^{3}$. The condition 
$g^{\dagger}\gamma^{0}g=\gamma^{0}$ implies a condition on the Lie algebra:
\beq
\gamma^{0}\lambda^{\dagger}\gamma^{0}\equiv \tilde{\lambda} =-\lambda
\eeq
where tilde denotes the reverse or ``rotor" operation defined by 
$\tilde{\lambda}=\lambda^{*}_{IJ...KL}\gamma^{L}\gamma^{K}...\gamma^{J}\gamma^{I}$ when 
$\lambda=\lambda_{IJ...KL}\gamma^{I}\gamma^{J}...\gamma^{K}\gamma^{L}$. The \textit{real}  Lie algebra
$su(2,2)\oplus u(1)$ is therefore spanned by the basis 
\beq
su(2,2)\simeq \left\{ \frac{1}{4}\gamma^{[I}\gamma^{J]}\ ,\  \frac{i}{2}\gamma^{I}\ ,\  
\frac{1}{2}\gamma^{I}\star\ ,\  i\star \ ,\ i \right\} .
\eeq
This algebra has two natural subalgebras: the basis $\left\{\frac{1}{4}\gamma^{[I}\gamma^{J]}\ ,\  
\frac{1}{2}\gamma^{I}\star\right\}$ spans the four dimensional anti-de Sitter Lie algebra, $AdS_{4}$, and the
basis 
$\left\{\frac{1}{4}\gamma^{[I}\gamma^{J]}\ ,\  
\frac{i}{2}\gamma^{I}\right\}$ spans the de Sitter Lie algebra, $dS_{4}$.

\subsection{The Lie algebra of the Majorana inner product}
Clearly the Dirac inner product does not single out a particular sign for the cosmological constant. We will now show
that the group that preserves \textit{both} the Dirac inner product and the Majorana inner product is locally
isomorphic to the de Sitter group, thereby singling out a positive cosmological constant.

The Majorana inner product crops up in the Majorana equation where it defines a chirality preserving mass term.
It is given by
\beq
\langle \phi, \psi\rangle_{Maj} =\phi^{\dagger}\gamma^{0}\mathbf{C}\psi^{*}\equiv \bar{\phi}\psi_{c}
\eeq
where $\mathbf{C}$, the charge conjugation operator, is itself an element of the Clifford algebra and is defined by
 $\mathbf{C}^{-1}\gamma^{I}\mathbf{C}=-\gamma^{I*}$.
Physically the charge conjugation operation $\psi\rightarrow \psi_{c}=\mathbf{C}\psi^{*}$ simply sends a spinor to its
conjugate, inverting its $U(1)$ charge and its chirality while preserving the spinorial transformation properties under
the Lorentz group. The transformations that preserve the inner product must
satisfy $g^{\dagger}\gamma^{0}\mathbf{C}g^{*}=\gamma^{0}\mathbf{C}$. As a Lie algebra condition this becomes
\beq
\mathbf{C}^{-1}\tilde{\lambda}\mathbf{C}=-\lambda^{*}.\label{MajCondition}
\eeq
We note that if an element $A$ satisfies (\ref{MajCondition}), then
so does $iA$. In this sense the Lie algebra is naturally defined over the complex field. However, for purposes of
comparison, we will assume the basis is a basis over the real numbers.
It can be easily verified that this Lie algebra, which we will denote by $maj_{4}$,
is spanned by:
\beq
maj_{4}\simeq \left\{\frac{1}{4}\gamma^{[I}\gamma^{J]}\ ,\  \frac{i}{4}\gamma^{[I}\gamma^{J]}\ ,\ 
\frac{i}{2}\gamma^{I}\ ,\  \frac{1}{2}\gamma^{I}\right\}.
\eeq
Now, clearly the intersection of $su(2,2)\oplus u(1)$ and $maj_{4}$ is spanned by 
$\left\{\frac{1}{4}\gamma^{[I}\gamma^{J]}\ ,\  \frac{i}{2}\gamma^{I}\right\}$. Thus, we see that
\beq
su(2,2)\oplus u(1) \bigcap maj_{4} = ds_{4}.
\eeq
Thus, the largest group that preserves both the Dirac and the Majorana inner products is locally isomorphic to the de
Sitter group. This result also has consequences for Majorana spinors. If the spinor is a Majorana spinor, which satisfies
$\psi=\psi_{c}$, the Majorana inner product reduces to the Dirac inner product. Thus, the largest group which can
preserve the mass term of a Majorana spinor is the de Sitter group. 

It should be noted that since the de Sitter group does not preserve chirality, and the standard model is chirally
asymmetric, if an underlying de Sitter symmetry exists, it must be in a symmetry-broken phase now. Here we appeal to the
left-right
symmetric formulations of the standard model \cite{Mohapatra:left-right}, whereby CP-symmetry is spontaneously broken.
This may allow for exact de Sitter symmetry in the standard model prior to the CP-violating phase.

The primary results quoted  above are independent of our choice of signature, though the generators are slightly
different. With a $(-,+,+,+)$ signature metric, the generators of the de Sitter group are
$\{\frac{1}{4}\gamma^{[I}\gamma^{J]},\frac{i}{2}\star\gamma^{I}\}$. It is straightforward to show that
$su(2,2)\oplus u(1) \bigcap maj(4)$ is spanned by this algebra when the signature is $(-,+,+,+)$.

\section{The Macdowell-Mansouri action}
We will now review the Macdowell-Mansouri formulation of gravity \cite{MMoriginal}. From the insights of the previous
section, it is natural to begin with the de Sitter group, and we will do so.
However, the reader should
be aware that all of the following works just as well when the underlying gauge group is the anti-de
Sitter group.
We will use our (somewhat non-standard) notation where the Clifford algebra is employed as a basis for various Lie
algebras.

The beauty of the Macdowell-Mansouri approach is that it unifies the spin-connection and the frame fields into a
single object that can be interpreted as a connection taking values in the de Sitter Lie algebra. Geometrically, when
we define, say, an $SO(3,1)$ connection, we first make a basis transformation so that the tangent space at each point
looks identical to Minkowski space. The connection is simply a rule for comparing vectors in the local
Minkowski frame at one point to vectors in the local Minkowski frame at another point. The heuristic picture is a
manifold with a different instance of the plane of Minkowski space tacked onto each point of the manifold. The
connection interpolates between different frames by rotating and boosting vectors in one frame in a pre-proscribed
manner as these vectors are moved around from frame to frame on the manifold. In
the presence of a positive cosmological constant, it is natural that the homogenous space tacked onto the manifold at
each point is deformed from Minkowski space to de Sitter space. This introduces extra degrees of freedom: as a vector
moves from frame to frame, we can rotate and boost it, but we can also ``translate" it by acting on it with the
pseudo-translation generators of the de Sitter group. All this can be defined in a rigorous manner by employing
Cartan's formulation of connections as shown recently in \cite{Wise:MMgravity}.

When a local basis for the Lie algebra is defined, the Cartan connection takes values in the Lie algebra of the de
Sitter group. Since this Lie algebra consists of the generators of the Lorentz group together with a generator that
transforms as a vector under the adjoint action of the Lorentz group, to the spin connection we must add a vectorial
piece. To this end, we define the de Sitter connection
\beq
\Lambda=\omega+\textstyle{\frac{i}{r_{0}}}e
\eeq
when the signature is $(+,-,-,-)$, or
\beq
\Lambda=\omega+\textstyle{\frac{i}{r_{0}}}\star e
\eeq
when the signature is $(-,+,+,+)$, where $\omega=\frac{1}{4}\gamma_{[I}\gamma_{J]}\omega^{IJ}$ is the spin connection,
$e=\frac{1}{2}\gamma_{I}e^{I}$
will be interpreted as a frame field, and $r_{0}=\sqrt{3/\lambda}$ is the de Sitter radius. Under a de Sitter
transformation, $g\in dS_{4}$, the connection transforms in the usual way:
\beq
\Lambda\stackrel{g}{\longrightarrow} \Lambda '=\Lambda-(D_{\Lambda}g)g^{-1}=g\Lambda g^{-1}-(dg)g^{-1}\,.
\eeq
Now suppose $g$ is an infinitesimal pseudo-translation generated by the vector field $\eta=\frac{1}{2}\gamma_{I}\eta^{I}$ so
that $g=1+i\eta$. Then the the connection transforms as
\beq
\Lambda\ \stackrel{1+i\eta}{\longrightarrow} \ \Lambda' =\Lambda -iD_{\Lambda}\eta
\eeq
which means that the spin connection and tetrad rotate into each other as follows:
\beqa
\omega &\stackrel{1+i\eta}{\longrightarrow}& \omega' =\omega-{\ts\frac{1}{r_{0}}}\,[\eta,e]\nn\\
e &\stackrel{1+i\eta}{\longrightarrow}& e'=e- r_{0}\,D_{\omega}\eta\,.
\eeqa
Consider the curvature $F=d\Lambda+\Lambda\wedge\Lambda$. In components it is given by
\beqa
F&=&d\omega+\omega\wedge\omega -{\ts\frac{1}{r_{0}^{2}}}\,e\wedge e+{\ts \frac{i}{r_{0}}}\,D_{\omega}e\nn\\
&=&R-{\ts\frac{1}{r_{0}^{2}}}\,e\wedge e+{\ts\frac{i}{r_{0}}}\,T\,.
\eeqa
Under an infinitesimal pseudo-translation, $g=1+i\eta$, the curvature transforms as
\beq
F\ \stackrel{1+i\eta}{\longrightarrow}\ F'=F+[i\eta,F]
\eeq
or
\beqa
R-{\ts\frac{1}{r_{0}^{2}}}\,e\wedge e  & \stackrel{1+i\eta}{\longrightarrow}&
R-{\ts\frac{1}{r_{0}^{2}}}\,e\wedge e-{\ts\frac{1}{r_{0}}}[\eta,T]\nn\\
T &\stackrel{1+i\eta}{\longrightarrow}& T+r_{0}\,[\eta, R-{\ts\frac{1}{r_{0}^{2}}}\,e\wedge e]\,.
\eeqa
Evidently, a de Sitter pseudo-translation mixes the curvature with the torsion and vice-versa. Using the equations of
motion $R-\frac{1}{r_{0}}e\wedge e=C$, and $T=0$, we see that under this transformation, the Weyl tensor and the
torsion transform by
\beqa
C  & \stackrel{1+i\eta}{\longrightarrow}&
C-{\ts\frac{1}{r_{0}}}[\eta,T]=C\nn\\
T &\stackrel{1+i\eta}{\longrightarrow}& T+r_{0}\,[\eta, C]=r_{0}\,[\eta, C]\,.
\eeqa
We see that torsion does not remain zero under a de Sitter pseudo-translation unless the Weyl tensor is zero (in which
case the equations describe de Sitter space, which must be invariant under the de Sitter group). 
Therefore, the Einstein equations cannot be
invariant under this type of transformation. Nevertheless, we will proceed under the knowledge that Einstein's equations
can only be achieved by a breaking of de Sitter invariance. 

The action we wish to construct should be built strictly out of the connection $\Lambda$ as the sole dynamical
variable, and it should reproduce the equations of motion of the Einstein-Cartan action. Motivated by the Yang-Mills
Lagrangian, we might expect that the action should be quadratic in the curvature, and contain a dual. However, the
dual operation itself contains metric information, which should only enter into the action via the $e$ terms of the
connection. There is another sort of dual one might try---the internal dual on the fiber indices. Fortunately we have
such a dual available to us since
the element of the Clifford algebra, $\star$, is an $so(3,1)$-invariant dual operator on the Clifford algebra.
Using this as the duality operator, the action we will consider is given by
\beq
S_{MM} = \alpha\int \star F\wedge F\,.
\eeq
Using the identity that the trace of two elements of the Clifford algebra is zero unless they are of the same
grade and expanding the curvature $F$ into its components, we have
\beq
S_{MM}=\alpha\int_{M} \star R\wedge R -{\ts\frac{2}{r_{0}^{2}}} \star e\wedge e\wedge R
+{\ts\frac{1}{r_{0}^{4}}}\star e\w e \w e \w e\,.
\eeq
If we now choose the coupling constant to be $\alpha==-\frac{r_{0}^{2}}{2k}=-\frac{3}{2k\lambda}$, the action reduces
to 
\beq
S_{MM}=-\frac{3}{2k\lambda}\int_{M} \star R\wedge R +\frac{1}{k}\int_{M} \star e\wedge e\wedge R
-{\ts\frac{\lambda}{6}}\star e\w e \w e \w e\,.
\eeq
We recognize the second two terms as precisely the Einstein-Cartan action with a cosmological constant. 
The first term is topological---using the Bianchi identity, the variation of it with respect to $\omega$ is
identically zero so it does not affect the equations of motion. In fact, it is a familiar topological invariant called
the Euler class. Fixing a global basis for the fiber-bundle, the term can be reduced to a total derivative:
\beq
\int_{M}\star R\wedge R =\int_{\partial M}\star \left(\omega\wedge d\omega 
+{\ts\frac{2}{3}}\omega\wedge\omega\wedge \omega\right)\,.
\eeq
The above term is not invariant under a large gauge transformation that cannot be deformed to the
identity\cite{Baez:Book}.
However,
it is invariant under small gauge transformations that do not change the global structure of the basis in the fiber
bundle. This boundary term, and especially the generalization of the term to include the Immirzi parameter, will play
an important role in understanding the Kodama state.

\subsection{Adding the Immirzi Parameter}
We now need to extend the Macdowell-Mansouri action to include the Immirzi parameter. This has been accomplished in 
the context of $BF$ theory in
\cite{Smolin:MMaction, Starodubtsev:MMgravity}, however some of the topological terms in the resultant action, in
particular the Nieh-Yan
class, are relics of the $BF$ formulation and do not enter in the minimal prescription that we will present in the next
section\cite{Randono:GKII}. 

We first recall how we add the Immirzi parameter to the Einstein-Cartan action to give the Holst action. Beginning with
the Einstein Cartan action, 
\beq
S_{EC}=\frac{1}{k}\int_{M} \star e\wedge e\wedge R
\eeq
we simply perturb the curvature by its dual:
\beq
R\ \longrightarrow \ R-\frac{1}{\beta}\star R
\eeq
and the action becomes the Holst action
\beq
S_{H}=\frac{1}{k}\int_{M} \star e\wedge e\wedge R+\frac{1}{\beta}e\wedge e\wedge R.
\eeq
We will try using same trick on the Macdowell-Mansouri action. The Macdowell-Mansouri action is given
by:
\beq
S_{MM}=-\frac{3}{2k\lambda}\int_{M}\star F\w F.
\eeq
The trick is to perturb the de Sitter curvature by its ``dual", 
\beq
F\ \longrightarrow \ F-\theta\star F,
\eeq
and possibly make appropriate adjustments to the coupling constant in order to regain the Holst action up to topological
terms. With the above substitution the Macdowell-Mansouri action becomes
\beqa
S_{MM+\beta}&=&\alpha\int_{M}\star (F-\theta\star F)\wedge (F-\theta\star F)\nn\\
&=& \alpha \int_{M} (1-\theta^{2})\star F\wedge F -\theta (\star F\wedge\star F-F\wedge F).
\eeqa
Making the identifications 
\beqa
\alpha(1-\theta^{2})=-\frac{3}{2k\lambda} & & \frac{2\theta}{1-\theta^{2}}=\frac{1}{\beta}
\eeqa
the action becomes 
\beq
S_{MM+\beta}=S_{topo}+S_{H+\lambda}
\eeq
where $S_{H+\lambda}$ is the Holst action with a positive cosmological constant and
\beq
S_{topo}=-\frac{3}{2k\lambda}\int_{M} \star R\wedge R+\frac{1}{\beta}R\wedge R\,.
\eeq
As before, these additional terms are topological. The first is again the Euler class, and the second term is the second
Chern class. These terms will play an important role in later sections.

 \section{A de Sitter invariant action}
Despite being constructed out of a de Sitter connection, the Macdowell-Mansouri action is not invariant under local de
Sitter transformations. This was expected by our previous argument that the Einstein equations are not de Sitter
invariant. Explicitly the non-invariance arises because the ``dual", $\star$ is only invariant under the Lorentz
subgroup and is not invariant under a pseudo-translation. Under this type of transformation, the action transforms by
\beq
S_{MM}\stackrel{1+i\eta}{\longrightarrow}S_{MM}+\frac{r_{0}}{k}\int_{M}\star[\eta,T]
\wedge\left(R-{\ts\frac{1}{r_{0}^{2}}}e\wedge e\right)\,.
\eeq
Thus, although the action is invariant on shell, it fails to be invariant throughout the whole phase space. It
appears, then, that we have gained little more than conceptual advantage in viewing gravity as a gauge theory of the
de Sitter group. We would now like to show that there are very natural extensions of the gravitational action that are
strictly de Sitter invariant.
Although de Sitter invariant modifications of the Macdowell-Mansouri action have been considered in the past by adding a
dynamical vector field in the
$SO(4,1)$ representation of the de Sitter group \cite{Smolin:MMaction}, the appearance of the vector field is rather {\it
ad hoc}
without physical justification. Here we show that the desired result can be obtained with more physical spinor fields.
The action we propose is motivated by the success of quadratic spinor techniques
\cite{Jacobson:QuadraticSpinor, Tung:Immirzi}.
We assume that the action consists of a gravitational piece and a matter piece: $S=S_{g}+ S_{m}$ where
\beq
S_{g}=-4\alpha \int \langle DD\phi, DD\psi\rangle
\eeq
where $\psi$ and $\phi$ transform in the fundamental representation of the de Sitter group and $D$ is the de Sitter
covariant derivative. The inner product in the action can be the Dirac inner product, the Majorana inner product or
any linear combination of the two. Since the de Sitter group preserves these two inner products, the action is clearly
de Sitter invariant.
All that we will require of the matter action is de Sitter symmetry, and that the full set of
equations of motion admit a stable ground state where the matter distribution is {\it homogenous} and {\it isotropic}.
Let's rewrite the above action in a form that is more amenable to analysis upon symmetry breaking. For definiteness we
set $\phi=\psi$ and take the inner product to be the Dirac inner product:
\beqa
S_{g} &=& 4\alpha \int \bar{\psi}F\wedge F \psi \\
&=& -4\alpha\int \psi\bar{\psi}F\wedge F 
\eeqa
where in the last line $\psi\bar{\psi}$ is recognized as a $4\times 4$ complex matrix and the trace in the integral is
assumed.

Now, the Fierz identity allows one to decompose bispinor matrices of the general form $\phi\bar{\psi}$ into a Clifford
basis:
\beq
\phi\bar{\psi}=\alpha_{\hat{I}}\Gamma^{\hat{I}}
\eeq
where $\Gamma^{\hat{I}}$ are the elements of the Clifford basis and $\alpha_{\hat{I}}$ are spin-currents. In our case,
the Fierz identity takes the form:
\begin{equation}
-4\psi\bar{\psi}=(\bar{\psi}\psi)\mathbf{1}-(\bar{\psi}\star\psi)\star
+(\bar{\psi}\gamma_{I}\psi)\gamma^{I}+(\bar{\psi}\gamma_{I}\star \psi)\gamma^{I}\star
-\frac{1}{2}(\bar{\psi}\gamma_{[I}\gamma_{J]}\psi)\gamma^{[I}\gamma^{J]}.
\end{equation}
Using this identity in our action we find that the action becomes:
\beq
S_{g}=\alpha \int -(\bar{\psi}\star\psi)\star F\wedge F + \bar{\psi}\psi F\wedge F +...
\eeq
where the remaining terms depend on the vector, axial-vector, and bivector currents. The effective action is obtained 
by evaluating the spin--currents on the assumed homogenous and isotropic ground state. The spinor fields are then
replaced by their expectation values in
this state. {\it Isotropy} means that the expectation values of any spin-currents that pick out a preferred direction in
space must be zero:
\beq
\langle\bar{\psi}\gamma^{I}\psi\rangle=\langle \bar{\psi}\gamma^{I}\star\psi\rangle
=\langle\bar{\psi}\gamma_{[I}\gamma_{J]}\psi\rangle=0.
\eeq
{\it Homogeneity} means that the remaining non-zero expectation values, $\langle\bar{\psi}\psi\rangle$ and 
$\langle\bar{\psi}\star\psi\rangle$, are constant. Such a vacuum expectation value necessarily breaks de Sitter 
invariance since the maximal subgroup of the de Sitter
group which preserves $\bar{\psi}\star\psi$ is the Lorentz group. With this in mind, the action reduces to:
\beq
-\alpha\int\langle\bar{\psi}\star\psi\rangle \star F\wedge F-\langle\bar{\psi}\psi\rangle F\wedge F.
\eeq
We now identify the gravitational coupling constants with the homogenous expectation value $\langle\bar{\psi}\star\psi\rangle$
and the theta parameter of the second Chern class with $\langle\bar{\psi}\psi\rangle$:
\beqa
\alpha \langle\bar{\psi}\star\psi\rangle &=& \frac{3}{16\pi G\lambda}\\
\alpha \langle\bar{\psi}\psi\rangle &=&\Theta .
\eeqa
Thus, the action reduces to the Macdowell-Mansouri action together with a topological theta-term.
\section{Concluding remarks}
We have given a geometric reason for positivity of the (bare) cosmological constant. Since the simplest extensions of
the standard model to include neutrino mass utilize both the Dirac and Majorana inner products, it is natural
to identify the spacetime symmetry group with the group that preserves both of these inner products, namely the de Sitter
group. For this reason, we will assume in the rest of the paper that the bare cosmological constant is positive.
However, the reader should be aware that the Kodama state can be defined even when the constant is negative. 
We have also shown that there are very natural extensions of the Einstein-Cartan action where gravity is a gauge theory
with exact local de Sitter symmetry. Although we will not use this particular formalism again in this work, the
underlying but broken de Sitter symmetry of Einstein-Cartan and Macdowell-Mansouri gravity will play an important
role in understanding the Kodama state.

\chapter{Hamiltonian General Relativity\label{HamiltonianGR}}

The Hamiltonian form of general relativity is the standard formalism in numerical relativity and canonical quantum
gravity. At first glance, the program seems to be against the spirit of relativity. After all, the lesson learned from
special and general relativity is that space and time should be treated on equal footing. The Hamiltonian approach
begins by breaking that common ground and splitting spacetime into {\it space} plus {\it time}. Despite this apparent
breaking of general covariance, when the full Hamiltonian theory is constructed the spacetime can be reconstructed and
the result is identical to ordinary general relativity. The relic of general covariance in the Hamiltonian formalism
lies in our freedom to choose how we would like to evolve the data. This freedom is manifest in the constraints that
generate gauge degrees of freedom. In numerical relativity the Hamiltonian formalism satisfies our intuitive notion of
events in the ``here and now" evolving to the ``there and then". In the quantum theory it is primarily useful in order
to appropriately constrain the phase space and define a set of fundamental commutators that we can trust to carry over
into the quantum theory. In this section we will review the classical formulation of Hamiltonian general relativity.

\section{The Hamiltonian formulation of General Relativity}
Since the starting point for modern canonical quantum gravity is a version of Einstein-Cartan gravity, we will present
the Hamiltonian construction of this theory. The major concepts closely parallel the ADM formulation of the
Einstein-Hilbert action\cite{ADM}. However, the precise form of the phase space and the constraints may unfamiliar to
those familiar with the standard ADM formulation.
Since the Holst action 
\beq
S_{H}=\frac{1}{k}\int_{M}\star e \w e\w R +\frac{1}{\beta} e\w e\w R-\frac{\lambda}{6}\star e\w e\w e\w e 
\label{HolstAction2}
\eeq
encompasses all of the three gravity formulations we wish to review (Einstein-Cartan, Ashtekar, and Barbero-Immirzi
for $\beta=\infty$, $\beta=\pm{i}$, and $\beta\in \mathbb{R}$, respectively) we will present the $3+1$ decomposition of
the above action for arbitrary $\beta$ and consider the special cases separately. 

The approach we will take follows the ADM formulation of $3+1$ gravity. The first step is to slice spacetime into
spacelike foliations representing the constant time slices of a congruence of observers. This corresponds to splitting
the manifold up into $M=\mathbb{R}\times \Sigma$ where $\Sigma$ is the spatial topology and $\mathbb{R}$ represents
the parameterization of time in our slicing. Not all spacetimes admit such a decomposition, so this choice amounts to
a restriction on the global topology of the spacetimes. Although there have been many attempts to avoid this
restriction we will not concern ourselves with this technicality---we are primarily interested in de Sitter space,
which does admit such a $3+1$ decomposition. The major insight of Arnowitt, Deser, and Misner is that even with a
given choice $M=\mathbb{R}\times \Sigma$, one is still free to choose different congruences of observers with a
different set of clocks\cite{ADM}. In particular, a given observer may observe time to run faster here and slower there,
as
he moves around through different points on consecutive slicings $\Sigma_{t}$. This freedom is encoded in the time
evolution vector field $\bar{t}$, which contains components orthogonal and parallel to the 3-manifold:
\beq
\bar{t}=N\bar{n}+\bar{N}.
\eeq
Here, $\bar{n}$ is the normal to the foliation, the ``lapse", $N$, is a measure of how much proper time elapses on a
clock in the rest frame of the observer as
he moves from $\Sigma_{t}$ to $\Sigma_{t+dt}$, and the ``shift", $\bar{N}$, measures the spatial coordinate distance the
observer
moves from one slice to the next. From this information, and the spatial three-metric $g_{ab}$ on each slice, one can
construct the full metric as follows:
\beq
ds^{2}=-(N\,dt)^{2}+g_{ab}(dx^{a}+N^{a}dt)(dx^{b}+N^{b}dt).
\eeq
We need to translate the above into tetrad language. From the definition of the tetrad we need
\beqa
g_{ab}&=&\eta_{IJ}\,e^{I}_{a}e^{J}_{b}\nn\\
g_{a0}&=&\eta_{IJ}\,e^{I}_{a}e^{J}_{0}=N_{a}\nn\\
g_{00}&=&\eta_{IJ}\,e^{I}_{0}e^{J}_{0}=-N^{2}+g_{ab}N^{a}N^{b}\,.
\eeqa
Solving the above for $e^{I}_{\mu}$ we have
\beqa
e^{I}_{a}&=& (e^{0}_{a}\,,\, E^{i}_{a})\nn\\
e^{I}_{0}&=& (N\,,\, N^{i}=E^{i}_{a}N^{a})\,.
\eeqa
We see from the above an interesting paradox---in contrast to the metric approach, the spatial triad, $E^{i}_{a}$, the
lapse $N^{a}$, and the shift $N$ do not uniquely determine the tetrad. Instead we need to specify some additional
information: the components $e^{0}_{a}$. The origin of this paradox comes from the fact that, prior to modding out the
gauge degrees of freedom, the tetrad $e^{I}_{\mu}$ contains more information than the metric $g_{\mu\nu}$. Just
counting the degrees of freedom, since the tetrad is not constrained to be diagonal it is an arbitrary (invertible but
otherwise arbitrary) $4\times 4$ matrix, so it has $16$ degrees of freedom. On the other hand, the metric must be
symmetric, so it has $(N^{2}+N)/2=10$ degrees of freedom. The additional degrees of freedom in the tetrad comes from
the Lorentz freedom in choosing a frame: given any set of frame fields that diagonalize the metric, we can always
perform a local boost or rotation at that point and the new frame will still diagonalize the metric. This accounts for
the extra degrees of freedom: three boosts and three rotations. In the context of the ADM variables, the metric
degrees of freedom (DOF's) are $DOF(g_{ab})+DOF(N^{a})+DOF(N)=6+3+1$. In the tetrad language, the shift and the lapse
still have $3+1=4$ degrees of freedom, but the triad has $DOF(E^{i}_{a})=9$ whereas the spatial metric has $6$ degrees
of freedom. This corresponds to our freedom to
{\it rotate} any spatial frame that diagonalizes the three metric. However, we are still left with our freedom to {\it
boost} the four-dimensional frame. This information must then be encoded in the remaining degrees of freedom,
$e^{0}_{a}$. Indeed, one can easily verify that given a frame where the components $e^{0}_{a}$ are zero at some point,
if we now make a local boost at that point, the new components $e'^{0}_{a}$ will not be zero. In
total then we have
\beqa
DOF(e^{I}_{\mu})&=&DOF(E^{i}_{a})+DOF(N^{a})+DOF(N)+DOF(e^{0}_{a})\nn\\
&=& 9+3+1+3\nn\\
&=& 16\ .
\eeqa

The next step in the ADM decomposition is to rewrite the action in terms of these new variables. Some of the variables
will have no conjugate momenta, and can therefore be treated as Lagrange multipliers. There is some ambiguity in the
way we perform the Legendre transform. After choosing a timelike vector field, $\bar{t}=\bar{\eta}+\bar{N}$ (where
$\eta=N\bar{n}$), we can either identify the $\mathbb{R}$ in $M=\mathbb{R}\times\Sigma$ with parameterization of the
integral curves of $\bar{\eta}$ or the integral curves of $\bar{t}$. This corresponds to a splitting of the action as
\beqa
S=\int_{M} \tilde{L}=\int_{\mathbb{R}}\tilde{d}t\int_{\Sigma}\tilde{L}(\bar{t})
\eeqa
or
\beqa
S=\int_{M} \tilde{L}=\int_{\mathbb{R}}\tilde{\eta}\int_{\Sigma}\tilde{L}(\bar{\eta})
\eeqa
where $\tilde{\eta}({\bar{\eta}})=1$ and $\tilde{d}t(\bar{t})=1$. The physical results will be the same for either case,
but the identification of Lagrange multipliers with physical fields will be different and the form of the
diffeomorphism constraints will be different. When one uses the $\bar{t}$ identification, one finds that the
diffeomorphism constraint has a piece which is not independent from the other constraints. Typically this is resolved by
subtracting off this piece, but we will simply avoid the issue altogether by identifying $\mathbb{R}$ with the
integral curves of $\bar{\eta}$. To this end, we proceed to decompose the action into pieces orthogonal and
perpendicular to $\bar{\eta}$. Making the identifications $\omega^{IJ}(\bar{\eta})=-\lambda$, and
$\eta^{I}=e^{I}(\bar{\eta})$, and using the identity
$R^{IJ}(\bar{\eta})=\mathcal{L}_{\bar{t}}\omega^{IJ}-D\omega^{IJ}(\bar{\eta})$, we have
\beq
S_{H}=\frac{1}{4k}\int_{\mathbb{R}}\tilde{\eta}\int_{\Sigma}
{P_{\star}}_{IJKL}\,e^{I}\w e^{J}\w\mathcal{L}_{\bar{t}}\omega^{KL}
-C_{D}(\bar{N})-C_{G}(\lambda)-C_{H}(\eta) \label{HamHolstV1}
\eeq
where  ${P_{\star}^{IJ}}_{KL}={\epsilon^{IJ}}_{KL}-\frac{1}{\beta}\delta^{IJ}_{KL}$ and we have (tentatively) defined
the diffeomorphism, Gauss, and Hamiltonian constraints
\beqa
C_{D}&=& \int_{\Sigma}{P_{\star}}_{IJKL}\,\mathcal{L}_{\bar{N}}\omega^{IJ}\w e^{K}\w e^{L} \\
C_{G}&=& \int_{\Sigma}-{P_{\star}}_{IJKL}\,D\lambda^{IJ}\w e^{K}\w e^{L} \\
C_{H}&=& \int_{\Sigma} {P_{\star}}_{IJKL}\,\eta^{I}e^{J}\w \left(R^{KL}
-{\ts\frac{\lambda}{3}}\,e^{K}\w e^{L}\right)\ .
\eeqa
We recall that $P_{\star}$ is an
invertible operator for all values of $\beta$ except $\beta=\mp i$ where it becomes a projection operator into the
left(right) tensor subspaces. At first glance it appears that we have succeeded in constructing a Hamiltonian version
of Einstein-Cartan gravity with dynamical variables $\omega^{IJ}$ and their conjugate momenta
$\pi^{IJ}=\frac{1}{4k}{P_{\star}^{IJ}}_{KL}e^{K}\w e^{L}$. However, this conclusion is a bit premature at this stage.
We can foresee problems with these definitions simply by counting the dynamical degrees of freedom of the momenta. An
arbitrary field $\pi^{IJ}_{ab}=\pi^{[IJ]}_{[ab]}$ would have $3\times 6=18$ degrees of freedom (recalling the forms
are pulled back to the three-manifold). However, from our definition, $\pi^{IJ}_{ab}$ only has the degrees of freedom
of $e^{I}_{a}$, which total to $3\times 4=12$. Thus, the momenta must be subject to some primary constraints,
sometimes referred to as the simplicity constraint. When this constraint is appropriately added to the full set
and the constraint algebra is computed, one finds that the algebra does not close due to the existence of
second class constraints stemming from the commutator of the simplicity constraint with the remaining constraints.
Alternatively, one could attempt to implement this constraint through the use of the Dirac bracket, and recompute to
the constraint algebra---this is the basis of the current ``covariant canonical" formulation of Hamiltonian general
relativity, which we will have more to say about later\cite{Livine:Covariant, Alexandrov:Covariant}. The standard
approach, however, is to avoid these issues
altogether by a partial gauge fixing. The partial gauge fixing, referred to as the time gauge, consists of choosing a
direction once and for all for the projection of $\bar{\eta}$ in the fiber, $\eta^{I}$ so that $\eta^{0}=N$ and $\eta^{i}=0$.
Since we used $\bar{\eta}$ to pull-back our fields to $\Sigma$, this means that $\eta_{I}e^{I}_{a}=0$, so that
$e^{0}_{a}=0$. From our previous discussion we found that the components $e^{0}_{a}$ were related to the local boost
degrees of freedom of the tetrad. Thus, we are in effect projecting the Lorentz group down to its rotation subgroup.
In other words, we are restricting ourselves to the group of transformations that preserves the vector
$\eta=(N,0,0,0)$, which is clearly the rotation subgroup. With this in mind, it is useful to rewrite all our variables
in 3--dimensional notation. Thus, we define the $so(3)$ valued spin connection, ${\omega^{ij}}$, and the remaining
components of the original spin connection transform like a tensor under $SO(3)$. Thus, we rewrite the components
$\phi^{*}_{\Sigma}{\omega^{i}}_{0}\equiv {K^{i}}$, where $\phi^{*}_{\Sigma}$ is the pull-back of the map of $\Sigma$ 
to $M$. The new variable $K^{i}$ is the ordinary extrinsic curvature when the time component of the torsion, $T^{0}$
is zero.
We will absorb the operator $P_{\star}$
into the connection itself by defining a new $so(3)$--valued connection $A^{ij}\equiv \omega^{ij}-\beta K^{ij}$, where
$K^{ij}={\epsilon^{ij}}_{k}K^{k}$. We are free to do this since one can always add a tensorial piece to any connection
and the new connection will still behave properly. The tetrad components are reduced to the frame field on the
three-space, $E^{i}_{a}\equiv e^{i}_{a}$, also referred to as the spatial triad. With
these substitutions, the original action becomes
\beq
S_{H}=-\frac{1}{2k\beta}\int_{\mathbb{R}}\tilde{\eta}\int_{\Sigma}E_{i}\w E_{j}\w \mathcal{L}_{\bar{t}} A^{ij}
-(constraints)\ .
\eeq
We can see immediately that we have solved one of the major problems with (\ref{HamHolstV1})---the issue of the
simplicity constraints on the momenta is resolved. To see this, simply count the degrees of freedom. The new momenta
conjugate to
$A^{ij}$ are ${\pi^{ij}}_{ab}=-\frac{1}{2k\beta}\Sigma^{ij}$ where $\Sigma^{ij}=E^{i}\w E^{j}$. An arbitrary tensor of
the form
${\pi^{ij}}_{ab}={\pi^{[ij]}}_{[ab]}$ has $3\times 3=9$ degrees of freedom. Since the momenta is built solely out of
the triad, and the triad $E^{i}_{a}$ also has $3\times 3=9$ degrees of freedom, there are no primary constraints on the
momenta. 

We are not quite done yet since there are still too many variables and too few constraints. If we attempt to rewrite the
action in terms of the new variables $A$, $\pi$, $\lambda$, and $\bar{N}$ we find that we are still left with terms
involving $\omega^{ij}$ and ${\omega^{i}}_{0}(\bar{\eta})$, which cannot be gotten rid of. To eliminate these variables,
we first impose two condition on the torsion:
\beqa
T^{0}(\bar{\eta})=0 &\quad \quad & ^{(3)}T^{i}=D_{\omega}E^{i}=0\ .
\eeqa
Since these components of the torsion vanish on the equations of motion, we are not changing the physical content of
the theory. The first condition is equivalent to setting the variable $\omega^{i0}(\bar{\eta})=0$. The second
condition on the 3-torsion amounts to replacing $\omega^{ij}$ with the
Levi-Civita connection $\Gamma^{ij}$, which is constrained to be torsion freeby definition. Since the Levi-Civita
connection
can be written as an explicit function of the triad as in (\ref{Levi-Civita}), which are in turn functions of the
momenta, we have successfully eliminated the extra variables. The action now becomes
\beq
S_{H}=-\frac{1}{2k\beta}\int_{\mathbb{R}}\tilde{\eta}\int_{\Sigma}\Sigma_{ij}\w \mathcal{L}_{\bar{t}} A^{ij}
-C_{D}(\bar{N})-C_{G}(\lambda)-C_{H}(N)
\eeq
where the (true) constraints are
\beqa
C_{D}(\bar{N})&=&-\frac{1}{2k\beta}\int_{\Sigma}\mathcal{L}_{\bar{N}} A^{ij}\w \Sigma_{ij} \\
C_{G}(\lambda)&=&-\frac{1}{2k\beta}\int_{\Sigma}-D_{A}\lambda^{ij}\w \Sigma_{ij}\\
C_{H}(N)&=&-\frac{1}{2k}\int_{\Sigma}\epsilon_{ijk}E^{i}\w
\left((1+{\ts\frac{1}{\beta^{2}}})R^{jk}-{\ts\frac{1}{\beta^{2}}}F^{jk}-{\ts\frac{\lambda}{3}}\Sigma^{jk}\right)\ .
\eeqa
In the above, $F^{ij}=dA^{ij}+{A^{i}}_{m}\w A^{mj}$, and
$R^{ij}=R^{ij}_{\Gamma}=d\Gamma^{ij}+{\Gamma^{i}}_{m}\w\Gamma^{mj}$ is understood to be an explicit function of the
momenta. This term complicates the Hamiltonian constraint severely, however, through the use of nested commutators it
has been shown that the Hamiltonian can be rewritten in a manageable form (see \cite{Thiemann:Book}). This form of the
constraints (in slightly
different notation) and the connection $A^{ij}=\Gamma^{ij}-\beta K^{ij}$, called the Ashtekar-Barbero connection, was
first constructed by Barbero in \cite{Barbero}.

\subsection{The evolution generated by the constraints}
With these variables, the symplectic structure defines a Poisson bracket. For any two integral functionals of $\Sigma$
and $A$, the Poisson bracket is given by
\beq
\left\{f,g\right\}=-\frac{1}{2k\beta}\int_{\Sigma}\frac{\delta f}{\delta A^{ij}}\w \frac{\delta g}{\delta \Sigma_{ij}}
-\frac{\delta g}{\delta A^{ij}}\w \frac{\delta f}{\delta \Sigma_{ij}}\ .\label{PB1}
\eeq
The fundamental commutators that results from this are
\beqa
\left\{A^{ij}\big|_{P}\,,\, \Sigma_{kl}\big|_{Q}\right\}&=&-2k\beta\ \delta^{ij}_{kl}\ \tilde{\delta}(P,Q)\nn\\
\left\{A^{ij}\big|_{P}\,,\,A^{kl}\big|_{Q}\right\}&=&0 \nn\\
\left\{\Sigma_{ij}\big|_{P}\,,\,\Sigma_{kl}\big|_{Q}\right\}&=& 0 \label{FundCommutators}
\eeqa
where $\tilde{\delta}(P,Q)$ is the delta-ditribution valued three-form defined by
$\int_{P\in\Sigma}f(P)\,\tilde{\delta}(P,Q)=f(Q)$. We note that despite $A=\Gamma-\beta K$ being dependent on the
momenta via $\Gamma[E]$, the components of the connection still commute under the Poisson bracket\footnote{We stress that
the components, viewed as functionals of the dynamical variables, commute under the {\it Poisson bracket}
commutator. This does not mean that they commute under the matrix commutator, where they satisfy the usual commutation
relations of the $su(2)$ Lie algebra.}.

It can then be shown that the diffeomorphism and Gauss constraints generate infinitesimal diffeomorphism and $SO(3,1)$
transformations in the following sense: given $f=f[A,\Sigma]$ the Poisson evolution under these constraints is
\beqa
f&\rightarrow& f+\left\{f,C_{D}(\bar{N})\right\}=f[A+\mathcal{L}_{\bar{N}}A\,,\,\Sigma+\mathcal{L}_{\bar{N}}\Sigma]\nn\\
f&\rightarrow& f+\left\{f,C_{G}(\lambda)\right\}=f[A-D_{A}\lambda\,,\, \Sigma+[\lambda, \Sigma]]\ .
\eeqa
The first equation describes the ordinary change of a functional under a small diffeomorphism, and the second is the
change under an small $SU(2)$ gauge transformation.
The action of the Hamiltonian constraint is considerably more complicated since it plays the dual role of generating
time reparameterizations (which is why the Hamiltonian constraint must vanish), and at the same time describing the
non-gauge dynamical evolution of the fields. Nevertheless, one can show that the vanishing of the constraints plus the
time evolution of the dynamical fields generated by the constraints under Hamilton's equations, reproduce the
appropriately gauge fixed Einstein-Cartan equations (when reality constraint are imposed if $\beta\in\mathbb{C}$). The
constraint algebra constructed from the commutators
of the constraints themselves closes in the sense that the commutator of any two constraints weakly vanishes on the
constraint submanifold.

\subsection{The special cases $\beta=\infty $ and $\beta=\mp i$}
Let us now consider the special cases where $\beta=\infty$ and $\beta=\mp i$. In the limit that $\beta\rightarrow \infty$,
the Holst action reduces to the Einstein-Cartan action. The phase space reduces to $(K^{i},
\frac{1}{2k}{\epsilon^{i}}_{jk}E^{j}\w E^{k})$
and the constraints reduce to
\beqa
C_{D}(\bar{N})&=&\frac{1}{2k}\int_{\Sigma}\epsilon_{ijk}\mathcal{L}_{\bar{N}}K^{i}\w E^{j}\w E^{k}\nn\\
C_{G}(\lambda)&=&\frac{1}{2k}\int_{\Sigma}\epsilon_{ijk}{\lambda^{i}}_{m}K^{m}\w E^{j}\w E^{k} \nn\\
C_{H}(N)&=& \frac{1}{2k}\int_{\Sigma}-\epsilon_{ijk}E^{i}\w \left(R^{jk}+K^{j}\w K^{k}
-{\ts\frac{\lambda}{3}}\Sigma^{jk}\right)\ .
\eeqa
Thus, the main advantage of the formulation of GR with a real Immirzi parameter is that the phase space consists of an
$so(3)$ connection and its conjugate momentum. We will see that this gives us significant advantage in the quantum
theory. The downside to this formalism is that it has been shown that the Ashtekar-Barbero connection
$A^{ij}=\Gamma^{ij}-\beta K^{ij}$, despite being a perfectly well-defined three-dimensional connection, cannot be
obtained as the pull-back of any four-dimensional connection\cite{Samuel:Connection}. Thus, the dynamical variables are
fundamentally
three-dimensional. Because of this it can sometimes be very difficult to extract spacetime information from data on
the spatial slices in these variables. In contrast, when $\beta\rightarrow\infty$, since $K^{i}$ has a well-defined
geometric interpretation,
it is relatively easy to reconstruct the spacetime from the dynamics of the variables on $\Sigma$. At the level of the
constraints, in terms of complexity there is no advantage of one formalism over the other. However, as we will 
see, implementing the Gauss and diffeomorphism constraints is much easier when the phase space consists of a
connection as a dynamical position variable.

The third approach, the Ashtekar formalism, follows from setting $\beta=\mp i$. From the Holst action it can be seen
that the choice corresponds to projecting the spin-conenction into its left(right) chiral subspace: $\omega\rightarrow
\frac{1}{2}(1\mp \gamma_{5})\omega$. In contrast to the real case, the three-dimensional connection, $A_{L/R}=\Gamma\pm
i K$, is the pull-back of the left(right) handed spin connection to $\Sigma$ \cite{Ashtekar:book}. Thus, we retain
much of
the geometric relation between spatial variables and spacetime. Since the phase space does consist of a connection, we
retain this advantage of the real formalism. In addition, the hamiltonian constraint simplifies drastically when
$\beta=\mp i$, where it reduces to
\beq
C_{H}(N)=-\frac{1}{2k}\int_{\Sigma}\epsilon_{ijk}E^{i}\w \left(F^{jk}
-{\ts\frac{\lambda}{3}}\Sigma^{jk}\right)\ .
\eeq
This simplification was the original reason for considering Ashtekar variables. In fact, it is unlikely that the
Kodama state would have ever been found without this dramatic simplification of the constraint. With all these
advantages, it is hard to see why one would not work with these variables. The downside to these variable comes from
the complexification of the phase space. One can show that when the triad are constrained to be real, general
relativity is regained---however, this is an additional constraint that must be put in by hand. This reality
constraint generally undermines the simplification of the Hamiltonian constraint. As we will see, the complexification
of the phase space is also the issue underpinning the problems associated with the Kodama state.

\chapter{Overview of Canonical Quantization\label{LQGIntro}}

In this section we will introduce some basics of Loop Quantum Gravity. Our introduction will be far from
exhaustive---our main purpose is to discuss the major concepts and results from the Loop approach that are relevant
for an understanding of the Kodama state. Thus, we will focus on the construction of spin network states as solutions
to the Gauss and diffeomorphism constraints, and the quantized area spectrum. For more in depth reviews, see
\cite{Ashtekar:book, Ashtekar:LQGReview1, Smolin:LQGReview1, Rovelli:book, Thiemann:Book}.
\section{The holonomy representation}
The distinguishing feature of canonical general relativity in the new variables, whether they be real as in the
Barbero-Immirzi approach or complex as in the Ashtekar formalism, is that the phase space consists of a connection as
``position" variable and its conjugate momentum. This allows us to compute the spectrum of the area operator and
construct a basis that allows for a rigorous construction of the kinematical 
Hilbert space. By ``kinematical", we are referring to the Hilbert space of states that are annihilated by both 
the Gauss and the diffeomorphism constraints. First, we will discuss perhaps the hallmark achievement of Loop Quantum
Gravity: the quantization of area and volume.

The fundamental commutation relations, (\ref{PB1}), that follow from the symplectic structure
of the Holst action in the Hamiltonian formalism are (\ref{FundCommutators}), repeated here:
\beqa
\left\{A^{ij}\big|_{P}\,,\, \Sigma_{kl}\big|_{Q}\right\}&=&-2k\beta\ \delta^{ij}_{kl}\ \tilde{\delta}(P,Q)\nn\\
\left\{A^{ij}\big|_{P}\,,\,A^{kl}\big|_{Q}\right\}&=&0 \nn\\
\left\{\Sigma_{ij}\big|_{P}\,,\,\Sigma_{kl}\big|_{Q}\right\}&=& 0 \ .
\eeqa
We will take these to be the fundamental commutators to carry over to the quantum theory without modification. As quantum
operator conditions, the commutators are
\beqa
\left[A^{ij}\big|_{P}\,,\, \Sigma_{kl}\big|_Q\right]&=&-2k\beta i\ \delta^{ij}_{kl}\  \tilde{\delta}(P,Q) \nn\\
\left[A^{ij}\big|_{P}\,,\,A^{kl}\big|_{Q}\right]&=&0 \nn\\
\left[\Sigma_{ij}\big|_{P}\,,\,\Sigma_{kl}\big|_{Q}\right]&=& 0 \ .
\eeqa
Because of the presence of the delta function, this operator commutator should be regularized.
Since the canonical position is a one-form and its conjugate momentum is a two-form, we should expect that the natural
regularization of the canonical commutation relations follows from somehow
integrating the connection along a one-dimensional extended line and integrating the momentum on a two-surface.
Fortunately, these
extended objects can be easily constructed. To regularize the position variable, we introduce holonomies. A holonomy
defined along a parameterized curve, $\gamma: [0,1]\rightarrow \Sigma$, defines the parallel transport of a
representation of the gauge group as we move the
object from the beginning to the end of $\gamma$. Naturally, the holonomy is constructed out of the connection. Since
the connections we are considering are non-abelian, we have to be careful about the ordering of the operators as the
representation is moved along the path. This is achieved by a path ordering which simply ensures that operators that
act closer to the end of the path always occur in the expression to the left of operators that act closer to the
beginning. The holonomy, denoted 
\beq
h_{\gamma}[A]=\mathcal{P}e^{\int_{\gamma}A}
\eeq
is itself an element of the group. When it acts on an element of the representation space we have to choose the
appropriate representation of the gauge group. Since we are concerned with the rotation group, the relevant gauge
group is $SO(3)$ or, more generally, its double cover $SU(2)$ whose representations are characterized by the spin of the
representation. Thus we will denote by $h^{(j)}_{\gamma}$ for holonomy along the path $\gamma$ in the spin $j$
representation of $SU(2)$.

Conceptually we can think of the curve as constructed out of infinitesimal pieces, $\delta \gamma$, whose union gives
the whole path $\gamma$ when they are glued end to end. If the path is parameterized by $s$, each of these small
pieces has a parameter length $ds$, and a tangent vector $\bar{\xi}=\frac{\partial x^{\mu}}{\partial
s}\frac{\partial}{\partial x^{\mu}}$.
The
infinitesimal holonomy along one of these pieces is an infinitesimal gauge transformation generated by the component
of the connection along the tangent vector to $\delta \gamma$
\beq
h_{\delta\gamma}[A]=1+\int_{\delta \gamma}A=1+ds\,i_{\bar{\xi}}A\ .
\eeq
Since $A$ takes values in the Lie algebra of the gauge group, the above is, indeed, an infinitesimal gauge
transformation with generator $ds\,i_{\bar{\xi}}A$. The holonomy along $\gamma$ is (loosely) constructed by composing
these transformations back to back along the full path $\gamma$:
\beqa
h_{\gamma}[A] &=& \left(1+ds\,i_{\bar{\xi}(x(1))}A(x(1))\right) \nn\\
& & \quad\cdots \left(1+ds\,i_{\bar{\xi}(x(ds))}A(x(ds))\right)\left(1+ds\,i_{\bar{\xi}(x(0))}A(x(0))\right)\ .
\eeqa
The holonomy is gauge covariant in the following sense. Given a gauge transformation, $g=g(x)$, that transforms the
connection by $A\rightarrow g A g^{-1}-(dg)g^{-1}$, the holonomy only transforms at its endpoints:
\beq
h_{\gamma}[A]\rightarrow g(x(1)) h_{\gamma}[A] g^{-1}(x(0))
\eeq
With this in mind, one can easily construct gauge invariant objects out of the holonomy. A simple example is a Wilson
loop. Suppose the beginning and endpoint of $\gamma$ are the same point, $x(1)=x(0)$, so the curve forms a loop. Then
the Wilson loop is defined by
\beq
W_{\gamma}[A]=Tr(h_{\gamma}[A])\ .
\eeq
From the transformation property of the holonomy given above, this is clearly a gauge invariant quantity. 

\section{Spin Networks}
Spin networks are another way of constructing gauge invariant functionals of $A$. Spin networks were invented by Roger
Penrose \cite{Penrose:SpinNetworks} as a visual aid in the combinatorics of angular momentum. Penrose suggested that
the spin networks might provide a combinatoric approach to quantum geometry, and this vision was first realized by
Rovelli and Smolin decades later in \cite{Rovelli&Smolin:SpinNetworks}.
Suppose we have a network of oriented curves
called edges whose endpoints always lie an a node where two or more edge meets. Take each curve and assign a group
element to it by the holonomy. Now assign to each edge a representation of the gauge group, called a coloring, by
labelling each edge with a spin, $j$. At each node the endpoints of two or more holonomies meet. Since the endpoint of
the holonomy transforms like a vector representation under the gauge group, the endpoint of an edge with spin $j$ can be
viewed as an element of the vector representation space $V_{(j)}$ if it is entering the node, or $V^{*}_{(j)}$ if
it is exiting. The total vector space of endpoints at the node is a tensor product 
\beq
V_{node}=V_{(j_{1})}\otimes V_{j_{(2)}}\otimes\cdots\otimes V^{*}_{(j_{n-1})}\otimes V^{*}_{(j_{n})}\ . 
\eeq
Now, the dual vector space $V^{*}_{node}$ may contain a set of elements of the gauge group that are invariant under gauge
transformations. The normalized versions of these special elements are called intertwiners. In the simplest example,
suppose
we have a node with only one edge coming in, and one edge going out. The dimension of the space of intertwiners is
zero if the edges have different spins, and exactly one if the edges have the same spin. The intertwiner is then
$\delta^{A}_{B}$, which is clearly gauge invariant. For a less trivial example, suppose we have
three edges, one spin-one going out and two spin-$\frac{1}{2}$ edges coming in. The normalized intertwiner is then
$\frac{1}{\sqrt{3}}{\sigma^{i}}_{AB}$ because of the identity
\beqa
{\lambda^{C}}_{A}{\sigma^{i}}_{CD}{\lambda^{D}}_{B}={\lambda^{i}}_{j}{\sigma^{j}}_{AB}&\rightarrow &
{\sigma^{i}}_{AB}={{\lambda^{-1}}^{i}}_{j}{\lambda^{C}}_{A}{\lambda^{D}}_{B}{\sigma^{j}}_{CD}
\eeqa

Since the intertwiners are invariant under
the gauge group, if we attach an intertwiner to each node of the graph by contracting its indices onto the indices of the
edges entering and exiting the edge, the full spin network will be a functional that is invariant under gauge
transformations. As an example consider the $\theta$-spin network consisting of two nodes with three edges connecting
them in the shape of the Greek letter $\theta$. Suppose two of the edges coming out of one node are spin-$\frac{1}{2}$
edges, and the other edge going into the node is a spin-$1$ edge.
The spin network functional is then
\beq
\Psi=\frac{1}{3}\left({\sigma_{i}}^{AB}\, {{h^{(1/2)}_{\gamma_{1}}}^{C}}_{A}\,
{{h^{(1/2)}_{\gamma_{2}}}^{D}}_{B}
\,{{h^{(1)}_{\gamma_{3}}}^{i}}_{j}\,{\sigma^{j}}_{CD}\right)\ .
\eeq
In the above, $\Psi$ can either be viewed as an explicit functional of the the connection $A$, or of the group
elements $h^{(j)}_{\gamma_{i}}$ corresponding to each edge. 

\section{Quantization of area and volume in quantum geometry}
The spin network states are much more than just a convenient way to build functionals of the connection. In fact, they
form a basis for the kinematical Hilbert space and they can naturally be interpreted as quantum geometry. To see this,
we have to look into the spectrum of the area operator. The conjugate momentum to $A^{ij}$ is the area two-form
$E^{i}\w E^{j}$. The natural geometric object to associate with this operator is a two-dimensional surface. Let
$\sigma$ be a two-dimensional surface embedded in $\Sigma$ that has the topology of a square in $\mathbb{R}^{2}$. In
other words, $\sigma:[0,1]\times [0,1]\rightarrow \Sigma$. Given a normal vector $n^{i}$, the area of the surface is
given by $\mathcal{A}(\sigma)=\int_{\sigma}\frac{1}{2}\epsilon_{ijk}n^{i}E^{j}\w E^{k}$. However, in a background
independent context the 
normal cannot be defined without knowing the metric, which is a dynamical quantum operator in the quantum theory. We
can get around this by temporarily fix the gauge so that we can define the directional area 
\beq
\mathcal{A}_{i}=\int_{\sigma}\frac{1}{2}\epsilon_{ijk}E^{j}\w E^{k}\ .
\eeq
This object has strange behavior under gauge transformations, which is why we needed to fix the gauge. However, in the
end, we will take the limit that the coordinate area (but not necessarily the true geometric area) of the surface goes to
becomes infinitesimally small so that
$\sigma$ is an infinitesimal surface. Then the directed are $\mathcal{A}_{i}$ transforms like a vector under
rotations. In this limit the absolute area of the small surface is
\beq
\mathcal{A}(\sigma)=+\sqrt{|\mathcal{A}_{i}\mathcal{A}^{i}|}\ .
\eeq
Now, let us suppose that the area element $\sigma$ pierces one edge of a spin network embedded in $\Sigma$. For
simplicity we will assume that the edge is not tangent to the surface and that the edge pierces the surface only once,
at $x_{0}$.
This splits the curve into $\gamma=\gamma_{1}\cup \delta\gamma \cup\gamma_{2}$, where $\delta \gamma$ is an
infinitesimal segment of the curve that contains the point if intersection.
If the representation of the holonomy of the edge is $h^{(j)}_{\gamma}$, the holonomy splits into three pieces
correspondingly:
\beq
h^{(j)}_{\gamma}=h^{(j)}_{\gamma_{2}}\left(1+\int_{\delta\gamma}A^{i}\,\tau^{(j)}_{i} \right)h^{(j)}_{\gamma_{1}}
\eeq
where $\tau^{(j)}_{i}$ is a basis for the Lie algebra in the spin $j$ representation. In the connection
representation, the operators we have defined become
\beqa
h^{(j)}_{\gamma}[\hat{A}]=h^{(j)}_{\gamma}[A] &\quad \quad &
\hat{\mathcal{A}}_{i}=ik\beta\int_{\sigma}\epsilon_{ijk}\frac{\delta}{\delta A_{jk}}\ .
\eeqa
Thus the action of the directed area element $\hat{\mathcal{A}}_{i}$ on the holonomy of the edge is
\beqa
\hat{\mathcal{A}}_{i}\,h^{(j)}_{\gamma}[A] &=& h^{(j)}_{\gamma_{2}}\,\left(\hat{\mathcal{A}}_{i}\,
h^{(j)}_{\delta\gamma}\right)\,h^{(j)}_{\gamma_{1}}\nn\\
&=& h^{(j)}_{\gamma_{2}}\,\left(
ik\beta\int_{\delta\gamma\times \sigma} \tilde{\delta}(x,x_{0})\ \tau^{(j)}_{i}\right) 
\,h^{(j)}_{\gamma_{1}}\nn\\
&=& h^{(j)}_{\gamma_{2}}\,\left(
\pm ik\beta\ \tau^{(j)}_{i}\right) 
\,h^{(j)}_{\gamma_{1}} \label{AreaOp1}
\eeqa
where the factor of $\pm$ depends on the relative orientation of the directed curve $\gamma$ and directed area element
defined by the two
form $E^{i}\w
E^{j}$---it is ``$+$" if they form a right handed system, and ``$-$" if they form a left-handed system. We note that we
can now take the limit as the coordinate size of $\sigma$ becomes infinitesimally small. As long as the surface still
pierces the edge at $x_{0}$ and it is not tangent to the edge at this point, the result is independent of $\sigma$, so
we can shrink the coordinate area of $\sigma$ as much as we want and the result will still give (\ref{AreaOp1}).
Now, consider the operator
\beqa
\hat{\mathcal{A}}^{2}&=&\hat{\mathcal{A}}_{i}\hat{\mathcal{A}}^{i}\,h^{(j)}_{\gamma}\nn\\
&=& h^{(j)}_{\gamma_{2}}\ \left(-(k\beta)^{2}\delta^{ij}\tau^{(j)}_{i}\tau^{(j)}_{j}\right)\ h^{(j)}_{\gamma_{1}}\ .
\eeqa
The operator $\delta^{ij}\tau^{(j)}_{i}\tau^{(j)}_{j}$ is the well known Casimir operator, which in the spin $j$
representation is:
\beq
-\delta^{ij}\tau^{(j)}_{i}\tau^{(j)}_{j}=j(j+1) \mathbb{I}\ .
\eeq
Because of this property, the operator is purely multiplicative
\beq
\hat{\mathcal{A}}^{2}=(k\beta)^{2}j(j+1)\ h^{(j)}_{\gamma}\ .
\eeq
Taking the the positive square root (which is unambiguous since the operator is diagonal) we have
\beq
\hat{\mathcal{A}}\,h^{(j)}_{\gamma}[A]=k\beta \sqrt{j(j+1)}\, h^{(j)}_{\gamma}[A]\ .
\eeq
This is the celebrated result of loop quantum gravity: the edges of a spin network
diagonalize the area operator, and the spectrum is discrete. In terms of the Planck length, the area spectrum is
\beq
\mathcal{A}(j)=8\pi l^{2}_{PL}\,\beta\,\sqrt{j(j+1)}\ .
\eeq
We note that the spectrum does depend on the Immirzi parameter. Thus, we have the peculiar property that whereas in
the classical theory the parameter did not affect the classical equations of motion at all, in the quantum theory the
discreteness at the Planck length is modulated by the parameter. Currently the parameter is believed to be fixed by
demanding that the Loop Quantum Gravity derivation of black hole entropy matches the Hawking
entropy\cite{Ashtekar:entropy}. Though there is no universal agreement on the precise value, all calculations have
yielded values of the parameter that are on the order of unity.

In addition to the area operator, the volume operator can be constructed essentially out of area operators. The rough
idea behind the construction of the operator is the following. Given a box of dimension $\Delta{x}=l$, $\Delta{y}=w$,
$\Delta{z}=h$, the volume can be written $V=l\times w\times h$. But, we can also compute the volume in terms of the
areas of the faces: $V=\sqrt{A_{x}\times A_{y}\times A_{z}}$. This is the essential idea behind the construction of
the volume operator, which is generally constructed by an appropriate regularization of the operator 
\beq
\hat{V}=+\sqrt{\frac{1}{3!}\big|\epsilon_{ijk}\hat{\mathcal{A}}^{i}\hat{\mathcal{A}}^{j}\hat{\mathcal{A}}^{k}}\big|\ .
\eeq
One can show that the action of the volume operator is nonzero only when it acts on a node of a spin-network, and then
only when the node has four or more edges entering or exiting it\cite{Loll:Volume}. Like the area operator, the spin
network states diagonalize the operator, and the spectrum is discrete.

\section{The kinematical Hilbert space}
In addition to diagonalizing the area and volume operators, the spin network state also provide a complete,
orthonormal basis of states spanning the kinematical Hilbert space, which we will sketch here (for more complete and
rigorous
reviews, see \cite{Rovelli:book, Ashtekar:LQGReview1, Thiemann:Book}).
By kinematical, we mean the set states
in the kernel of the Gauss and diffeomorphism constraints with an appropriate inner product defining a Hilbert space.
In the position, or connection representation, the kinematical states must
be gauge and diffeomorphism invariant functionals of the connection $A$. We will adopt the notation that the Hilbert 
space of states that satisfy, say, the Gauss constraint is $\mathcal{H}_{G}$. The kinematical Hilbert space is
$\mathcal{H}_{kin}=\mathcal{H}_{DG}$, and the physical Hilbert space is $\mathcal{H}_{phys}=\mathcal{H}_{DGH}$.

In the connection representation, the wave function lives in the space of functionals of a smooth connection, denoted
$\mathcal{A}$ (not to be confused with the area operator). Let us restrict ourselves to the subspace of $\mathcal{A}$ consisting
of functionals that only depend on the value of the holonomy of the connection along a set of curves, called a graph, $\Gamma=\{\gamma\}$
embedded in $\Sigma$. Any such functional, $f_{\Gamma}$, called a cylindrical function, can be thought of either as a
functional of the connection,
$f_{\Gamma}=f_{\Gamma}\left(h_{\gamma_{1}}[A], h_{\gamma_{2}}[A],\dots, h_{\gamma_{n}}[A]\right)$, or simply a functional
of the group
elements $f_{\Gamma}=f_{\Gamma}(h_{\gamma_{1}}h_{\gamma_{2}},\dots,h_{\gamma_{n}})$. The inner product of two functionals
can therefore be written either
\beq
\langle f_{\Gamma}| g_{\Gamma}\rangle =\int d\mu_{0}\, f^{*}(h_{\gamma_{1}}[A],\dots,h_{\gamma_{n}}[A])
g(h_{\gamma_{1}}[A],\dots,h_{\gamma_{n}}[A]) 
\eeq
where $d\mu_{0}$ is some measure (that we will make more precise later), or 
\beq
\langle f_{\Gamma}| g_{\Gamma}\rangle =\int dh_{\gamma_{1}}\dots dh_{\gamma_{n}}
 \,f^{*}(h_{\gamma_{1}},\dots,h_{\gamma_{n}})
g(h_{\gamma_{1}},\dots,h_{\gamma_{n}})\label{HaarInnerP}
\eeq
where $dh_{\gamma_{i}}$ is the Haar measure on $SU(2)$. The set of all such functionals (defined on all possible
graphs $\Gamma$) with the given inner product forms a Hilbert space, $\mathcal{H}$. The remarkable property of this
Hilbert space is the direct connection with the functionals of the connection, $A$. In lattice QCD, the inner product
on the lattice is defined analogously to (\ref{HaarInnerP}). However, there we know that the lattice theory is only truly a
discrete approximation to the true continuous theory based on the connection. In the present case, since we are
considering the space of all possible cylindrical functions on all possible graphs, the space is not an approximation
to the continuum at all, but actually contains exactly the same information. More precisely, the Hilbert space is
identical to the space of square integrable functions on a generalization of the space of smooth connections (see
\cite{Thiemann:Book}):
\beq
\mathcal{H}=L^{2}(\overline{\mathcal{A}}, d\mu_{0})\ .
\eeq
The measure on this space, $d\mu_{0}$, called the Ashtekar-Lewandowski measure has been constructed exactly by
considering the
projective limit of nested graphs, and the $\overline{\mathcal{A}}$ is a generalization of the space of connections to
include distributional connections.
Thus, we have a concrete correspondence between
one-dimensional excitations of the connections on one hand and the space of functionals of smooth connections on
the other.

Returning to the inner product (\ref{HaarInnerP}), we need to construct an orthonormal basis of states. Given a
particular graph $\Gamma$, define $\tilde{\mathcal{H}}_{\Gamma}$ to be the set of square integrable cylindrical
functionals
defined on the graph, $\Gamma$. Since we can view the elements of this space as functions of the gauge group, we have
$\tilde{\mathcal{H}}_{\Gamma}=L_{2}[SU(2)^{n}]$, where $n$ is the number of edges in the graph. A major result from the
study
of integration on group manifolds is the Peter-Weyl theorem, which tells us that the irreducible representations of the
gauge group form an orthonormal basis with respect to the Haar measure. Thus, if we label each edge of the graph with
a spin, $j$, the two functions in the inner product will be orthogonal unless they have the same edge labellings. The
proper subspace consisting of graphs whose edges are only labelled with {\it non-zero} spins forms an orthonormal basis
on $\mathcal{H}$. Clearly, we are getting closer and closer to spin networks as basis states. 

Finally, we wish to impose the Gauss and diffeomorphism constraints. The Gauss constraint simply generates $SU(2)$
gauge transformations, thus,
we will search for the space of gauge invariant functionals in $\mathcal{H}$, denoted by $\mathcal{H}_{G}$. This is
where spin networks enter in their full glory. The spin network functionals are gauge invariant by construction, and
under the given inner product, the spin networks form an orthonormal basis spanning
$\mathcal{H}_{G}$. Two functions are orthogonal if they are defined on different graphs, $\Gamma$, or if they have
different edge and intertwiner labellings. 

The diffeomorphism constraint is generally implemented via the inner product. A diffeomorphism
invariant inner product should give zero unless the two spin networks lie on graphs that are in the same knot class.
Suppose one had a measure over the space of all diffeomorphisms $\mathcal{D}\phi$. Then one could define a diffeomorphism
invariant inner product by integrating over all possible diffeomorphisms:
\beqa
\langle \Phi_{\Gamma}|\Psi_{\Gamma'}\rangle_{\mathcal{H}_{DG}}&=&
\int \mathcal{D}\phi \ \langle\Phi_{\Gamma}|U_{\phi}|\Psi_{\Gamma'}\rangle_{\mathcal{H}_{G}}\nn\\
&=&\int \mathcal{D}\phi \ \langle\Phi_{\phi\Gamma}|\Psi_{\Gamma'}\rangle_{\mathcal{H}_{G}} \label{DiffonSN}
\eeqa
One can then think of the diffeomorphism invariant state as the state, $|P_{diff}\Phi_{\Gamma}\rangle$, that satisfies
\beq
\langle P_{diff}\Phi_{\Gamma}|\Psi_{\Gamma'}\rangle =\langle \Phi_{\Gamma}|\Psi_{\Gamma'}\rangle_{\mathcal{H}_{DG}}
\eeq
Loosely speaking, the dual of this state is the group averaged state 
\beq
\langle P_{diff}\Phi_{\Gamma}|=\int \mathcal{D}\phi \ \langle\Phi_{\phi\Gamma}|\ .
\eeq
Although a measure over the set of diffeomorphisms in closed form may not be tractable in practice, the spin network
states solve
this problem in a novel way---they reduce the problem to the problem of determining when two knots (or graphs when we
include nodes) are in the same knot class. This problem may be tractable, though at the present time no algorithm
exists that can uniquely determine the knot class of a given knot imbedded in a three-manifold. We will see in the next
section, that one of the best available
knot invariants, the Kauffman bracket, does distinguish a large class of knots, and is directly related to the Kodama
state.

In total, then, the spin network states form an orthonormal basis of states that span the kinematical Hilbert space
$\mathcal{H}_{DG}$. In addition, there is a duality between the spin network state and functionals on the space of
generalized connections, $\overline{\mathcal{A}}$. This duality is displayed elegantly by the Kodama state, and it
allows for a semi-classical interpretation in a quantum theory whose elementary building blocks are Planck
scale discrete geometries. In our discussion above, we have completely ignored the Hamiltonian constraint, which
encodes the non-gauge dynamics of the gravitational degrees of freedom. Solving this constraint in general is still an
open problem, and has spawned various approaches including the Master Constraint Program\cite{Thiemann:Book} and Spin
Foam models. We will see that the Kodama state does solve a version of the Hamiltonian constraint.

\chapter{The Original Kodama State\label{KState0}}

We are now in a position to discuss the original form of the Kodama state\cite{Kodama:original, Kodama:original2,
Smolin:kodamareview}.
The state was originally derived using
Ashtekar variables with the Immirzi parameter, $\beta$, equal to $-i$. We recall from our discussion of the canonical
theory that for this special case, the phase space consists of a complex valued connection $A^{ij}=\Gamma^{ij}+i
K^{ij}$, and its conjugate momentum, $\pi^{ij}=\frac{1}{2ki}\Sigma^{ij}=\frac{1}{2ki}E^{i}\w E^{j}$. In contrast to
the {\it real} Ashtekar Barbero connection, in this case the connection is the pull-back to the three-manifold of a
spacetime connection: the left--handed spin connection, which defines the parallel transport of left chiral spinors.
The Poisson bracket of the fundamental dynamical variables is
\beq
\left\{A^{ij}\big|_{P}\,,\, \pi_{kl}\big|_{Q}\right\}= \delta^{ij}_{kl}\, \tilde{\delta}(P,Q) \ \longrightarrow \ 
\left\{A^{ij}\big|_{P}\,,\, \Sigma_{kl}\big|_{Q}\right\}= 2ki\,\delta^{ij}_{kl}\,\tilde{\delta}(P,Q)\ .
\eeq
 We take the above to be the fundamental
commutator to carry over to the quantum theory as an operator commutator without modification so that
\beq
\left[\hat{A}^{ij}\big|_{P}\,,\, \hat{\Sigma}_{kl}\big|_{Q}\right]= -2k\,\delta^{ij}_{kl}\,\tilde{\delta}(P,Q)\,.
\eeq
We note that the factor of $i$ occurs in the classical commutator due to the complexification of the phase space, and
it cancels with the $i$ from quantization in the quantum commutator. We
choose to work in a representation where $\hat{A}$ is multiplicative. With this choice, the fundamental operators are
\beqa
\hat{A}^{ij}=A^{ij} &\quad\quad & \hat{\Sigma}_{kl}=2k\frac{\delta}{\delta A^{kl}}\ .
\eeqa
We recall that the Hamiltonian constraint simplifies in these variables to 
\beq
C_{H}(N)=-\frac{1}{2k}\int_{\Sigma}\epsilon_{ijk}E^{i}\w \left(F^{jk}
-{\ts\frac{\lambda}{3}}\Sigma^{jk}\right)\ .
\eeq
To obtain the Kodama state, we first have to choose an operator ordering for the constraints. In this representation, the
choice of operator ordering for the Gauss and diffeomorphism constraints is nearly unambiguous from a geometric
perspective. Assuming the wave function is an integral functional of $A$, the change in $\Psi=\Psi[A]$ under an
infinitesimal diffeomorphism or $SO(3)$ gauge transformation is simply
\beq
\Psi[A]\longrightarrow \Psi[A+\delta A]=\Psi[A]+\int_{\Sigma}\delta A^{ij}\wedge \frac{\delta \Psi}{\delta A^{ij}}
\eeq
where $\delta A=\mathcal{L}_{\bar{N}}A$ for a diffeomorphism, and $\delta A= -D_{A}\lambda$ for a local gauge
transformation. Thus, the natural choice of ordering for the corresponding constraints is
\beqa
\hat{C}_{D}(\bar{N})&=&\int_{\Sigma}\mathcal{L}_{\bar{N}} A^{ij}\w \frac{\delta}{\delta A_{ij}} \nn\\
\hat{C}_{G}(\lambda)&=&\int_{\Sigma}-D_{A}\lambda^{ij}\w \frac{\delta}{\delta A_{ij}}\ .
\eeqa
With this choice, the operator conditions, $\hat{C}_{D}(\bar{N})\Psi[A]=0$ and $C_{G}(\lambda)\Psi[A]=0$, implies that
the functional $\Psi[A]$ must be invariant under (small) diffeomorphisms and local gauge transformations. 

The geometric picture is not so clear for the Hamiltonian constraint. The standard choice that allows us to define the
Kodama state is
\beqa
\hat{C}_{H}(N)&=&-\frac{1}{2k}\int_{\Sigma}\epsilon_{ijk}\hat{E}^{i}
\left(\hat{F}^{jk}-{\ts\frac{\lambda}{3}}\hat{\Sigma}^{jk}\right)\ .
\eeqa
We will look for a state that is in the kernel of the operator in parentheses above. Thus, we need a state that
satisfies
\beq
F^{ij}\,\Psi[A]=\frac{2k\lambda}{3}\frac{\delta }{\delta A_{ij}}\Psi[A]\ . \label{F=EE}
\eeq
With this in mind, we introduce the Chern-Simons functional:
\beq
I[A]=\int_{\Sigma} Y[A]=\int_{\Sigma}{A^{i}}_{j}\w d{A^{j}}_{i} 
+\frac{2}{3}{A^{i}}_{j}\w {A^{j}}_{k}\w {A^{k}}_{i}\ .
\eeq
This integral has some striking properties that we will make use of throughout this work. Most importantly for now,
the functional derivative is 
\beq
\frac{\delta}{\delta A_{ij}}I[A]=-2 F^{ij}\,I[A]\,.
\eeq
Using this property, if we define the state
\beq
\Psi[A]=\mathcal{N}e^{-\frac{3}{4k\lambda}\int_{\Sigma}Y[A]},
\eeq
where $\mathcal{N}$ is a normalization constant, the state clearly satisfies (\ref{F=EE}).
In fact, any state that satisfies property (\ref{F=EE}) will also be a solution to all the constraints. To see this,
we first note that the Hamiltonian constraint is clearly satisfied with our choice of operator ordering (this is why we
chose this ordering). The Gauss constraint is satisfied by the Bianchi identity:
\beqa
\hat{C}_{G}(\lambda)\, \Psi[A] &=& -\frac{3}{2k\lambda}\int_{\Sigma} D_{A}\lambda^{ij}\w F^{ij}\,\times\,\Psi[A] \nn\\
&=& \frac{3}{2k\lambda}\int_{\Sigma} \lambda^{ij}\w D_{A} F^{ij}\,\times\,\Psi[A] \nn\\
&=& 0\ .
\eeqa
Similarly, the diffeomorphism constraint is satisfied since
\beqa
\hat{C}_{D}(\bar{N})\,\Psi[A]&=& -\frac{3}{2k\lambda}
\int_{\Sigma}\mathcal{L}_{\bar{N}}A_{ij}\w F^{ij}\,\times\,\Psi[A]\nn\\
&=& \frac{3}{4k\lambda}\int_{\Sigma}\mathcal{L}_{\bar{N}}Y[A]\,\times\,\Psi[A]\\
&=& 0\,,
\eeqa
where the last line follows from the diffeomorphism invariance of the Chern-Simons functional. Thus,
the state satisfies all the constraints! 

\subsection{The physical interpretation of the Kodama state}
We now turn to the physical interpretation of the state. The key property of the state is that it satisfies
\beq
\widehat{F^{ij}}\,|\Psi \rangle ={\ts\frac{\lambda}{3}}\widehat{E^{i}\w E^{j}}\,|\Psi\rangle \,.
\eeq
It is natural to interpret $|\Psi\rangle$ as the quantum state corresponding to the classical spacetime whose initial
data on $\Sigma$ satisfies
\beq
F^{ij}[A]={\ts\frac{\lambda}{3}}E^{i}\w E^{j}\,.
\eeq
We recall that in order to regain general relativity we need to impose reality conditions on the triad $E^{i}$. Let us
assume that this constraint has been succesfully imposed. Then we can decompose the equation above into its real and
imaginary parts. From the identity $F^{ij}=R^{ij}-iD_{\Gamma}K^{ij}-{K^{i}}_{k}\w K^{kj}$, the result is
\beqa
R^{ij}+K^{i}\w K^{j}&=&{\ts\frac{\lambda}{3}}E^{i}\w E^{j} \\
D_{\Gamma}K^{i}&=&0\ .
\eeqa
But, from our previous discussion, we recognize this as the defining condition, (\ref{dScondition0.1}), in the time gauge
on the initial data of de Sitter space!
Thus, to recap, we have an exact quantum state that satisfies all of the constraints of canonical quantum gravity (in
an appropriate operator ordering), that satisfies a set of operator equations whose classical analogues define de
Sitter space. Clearly we have a candidate for a non-perturbative state representing quantum de Sitter space.

\subsection{The Problems}
Despite these nice features, the state is riddled with problems, which we list here, many of which were brought up in
\cite{Witten:note} by analogy with a corresponding Chern-Simons state in Yang-Mills theory.
\begin{itemize}
\item \textbf{Non-normalizability:} Perhaps the most serious objection to the state, from which many of the other
problems may follow, is that the state does not appear to be normalizable. It is obviously not normalizable under the
kinematical inner product, where
one simply integrates $|\Psi|^{2}$ over all values of the complex Ashtekar connection since the integrand is unbounded.
It
has been suggested by analogy with the corresponding Chern-Simons state in Yang-Mills theory, that if an inner-product
that normalized the Kodama state existed, then the spectrum of excitations around the state would necessarily contain
non-normalizable modes.
This has been substantiated in \cite{Smolin:linearkodama} where it was shown that linearized perturbations around the
Kodama state are non-normalizable under a linearized inner product. The state is not known to
be normalizable under a physical inner product defined by, for example, path integral methods. 
\item \textbf{Negative Energies:} It has been argued, again by analogy with the Yang-Mills Chern-Simons state, that the
Kodama state necessarily contains negative energy sectors. The expectation is that if the
energy of positive helicity graviton-like excitations to the state is positive, then the negative energy states would
necessarily have negative energy. It is this cancellation of positive and negative energy modes that allows the QCD
state to have zero energy, and this could be the mechanism that allows the Kodama state to be in the kernel of the
Hamiltonian constraint.
\item \textbf{CPT Violation:} The states are not invariant under CPT. This is particularly poignant
objection in view
of the CPT theorem of perturbative quantum field theory, which connects CPT violation with Lorentz violation. It is not
known if the result carries over to non-perturbative quantum field theory, but it has yet to be demonstrated that the
Kodama state does not predict Lorentz violation. In the analogy with the QCD Chern-Simons state, the cause of this
failure is that $CP$ interchanges positive and negative helicity states, but the time reversal operator acts trivial
since the QCD state is real. As we will see, the state is only real
because of the complexification necessary
in the construction of the state so that the $i$ from complexification of the phase space cancels with the $i$ from
quantization. A similar result holds for the Kodama state.
\item \textbf{Non-Invariance Under Large Gauge Transformations:} Although the state is invariant under the small gauge
transformations generated by the quantum constraints, it is not invariant under large gauge
transformations where it changes by a factor related to the winding number of the map from the manifold to the gauge
group. However, it has been argued that the non-invariance of the Kodama state under large gauge transformations give
rise to the thermal properties of de Sitter spacetime\footnote{Paradoxically, we will argue the opposite: that demanding
invariance of the generalized states we will present under large gauge transformations gives rise to evidence of
cosmological horizons, which in turn should give rise to the thermal nature of de Sitter space.}\cite{Soo:thermalkodama}.
Thus, non-invariance under large gauge transformations could be a problem or a benefit, but it is deserving of mention.
\item \textbf{Reality Constraints:} The Lorentzian Kodama state is a solution to the quantum constraints in the Ashtekar
formalism where the connection is complex. To obtain classical general relativity one must implement reality
conditions which ensure that the metric is real. In addition, the interpretation of the state as quantum de Sitter
space relied on the reality conditions to separate corresponding initial data condition into its real and imaginary
parts. It is an open problem as to how to implement these constraints
on a general state. Generally it is believed that the physical inner product will implement the reality constraints,
but this could change the interpretation of the state considerably.
\end{itemize}

Thus, we have seen that the Kodama state bears the promise of a non-perturbative definition of quantum de Sitter space,
but it fails some key tests that a physical quantum state must pass. The main purpose of this work is to generalize the
state to address these problems. We will show that there is good evidence that the generalized state
we will construct resolves most of these problems.

\chapter{Generalizing the Kodama State: Construction\label{KState1}}

We are now in a position to discuss the generalization of the Kodama state in an attempt to resolve the problems
associated with the original state. Much of this chapter will closely follow \cite{Randono:GKI} with some important
changes toward the end regarding the solutions to the Gauss and diffeomorphism constraints and the zero curvature
Kodama state.
First
we will discuss the root of the problems associated with the original Kodama state, which will suggest a resolution.
\section{Tracking down the root of the problem}
Many of the problems we listed associated with the Kodama state can be tracked down to the
complexification of the phase space necessary in the construction of the state. To see this, one can simply appeal to the
Euclidean
version of the state. In the Euclidean formalism, the gauge group $SO(4)$ splits into two left and right pieces as in
the complex theory. This corresponds to replacing the spin connection, $\omega$, with the left(right) handed
connection $\omega_{L/R}=\frac{1}{2}(1\mp \star)\omega $, where $\star$ is the internal dual on the $SO(4)$
representation space. In contrast to the Lorentzian sector where the projector, $\frac{1}{2}(1\mp i\star)$, is complex,
in the Euclidean sector everything is real. Pulling back the left handed part of the spin connection to the
three-manifold in the Euclidean theory,
the canonical variables in the Ashtekar formalism consist
of a \textit{real} $SO(3)$ connection and its \textit{real} conjugate momentum. The analogous state in the Euclidean
theory is:
\begin{equation}
\Psi[A]=\mathcal{N}e^{-i\frac{3}{4k\lambda}\int Y_{CS}[A]}.
\end{equation}
Notice, the state is pure phase, because of the reality of the connection and the factor of $i$ in front of the
integral. Comparing the Lorentzian
state with the Euclidean state sometimes gives us insight into which properties of the Lorentzian state stem from
the complexification of the phase space. 

The Lorentzian signature state was clearly non-normalizable under the natural inner product $\int\mathcal{D}A
|\Psi[A]|^{2}$ since the integrand is unbounded. However, if we consider the Euclidean state, 
the integrand is bounded
since the state is a pure phase. The ordinary 
momentum states of non-relativisic quantum mechanics share this pure phase property, and we know that they
are strictly normalizable on a compact manifold, and they are delta-function normalizable on a non-compact manifold. 
Thus, one might expect the Euclidean state to be either delta-function normalizable or strictly normalizable by some
regularization procedure. In fact, it has been shown that linearized perturbations to the Euclidean state are delta-function
normalizable under a linearized inner product\cite{Smolin:linearkodama}. The details of the calculation make it clear
that the reason for the delta-function normalizability is the pure phase nature of the Euclidean Kodama state. In
addition, the state is CPT invariant due to the factor of
$i$ in the argument which inverts under time reversal canceling the action of parity. Although it is not known if the
state has negative energies, one cannot appeal to the analogy with the Yang-Mills Chern Simons state where a
positive energy sector will become a negative energy sector under CPT reversal, because the action of CPT is now
trivial. Since the state is now pure phase, the level of the
Chern-Simons theory is real. Thus, by fine tuning Newton's constant or the cosmological constant (within observational
error), one can make the level an integer, in which case the state is invariant under large gauge transformations.
Finally, there are no reality conditions in the Euclidean theory since the connection and its conjugate momentum are
real. 

\section{A possible resolution}
In the previous section we saw that the Euclidean state appears to be free of most of the known problems associated
with the Lorentzian
state. However, the real world is Lorentzian: \textit{can one salvage the Lorentzian Kodama state despite all of these
problems?} 

The above properties of the Euclidean state suggest that the problems associated with the Lorentzian Kodama state
are rooted in the complexification of the phase space. The phase space is complex because of a particular choice for a
free parameter, the Immirzi parameter $\beta$, which is chosen to be the unit imaginary, $-i$, in the complex
Ashtekar formalism. Modern formulations of Loop Quantum Gravity assume that $\beta$ is an arbitrary \textit{real}
number\cite{Barbero}. The parameter is currently believed to be fixed by demanding consistency with the spin network
derivation of the entropy of an isolated
horizon, and Hawking's formula for the entropy of a static, spherically symmetric black hole\cite{Ashtekar:entropy}. The
first few sections of this chapter will be devoted to generalizing the state to real values
of the Immirzi parameter. The discussion will initially follow along the the lines of \cite{Randono:GK}, and then will
diverge, addressing some deficiencies of that initial attempt at generalizing the Kodama state. We
will show that generalizing the state opens up a large Hilbert space of states each
parameterized by a particular configuration of the three-dimensional Riemannian curvature. By exploiting an analogy
between these states and the ordinary momentum eigenstates of single particle quantum mechanics we will show that
under a kinematical inner product the
states are delta function normalizable and orthogonal unless they are parameterized by the same 3-curvature modulo
$SU(2)$ gauge and diffeomorphism transformations. Using this property we will show that the states can be used to
construct a natural Levi-Civita curvature operator. When this operator is used in the Hamiltonian constraint, all
of the states are annihilated by the Hamiltonian constraint. These states then serve as a basis for a Hilbert space
$\mathfrak{h}\subset \mathcal{H}_{H}$. We then show that there is at least one state in $\mathfrak{h}$ which also
satisfies the remaining constraints, and is therefore in the physical Hilbert space.

\section{Chiral Asymmetric Extension of the Kodama State }
\subsection{Chirally Asymmetric Gravity}
Following along the lines of \cite{Randono:GK}, we begin the construction of the states using a chirally
asymmetric, complex action. This will allow us to make headway in generalizing the state to arbitrary \textit{imaginary}
values of the Immirzi parameter. Later we will analytically extend the states to real values of the Immirzi parameter.
The starting point for the construction of the generalized Kodama states is the Holst action with a cosmological
constant, (\ref{HolstAction2}). At this stage we will
take the Immirzi parameter to be purely imaginary and later analytically extend to real values. The reason we begin
with imaginary $\beta$ is because in this case the Holst action splits into two independent left and right handed components.
In this sense, imaginary values of $\beta$ can not only be interpreted as a measure of parity violation, but more
specifically they measure the degree of \textit{chiral} asymmetry built into the framework of gravity. To see this, we
introduce the left and right handed chiral projection operators $P_{L/R}=\frac{1}{2}(1\mp i\star)$, and define the
chirally asymmetric Einstein-Cartan action (writing $\Sigma \equiv e\w e$):
\beqa
S &=& \frac{1}{k}\int_{M}2(\alpha_{L} P_{L}+\alpha_{R}P_{R})
\star \Sigma \w \left(R-\ts{\frac{\lambda}{6}}\Sigma\right) \nn \\
&=& \frac{2}{k}\int_{M}\alpha_{L}\star\Sigma_{L}\w \left(R_{L}-\ts{\frac{\lambda}{6}}\Sigma_{L}\right)
+ \alpha_{R}\star\Sigma_{R}\w \left(R_{R}-\ts{\frac{\lambda}{6}}\Sigma_{R}\right) \nn\\
&=& \frac{1}{k}\int_{M}(\alpha_{L}+\alpha_{R})\star \Sigma \w \left(R-\ts{\frac{\lambda}{6}}\Sigma\right)
 + i(\alpha_{L}-\alpha_{R})\Sigma \w R.
\eeqa
The coupling constants $\alpha_{L}$ and $\alpha_{R}$ are assumed to be real. The last line above is the Holst action if
we make the identifications $\alpha_{L}+\alpha_{R}=1$ and 
\beq
\beta=\frac{-i}{\alpha_{L}-\alpha_{R}}.
\eeq
We note that in the limiting case when $\alpha_{R}=0$, so that $\beta=-i$, we recover the left handed Einstein-Cartan
action
whose phase space consists of the complex left-handed Ashtekar action and its conjugate momentum. The advantage of
this formalism is that the action splits into two components that, prior to the implementation of reality constraints,
can be treated independently. The reality constraint requires that $e^{I}$ and $\omega^{IJ}$ are real. This implies the
constraints $\Sigma^{IJ}_{L}=\overline{\Sigma^{IJ}_{R}}$ and $\omega^{IJ}_{L}=\overline{\omega^{IJ}_{R}}$. We will
proceed to construct the quantum constraints and a generalization of the Kodama state initially assuming all left
handed variables are independent of right handed variables. Later we will impose the above reality
constraints.

Proceeding to construct the constraints assuming left and right handed variables are independent we find that the
constraint algebra spits into two independent copies which differ by handedness and by the relative coupling constants
$\alpha_{L}$ and $\alpha_{R}$. As usual we demand that the manifold has topology $\mathbb{R}\times \Sigma$ where 
$\Sigma$ is the spatial topology. Gauge fixing to the time gauge is not strictly necessary in the complex formalism
since the complex variables naturally pull-back to the 3-manifold.
However, we will work in the time gauge in order to make contact with the real Ashtekar-Barbero
formalism where gauge fixing is necessary. The left and right handed connections pullback naturally to $\Sigma$ to
form the canonical position variables $A^{ij}_{L}=\omega^{ij}+i K^{ij}$ and 
$A^{ij}_{R}=\omega^{ij}-i K^{ij}$. At this level it is not necessary to fix the three-torsion to zero since the
variables $\omega^{ij}$ only occur in the Lagrangian via $A^{ij}_{L}$ and $A^{ij}_{R}$. The canonical
momenta to $\omega^{ij}_{L}$ and $\omega^{ij}_{R}$ are the two forms $-\frac{i\alpha_{L}}{2k}\Sigma^{L}_{ij}$ and 
$\frac{i\alpha_{R}}{2k}\Sigma^{R}_{ij}$. Eventually we will want 
$\Sigma^{ij}_{L}=\Sigma^{ij}_{R}=E^{i}\w E^{j}$ where $E^{i}_{a}\equiv e^{i}_{a}$ is the spatial triad in the time gauge,
but for now we are treating the two as independent. With this, the canonical commutation relations are
\beqa
\left\{A^{ij}_{L}|_{P}, \Sigma^{L}_{kl}|_{Q}\right\}
&=&-\frac{2k}{\alpha_{L}}\ \delta^{[i}_{m}\delta^{j]}_{n}\ \delta(P,Q)\nn \\
\left\{A^{ij}_{R}|_{P}, \Sigma^{R}_{kl}|_{Q}\right\}
&=& \frac{2k}{\alpha_{R}}\ \delta^{[i}_{m}\delta^{j]}_{n}\ \delta(P,Q)\nn\\
\left\{A^{ij}_{L}|_{P}, \Sigma^{R}_{kl}|_{Q}\right\}&=&\left\{A^{ij}_{R}|_{P}, \Sigma^{L}_{kl}|_{Q}\right\}=0
\label{CCR1}
\eeqa

Each of the constraints contain two independent left and right handed components:
\beqa
C_{H}(N)&=&\alpha_{L}\int_{\Sigma}N \left(*\Sigma^{L}_{ij}\w\left(R^{ij}_{L}-{\ts
\frac{\lambda}{3}}\Sigma^{ij}_{L}\right)\right)
\ +\ (L\rightarrow R) \\
C_{G}(\lambda_{L},\lambda_{R})&=&\alpha_{L}\int_{\Sigma}D_{L}\lambda^{L}_{ij}\w \Sigma^{ij}_{L}\ -\ (L\rightarrow R)
\\
C_{D}(\bar{N})&=&\alpha_{L}\int_{\Sigma} \mathcal{L}_{\bar{N}}A^{ij}_{L}\w \Sigma^{L}_{ij}\ -\ (L\rightarrow R)
\eeqa
As usual, we expect that the Hamiltonian constraint, $C_{H}$, generates time reparameterizations through the lapse, $N$,
the Gauss constraint, $C_{G}$, generates infinitesimal $SU_{L}(2)\times SU_{R}(2)$ transformations with $\lambda_{L}$
and $\lambda_{R}$ as generators, and
the diffeomorphism constraint, $C_{D}$, generates infinitesimal three-dimensional diffeomorphisms along the vector field
$\bar{N}$.
We now need to promote the constraints to quantum operators. We will work in the connection representation where the
momenta are functional derivatives: 
\beqa
\Sigma^{L}_{ij}=\frac{2k}{\alpha_{L}}\frac{\delta}{\delta \omega^{ij}_{L}} 
& & \Sigma^{R}_{ij}=-\frac{2k}{\alpha_{R}}\frac{\delta}{\delta \omega^{ij}_{R}}.
\eeqa
Since the left and right handed variables are independent the Hilbert space also splits into two copies: $\mathcal{H}_{R}\times
\mathcal{H}_{L}$. Thus we will look for solutions of this form. With the operator ordering given above, the
constraints immediately admit the Kodama-like solution:
\beq
\Psi[A_{L}, A_{R}]=\mathcal{N}\exp\left[-\frac{3}{4k\lambda}\left(\alpha_{L}\int_{\Sigma}Y_{CS}[A_{L}]
-\alpha_{R}\int_{\Sigma}Y_{CS}[A_{R}]\right)\right]. \label{LRKodama}
\eeq
where the $Y[A]=A\wedge dA +\frac{2}{3}A\w A\w A $ is the Chern-Simons three-form and the
implied trace is in the adjoint representation of $su(2)$. Here we have used the fundamental identity 
\beq
\frac{\delta}{\delta A_{ij}}\int_{\Sigma}{A^{p}}_{q}\w d{A^{q}}_{p}+
\frac{2}{3}{A^{p}}_{q}\w{A^{q}}_{r}\w{A^{r}}_{p}=-2F^{ij}.
\eeq
We note that in the limit that $\alpha_{L}=1$ and $\alpha_{R}=0$, we regain the original form of the Kodama state. 

\subsection{Imposing the Reality Constraints}
We now need to impose the reality constraints $\Sigma_{L}=\overline{\Sigma_{R}}$ and 
$A_{L}=\overline{A_{R}}$. Imposing the constraints on the position variables is easy since these
are just multiplicative operators. We define the real and imaginary parts of $A_{L}$ by\footnote{Having fixed our
index conventions in the previous sections, in the remaining sections we will drop all indices. Unless stated
otherwise, we will work in the adjoint representation of $SU(2)$.}
\beqa
\omega\equiv Re(A_{L})&=&\frac{1}{2}(A_{L}+A_{R})\\
K\equiv Im(A_{L}) &=& \frac{1}{2i}(A_{L}-A_{R}).
\eeqa
It then follows that $A_{L}=Re(A_{L})+iIm(A_{L})=\omega+iK$ and 
$A_{R}=\overline{A_{L}}=\omega-iK$. The constraint on the momentum variables is slightly more subtle due to the
partial gauge fixing we have employed. Without gauge fixing we would have $\Sigma^{ij}_{L}=e^{i}\w e^{j}
+i{\epsilon^{ij}}_{k}e^{i}\w e^{0}$, but in the time gauge $e^{0}_{a}=0$ so $\Sigma^{ij}_{L}=E^{i}\w E^{j}$ is real. To
implement this in the quantum theory we define
\beqa
\Sigma \equiv Re(\Sigma_{L}) &=& \frac{1}{2}(\Sigma_{L}+\Sigma_{R}) \\
C_{\Sigma}\equiv Im(\Sigma_{L})&=& \frac{1}{2i}(\Sigma_{L}-\Sigma_{R})=0.
\eeqa

We now need to add the constraint $C_{\Sigma}$ into the full set of constraints. We encounter a problem when
evaluating the full set of commutators---the constraint algebra no longer closes. In particular, we find that the
commutator between the Hamiltonian constraint and $C_{\Sigma}$ yields a second class constraint proportional to the
torsion of $\omega$:
\beq
\{C_{H}, C_{\Sigma}\} \sim D_{\omega}*\Sigma=T.
\eeq
Typically this second class constraint is solved at the classical level by replacing the unconstrained $SU(2)$ spin
connection $\omega$ with the torsion-free Levi-Civita connection, $\Gamma=\Gamma[E]$, where $\Gamma$ is a solution to the
torsion
condition $dE^{i}=-{\Gamma^{i}}_{k}\w E^{k}$. In our context this implies that the left and right spin connections are
replace by $A_{L}=\Gamma+iK$ and $A_{R}=\Gamma-iK$. With these replacements, the left and right handed connections will
no longer commute: $\{\omega_{L}, \omega_{R}\}\neq 0$. This is our first indication that something will go wrong with
this initial attempt at generalizing the Kodama state when the full set of constraints is employed. We will see that
we can avoid this issue entirely by a proper reinterpretation of the problem.

However, there is another, potentially more serious problem associated with the introduction of the constraint
$C_{\Sigma}$. In particular, the generalized state we have constructed does not satisfy the quantum constraint
$C_{\Sigma}\Psi = 0$. To illustrate the problem it is useful to redefine the basis of our phase space such that
$\Sigma=\frac{1}{2} (\Sigma_{L}+\Sigma_{R})$ and $C_{\Sigma}$ are the new canonical momenta up to numerical
coefficients. 
The associated canonical position variables are 
\beqa
A_{-\frac{1}{\beta}}&\equiv&\alpha_{L}A_{L}+\alpha_{R}A_{R}=\Gamma+{\ts\frac{1}{\beta}}K \\
A_{\beta}&\equiv&\frac{\alpha_{L}A_{L}-\alpha_{R}A_{R}}{\alpha_{L}-\alpha_{R}}=\Gamma-\beta K, 
\eeqa
which can be seen from the canonical commutation relation that follow directly from \ref{CCR1},
\beqa
\{A_{-\frac{1}{\beta}}, C_{\Sigma}\}&=& i2k\ \tilde{\delta} \nn\\
\{A_{\beta}, \Sigma \}&=& -i2k\beta\ \tilde{\delta} \nn\\
\{A_{-\frac{1}{\beta}}, \Sigma\}&=& 0 \nn\\
\{A_{\beta}, C_{\Sigma}\} &=& 0 \label{CCR2}.
\eeqa
We recognize $A_{\beta}$ and $\Sigma=E\w E$ as the Ashtekar-Barbero connection and its momentum that emerge in the real
formulation of LQG. The reason for introducing these variables is that the constraint $C_{\Sigma}\Psi=0$ takes a
particularly simple form. In the connection representation, $C_{\Sigma}=2k \frac{\delta}{\delta A_{-1/\beta}}$.
Thus, we must have\footnote{The limiting case when $\beta \rightarrow \mp i$ must be treated separately here because
in those cases we have an initial primary constraint that $\Sigma_{R/L}=0$.} 
\beqa
C_{\Sigma}\Psi=2k \frac{\delta}{\delta A_{-1/\beta}}\Psi =0 \ &\longrightarrow& \ \Psi=\Psi[A_{\beta}].
\eeqa
That is, the wavefunction can only be a function of the Ashtekar-Barbero connection $A_{\beta}$ and is independent of
$A_{-1/\beta}$.

Now we need to check that the state (\ref{LRKodama}) is only a function of $A_{\beta}$. To do so, we express the
state in terms of the $A_{\beta}$ and $A_{-1/\beta}$. We rewrite the state in a form that will be convenient for
later use:
\begin{equation}
\Psi[A]=\mathcal{N}\exp\left[\frac{-3i}{4k\Lambda \beta^{3}}\int_{\Sigma}Y_{CS}[A]-(1+\beta^{2})Y_{CS}[\Gamma] 
+2\beta(1+\beta^{2})Tr(K\wedge R_{\Gamma})\right]. \label{GK1}
\end{equation}
Here $\Gamma$ and $K$ are explicit functions of both $A_{\beta}$ and $A_{-1/\beta}$, given by
\beqa
\Gamma &=& \frac{A_{\beta}+\beta^{2}A_{-1/\beta}}{1+\beta^{2}}\\
K &=& \frac{1}{\beta}(\Gamma-A_{\beta}).
\eeqa
We see that the state is explicitly a function of \textit{both} $A_{\beta}$ and $A_{-1/\beta}$. Thus,
$C_{\Sigma}\Psi\neq 0$.

\subsection{Resolution}
The problems we have encountered with this initial attempt at generalizing the Kodama state are twofold. First, we
encounter a second-class constraint whose solution requires that we introduce the torsion-free spin connection
$\Gamma=\Gamma[E]$. This means that the left and right handed variables will no longer commute, or in the new
variables, $A_{\beta}$ and $A_{-1/\beta}$ will no longer commute. Second, we find that the reality constraint on the
momentum requires that the wave function is a functional of $A_{\beta}$ only, which is not true for our left-right
asymmetric state. We can recast the problem in a slightly more intuitive way by eliminating $A_{-1/\beta}$ in favor of
the momentum $\Sigma$. That is, we explicitly write $A_{-1/\beta}=\frac{1}{\beta^{2}}((1+\beta^{2})\Gamma-A_{\beta})$
and treat $\Gamma[E]$ as an explicit function of the momentum conjugate to $A_{\beta}$. Then the problem can be
restated, \textit{why is the wavefunction an explicit function of both position and momentum variables?} The problem
of defining the commutator of $A_{\beta}$ and $A_{-1/\beta}$ is transmuted into the problem of defining the operator
$\Gamma[E]$ which occurs explicitly in the Hamiltonian through the Levi-Civita curvature, $R_{\Gamma}$, or the extrinsic
curvature, $\frac{1}{\beta}(\Gamma-A_{\beta})$, depending on how one writes the constraints. We will see that we can
address
both of these problems by analytically extending the state to real values of the Immirzi parameter, $\beta$. This will
allow us to exploit an analogy between the generalized Kodama state and the non-relativistic momentum eigenstates, which
will suggest a reinterpretation of the explicit momentum dependence of the state and at the same time suggest a
natural definition of the Levi-Civita curvature operator $R_{\Gamma}$. This will be the subject of the rest of the
this chapter.

\section{The Generalized Kodama States}
\subsection{Properties of the real state}
We now consider the state (\ref{GK1}) when the Immirzi parameter $\beta$ is taken to be a non-zero, but otherwise
arbitrary real number. Modern formulations of Loop Quantum Gravity begin with arbitrary real values of $\beta$ in the
canonical construction because the analysis of real $SU(2)$ connections is better understood than that for complex
connections. In addition, it is believed that thermodynamic arguments will eventually fix the value of the Immirzi
parameter unambiguously. For our purposes, taking $\beta$ to be real changes the properties of the generalized Kodama state considerably.

We first address the issue of the explicit momentum dependence of the state. We appeal to a similar situation in
ordinary single particle quantum mechanics. The generalized Kodama state for real values of $\beta$ shares many
properties in common with the ordinary momentum eigenstates. First
of all, both states are pure phase. This means that they are bounded, which has implications for the inner product.
Whereas the complex Kodama state is unbounded, which implies that the state is non-normalizable under a na\"{i}ve inner
product, the real state is pure phase and therefore may be normalizable in the strict sense if the phase space is
compact, or delta-function normalizable if the phase space is non-compact. Secondly, the momentum eigenstates share
the property in common with the generalized Kodama state that they ostensibly depend explicitly on both the momentum
and position variables. Of course, the role of the momentum in the momentum eigenstates is very different from the
role of the position variables. The state $\Psi_{p}(x)=\mathcal{N}e^{i p\cdot x-iEt}$ is
explicitly a function of the position variable only, but it is \textit{parameterized} by the momentum $p$.
That is, the momentum eigenstates form a large family of orthogonal states distinguished by a particular value of
$p$, and together with the ordinary inner product, they span the Hilbert space. This is the interpretation we will adopt
for the role of the momentum in the generalized Kodama state. In particular, we will show that at this level, the
states form an infinite class of states which span a Hilbert space $\mathfrak{h}$. The definition of 
his space and its relation of the
Hilbert space, $\mathfrak{h}$ to the ordinary kinematical and physical Hilbert spaces will become more clear shortly.

To see this explicitly, we rewrite the state (\ref{GK1}) in a more suggestive form by absorbing irrelevant factors which
depend only on the momentum through $\Gamma[E]$ into the normalization constant. The state becomes:
\begin{equation}
\Psi_{R}[A]=\mathcal{P}\exp\left[i\kappa\int_{\Sigma} A\wedge R-\frac{1}{2(1+\beta^{2})}Y_{CS}[A] \right].
\end{equation}
Here we see explicitly, $A$ plays the role of the position variable $x$, the Levi-Civita curvature
$R=d\Gamma +\Gamma\w\Gamma$ plays the role of the momentum, $\kappa=\frac{3(1+\beta^{2})}{2k\lambda\beta^{3}}$ 
is
simply a scaling factor, we have a dimensionless energy $\frac{1}{2(1+\beta^{2})}$, and the
Chern-Simons term $\int Y_{CS}[A]$ plays the role of the time variable. We note that it has been independently suggested
that the Chern-Simons invariant is a natural time variable on the canonical phase space\cite{Soo:thermalkodama}. With
this interpretation, the
generalized state is not a single state at all, but a large class of states parameterized by a specific configuration of
the three-dimensional Levi-Civita curvature, $R$. 

\subsection{The na\"{i}ve inner product}
We can push the analogy further by considering the inner product between two states with different curvature
configurations $\langle\Psi_{R'}|\Psi_{R}\rangle$. The analogue of this is the inner product of two momentum states:
\beqa
\langle p'|p\rangle&=&\int d^{n}x \ \Psi^{*}_{p'}[x,t]
\Psi_{p}[x,t] \nn\\
&=& \mathcal{P}[p', p]\int d^{n}x \ \exp[-i(p'-p)\cdot x]\nn\\
&\sim& \delta^{n}(p'-p).
\eeqa
Following along these lines, we define a na\"{i}ve inner product:
\beqa
 \langle\Psi_{R'}|\Psi_{R}\rangle_{ naive} &=& \mathcal{P}[\Gamma',\Gamma]\int_{\Sigma}\mathcal{D}A\ 
  \Psi^{*}_{R'}[A]\Psi_{R}[A]\nn\\
 &=& \int \mathcal{D}A\ \exp\left[-i\kappa \int_{\Sigma} A\w (R'-R)\right] 
 \eeqa
 Formally integrating over the space of connections we have
 \beq
 \langle \Psi_{R'}|\Psi_{R}\rangle_{naive} \sim \delta(R'-R).
 \eeq
 Thus, under this na\"{i}ve inner product, two states are orthogonal unless they are parameterized by the same configuration
of the Levi-Civita curvature. The deficiency of this inner product is that it is not gauge invariant. If the two
fields $R'$ and $R$ represent the same curvature written in a different gauge, either $SU(2)$ or diffeomorphism, they
will be orthogonal. Thus, in order to construct a proper inner product on $\mathfrak{h}$ we need to modify the inner
product to make it gauge invariant.

\subsection{Gauge covariance and the kinematical inner product}
In order to define a gauge invariant inner product on $\mathfrak{h}$ we first need to discuss the gauge properties of the
generalized states. The set of states ${\Psi_{R}}$ are not strictly speaking $SU(2)$ gauge or diffeomorphism invariant.
The reason is because of the presence of the parameter $R$ in the argument which acts like an effective ``background"
against which one can measure the effect of a gauge transformation or diffeomorphism. This is not unfamiliar. We
recall from equation (\ref{DiffonSN}) we
encountered the same difficulty with the spin-network states where the graph serves as a ``background" against which one
can measure the effect of a diffeomorphism shifting the connection, $A$ (see e.g. \cite{Rovelli:book}). In the spin
network states, the action of a
one-parameter diffeomorphism $\phi_{-\bar{N}}$ on the connection configuration, $A$, is equivalent to shifting the graph
in the opposite direction
by $\phi_{\bar{N}}$. Similarly, one can show that the combined effect of an SU(2) gauge transformation and diffeomorphism
on the field configuration which we will denote by $\phi_{\{g^{-1},-\bar{N}\}}A$ is equivalent to the inverse
transformation
on the curvature denoted by $\phi_{\{g, \bar{N}\}}R$. Thus, under the action of the Gauss and diffeomorphism constraint,
the state transforms as follows:
\begin{equation}
\Psi_{R} \rightarrow  
\hat{U}_{\phi}(g^{-1},-\bar{N}) \Psi_{R} =
 \Psi_{\phi_{\{g,\bar{N}\}}R}.
\end{equation}

The strategy with the spin network states is to implement the diffeomorphism symmetry via the inner product
where the diffeomorphism symmetry is manageable, and this is the strategy we will also adopt. To make the inner product
gauge invariant, we introduce the measure
$\mathcal{D}\phi_{\{g,\bar{N}\}}$ over the set of all SU(2) gauge transformations (which may be accomplished by the
Haar measure) and the set of all diffeomorphisms. Although a measure over the set of all diffeomorphisms is undefined,
the end result may still be manageable due to the specific form of the integrand. As we have seen, this is true in the
inner product on spin network states, where the problem of defining a measure over the group of diffeomorphisms is
relegated to the problem of determining when two graphs are in the same equivalence class of knots. A similar result
applies here. To see this, we define the kinematical inner product as follows:
\beq
\langle\Psi_{R'}|\Psi_{R}\rangle_{kin} = 
\int \mathcal{D}\phi_{\{g,\bar{N}\}}\langle \Psi_{R'}|U_{\phi}(g,\bar{N})|\Psi_{R}\rangle_{naive}\ .
\eeq
From the gauge covariance of the states $\Psi_{R}$ we have:
\beqa
\langle\Psi_{R'}|\Psi_{R}\rangle_{kin} &=& 
\int \mathcal{D}\phi_{\{g,\bar{N}\}}\langle\hat{\phi}_{\{g^{-1},-\bar{N}\}}\Psi_{R'}|\Psi_{R}\rangle \nn\\
&=& \int \mathcal{D}\phi_{\{g,\bar{N}\}}\langle\Psi_{\phi_{\{g,\bar{N}\}}R'}|\Psi_{R}\rangle \nn\\
&\sim& \int \mathcal{D}\phi_{\{g,\bar{N}\}}\delta(\phi_{\{g,\bar{N}\}}R'-R)\nn\\
&=& \delta(\mathcal{R'}-\mathcal{R})
\eeqa
where in the last line $\mathcal{R'}$ and $\mathcal{R}$ are elements of the equivalence class of curvatures modulo
$SU(2)$-gauge and diffeomorphism transformations. Thus, the problem of defining a measure over the set of diffeomorphisms
is reduced to the problem of determining when two curvatures are gauge related---a problem that is all too familiar from
classical General Relativity. The states $\Psi_{R}$ and
$\Psi_{R'}$ are orthogonal unless there is a diffeomorphism and/or SU(2) gauge transformation relating $R$ and $R'$.

Thus, we see that, at this level the generalized Kodama states form an infinite class of states that are delta-function
normalizable with respect to a very natural inner product. We will define the space that they span with the induced
inner product to be the Hilbert space $\mathfrak{h}$.

\subsection{Levi-Civita curvature operator}
Continuing the analogy with the momentum eigenstates we proceed to define a Levi-Civita curvature operator on
$\mathfrak{h}$.
We recall the momentum operator can be defined in terms of the momentum eigenstates:
\beq
\hat{p}=\int d^{n}p'\ p'|p'\rangle\langle p'|.
\eeq
By construction, the states $|p\rangle$ are then eigenstates of $\hat{p}$.

Since the generalized states $|\Psi_{R}\rangle$ represent a family of orthogonal states parameterized by the curvature
configuration $R$, it is natural to define a curvature operator such that the states are curvature eigenstates. Analogous to
the momentum operator, we define the operator in its diagonal form as follows (writing $\phi=\phi_{\{g,\bar{N}\}}$):
\begin{equation}
\int_{\Sigma}\alpha\wedge\hat{R}_{\Gamma}=\int \mathcal{D}\phi\mathcal{D}\Gamma '
\left[\left(\int_{\Sigma}\lambda\wedge\phi R'_{\Gamma '}\right) |\Psi_{\phi R'}\rangle
\langle\Psi_{\phi R'}|\right]
\end{equation}
where $\alpha$ is an arbitrary $su(2)$ valued one-form serving as a test function, and $\mathcal{D}\Gamma '$ is an
appropriate measure to integrate over all values of the Levi-Civita 3-curvature
$R'_{\Gamma '}$. When operating on a state $|\Psi_{R}\rangle$ it is understood that the intermediate inner product is
the 
na\"{i}ve inner product. That is,
\beqa
& &\int_{\Sigma}\alpha \wedge \hat{R}\ |\Psi_{R}\rangle \nn\\
& &=\int \mathcal{D}\phi \mathcal{D}\Gamma '
\left[\left(\int_{\Sigma}\alpha\wedge\phi R'_{\Gamma '}\right) |\Psi_{\phi R'}\rangle
\langle\Psi_{\phi R'}|\Psi_{R}\rangle_{naive}\right]\nn\\
&& =\int \mathcal{D}\phi \mathcal{D}\Gamma '
\left[\delta(\phi R'-R)\ \left(\int_{\Sigma}\alpha\wedge\phi R'_{\Gamma '}\right)
|\Psi_{\phi R'}\rangle \right] \nn\\
&& =\int_{\Sigma}\alpha\wedge R_{\Gamma}\ 
|\Psi_{R}\rangle
\eeqa

Thus, with this definition, the states
$|\Psi_{R}\rangle$ are eigenstates of the curvature operator $\hat{R}_{\Gamma}$:
\beq
\int_{\Sigma}\alpha\wedge \hat{R}\ |\Psi_{R}\rangle=\int_{\Sigma}\alpha\wedge R \ |\Psi_{R}\rangle. \label{Roperator}
\eeq

\subsection{The Hamiltonian constraint}
We now address the issue of the Hamiltonian constraint. The beauty of the complex Ashtekar formalism is that the
Hamiltonian constraint simplifies to the point where it is solvable, admitting the Kodama state as a quantum solution to
the Hamiltonian constraint. Our partial parity violating version of the Ashtekar action held the promise of a
simplified Hamiltonian until the reality constraints were imposed, which introduced second class constraints on the
torsion. When solved, the constraint implies that the left and right handed connections no longer commute because
they both
contain a term $\Gamma[E]$. The real formulation of the Holst action, is plagued with the same problem. Although the
phase space consists of just the connection $A$ and its conjugate momentum, the Hamiltonian constraint explicitly
contains terms involving $\Gamma[E]$. Depending on how one writes the constraint, they enter via extrinsic curvature
terms, $K=\frac{1}{\beta}(\Gamma-A)$, or through the Levi-Civita curvature, $R=d\Gamma+\Gamma\w\Gamma$. The standard
representation of the Hamiltonian constraint is\footnote{The term involving $*\Sigma\w D_{\Gamma}K$ may be unfamiliar
since it is usually not included in the constraint. The term does explicitly occur in the Hamiltonian decomposition,
but it can be integrated away using the fact that $\Gamma$ is torsion free so $D_{\Gamma}*\Sigma=0$. We will keep the
term explicitly because it simplifies the algebra in the next step.}
\beq
C_{H}=\int_{\Sigma}*\Sigma\w\left(F+(1+\beta^{2})({\ts \frac{1}{\beta^{2}}}D_{\Gamma}K
-K\w K)-{\ts \frac{\lambda}{3}}\Sigma \right),
\eeq
where $F=F[A]$ is the curvature of $A$. Because of the complexity of this constraint, it appears to be very difficult
to determine if our generalized Kodama
states are in the kernel of the corresponding quantum operator. However, the constraint can be rewritten by
substituting the extrinsic curvature terms in favor of the Levi-Civita curvature. In this form, discussed previously, the
constraint then takes the form
\beq
C_{H}=\int_{\Sigma} *\Sigma\w\left((1+{\ts \frac{1}{\beta^{2}}})R-{\ts\frac{1}{\beta^{2}}}F
-{\ts \frac{\lambda}{3}}\Sigma \right).
\eeq
This form is particularly convenient for our purposes because we have already suggested a form for the Levi-Civita
curvature operator on $\mathfrak{h}$. In the standard Kodama operator ordering where $*\Sigma$ is placed on the far left,
the full
set of generalized Kodama states are in the kernel of the Hamiltonian by virtue of being in the kernel of the quantum
operator
\beq
\int_{\Sigma} \alpha\w \left((1+{\ts \frac{1}{\beta^{2}}})\hat{R}-{\ts\frac{1}{\beta^{2}}}\hat{F}
-{\ts \frac{\lambda}{3}}\hat{\Sigma} \right)
 \eeq
where $\alpha$ is a test function. To see this, in the connection representation $\Sigma$ is a differential operator
which acts on $\Psi_{R}[A]$ by:
\beq
-{\ts\frac{\lambda}{3}}\Sigma \ \Psi_{R}[A]=i2 k\beta {\ts\frac{\lambda}{3}}\ \frac{\delta \Psi_{R}[A]}{\delta A} 
=\left({\ts \frac{1}{\beta^{2}}}F-(1+{\ts \frac{1}{\beta^{2}}})R\right)\ \Psi_{R}[A]\ .
\eeq
The curvature $F$ cancels since $\hat{F}$ is multiplicative in the connection representation.
We are left with 
\beq
(1+{\ts \frac{1}{\beta^{2}}}) \int_{\Sigma}\alpha\w (\hat{R}-R)\ \Psi_{R}[A]\ ,
\eeq
which vanishes by (\ref{Roperator}). Thus, for any curvature configuration, $R$, with the standard Kodama operator
ordering we have
\beq
\hat{C}_{H}\ |\Psi_{R}\rangle =0 . 
\eeq
Thus, we have shown that with the given choice of operator ordering and the curvature operator defined by the
generalized Kodama states,
$\mathfrak{h}$ is a subset of the Hilbert space spanned by solutions to the Hamiltonian constraint:
\beq
\mathfrak{h}\subset \mathcal{H}_{H}\ .
\eeq

\section{The action of the remaining constraints on the generalized Kodama states}
The remaining constraints still need to be addressed. We will show that $\mathfrak{h}$ contains at least one state
that solves all of the constraints and is therefore an element of $\mathcal{H}_{phys}=\mathcal{H}_{GDH}$. 

We will work with the smeared form of the Gauss and diffeomorphism constraints.
In the connection representation the action of the Gauss constraint on a Kodama state is
\beq
\int_{\Sigma}D_{A}\lambda_{ij}\w(\hat{\Sigma}^{ij})\  \Psi_{R}[A]
 = {\ts\frac{3}{\lambda}}\int_{\Sigma}D_{A}\lambda_{ij}\w\left((1+{\ts\frac{1}{\beta^{2}}})\hat{R}^{ij}
 -{\ts\frac{1}{\beta^{2}}}F^{ij}\right)\ \Psi_{R}[A]\,.
\eeq
Integrating by parts and using the Bianchi on the term involving $F$ on the right, the vanishing of the Gauss
constraint on a generalized Kodama state reduces to
\beq
\int_{\Sigma}D_{A}\lambda_{ij}\w\hat{R}^{ij}\ \Psi_{R}[A]=0\,.
\eeq
A similar result holds for the diffeomorphism constraint whose action on a Kodama state is
\beq
\int_{\Sigma}\mathcal{L}_{\bar{N}}A_{ij}\w \Sigma^{ij}\ \Psi_{R}[A]=
{\ts\frac{3}{\lambda}}\int_{\Sigma}\mathcal{L}_{\bar{N}}A_{ij}\w\left((1+{\ts\frac{1}{\beta^{2}}})\hat{R}^{ij}
 -{\ts\frac{1}{\beta^{2}}}F^{ij}\right)\ \Psi_{R}[A]\,.
\eeq
As in the previous case we can eliminate the term on the right involving $F$, this time by employing the
identity
\beq
\int_{\Sigma}-2\mathcal{L}_{\bar{N}}A_{ij}\w F^{ij}
=\int_{\Sigma}\mathcal{L}_{\bar{N}} Y_{CS}[A],
\eeq
which vanishes since the Chern-Simons action is invariant under small diffeomorphisms. In total then action of the Gauss
and diffeomorphism constraints on a generalized Kodama state reduces to
\beqa
C_{G}(\lambda)\,\Psi_{R}[A] &=& \int_{\Sigma}D_{A}\lambda_{ij}\w\hat{R}^{ij}\ \Psi_{R}[A] \nn\\
C_{D}(\bar{N})\,\Psi_{R}[A]&=& \int_{\Sigma} \mathcal{L}_{\bar{N}}A_{ij}\w \hat{R}^{ij}\ \Psi_{R}[A]\,.
\eeqa
These constraints must be solved for the state to be in $\mathcal{H}_{phys}$. There is at least one state which
satisfies all three constraints. The two constraints above are clearly solved by the state $\Psi_{R=0}[A]$ since
$\hat{R}\,|\Psi_{R=0}\rangle=0$. Thus, we have
\beqa
C_{\{GDH\}}\,|\Psi_{R=0}\rangle =0 \ &\longrightarrow &\  |\Psi_{R=0}\rangle \in \mathcal{H}_{phys}\,.
\eeqa

The zero curvature state itself has some remarkable properties. In particular, the state is pure Chern-Simons in the
connection representation:
\beq
\langle A|\Psi_{R=0}\rangle=\mathcal{P}\exp\left[-\frac{3i}{4k\lambda\beta^{3}}\int_{\Sigma}Y_{CS}[A]\right]\,.
\eeq
Since not every three-topology admits a flat Levi-Civita connection, there are some limitations on the topology of
$\Sigma$ for this state to exist. The form of the state in the connection representation makes it clear that the state
is invariant under (small) gauge transformations and diffeomorphism, since the Chern-Simons action is invariant under
these transformations. Indeed, it can easily be checked that the state is annihilated by the Gauss and diffeomorphism
constraints. It is well known that the state is not invariant under large gauge transformations, gauge transformations
whose global structure cannot continuously be deformed to the identity,
unless the prefactor is\footnote{We recall that in our definition of the Kodama state, we have used the trace in the
adjoint representation of $SU(2)$. To obtain the level, one generally uses the trace in the fundamental
representation. The difference is a factor of two: $\int Tr_{Adj}(A\w dA) =2\int Tr_{fund}(A\w dA)$. Thus, the relevant
prefactor in the Kodama state is $2\times \frac{3}{4k\lambda\beta^{3}}$.}
\beqa
\frac{3}{2k\lambda\beta^{3}}=\frac{\kappa}{4\pi}\ &\longrightarrow &\ \kappa=\frac{3}{4 G\lambda\beta^{3}}
\eeqa
for some integer $\kappa$, called the level of the Chern-Simons theory. For realistic modern day values, the factor $G\lambda
\simeq 10^{-120}$ and $\beta$
is believed to be of order unity. Thus, $\kappa\simeq 10^{120}$ is enormous, and we have considerable freedom in fine
tuning
$G$, $\lambda$, or $\beta$ within experimental error so that $\kappa$ is an integer. So this so-called ``pre-quantization"
condition on the Chern-Simons theory does not realistically put severe restrictions on the values of the various
physical constants, and we can safely assume that the level is an integer. With this assumption, the state has the
remarkable property that its functional form is well known in both the connection and the spin-network representation.
In particular, let $\Gamma$ be a spin network with edges and intertwiners labeled by representations of $SU(2)$. We
loosely define the spin-network representation of any state to be
\beq
\Psi[\Gamma]=\langle A|\Psi\rangle =\int \mathcal{D}A\ \Psi[A] \,h_{\Gamma}[A]\,
\eeq
where $h_{\Gamma}[A]$ is the holonomy representation of the spin network.
For the $R=0$ Chern-Simons state above, we have
\beq
\Psi_{R=0}[\Gamma]=\int \mathcal{D}A \exp\left[-\frac{3i}{4k\lambda\beta^{3}}\int_{\Sigma}Y_{CS}[A]\right]\,
h_{\Gamma}[A]
\eeq
The need to define the above integral more rigorously requires a framing of the spin network and a deformation of the
gauge group to $SU_{q}(2)$ where the deformation parameter is $q=e^{\frac{2\pi i}{\kappa+2}}$ and $\kappa$ is given
above.
The celebrated result of Witten is that the functional integral can be computed and the result is the Kauffman bracket
(with parameter $q$) of the framed spin-network\cite{Witten:knots}:
\beq
\Psi_{R=0}[\Gamma]=\langle\Gamma |\Psi_{R=0}\rangle = K_{\Gamma}(q)\,.
\eeq
Since the Kauffman bracket\cite{Kauffman:q-deformation} is a well known knot invariant that is invariant under diffeomorphisms (i.e. the
Reidemeister moves of regular isotopy), and the edge labels of the spin network are invariant under gauge
transformations, it is clear
from the spin-network representation as well that the state must satisfy the Gauss and diffeomorphism constraints. The
q-deformation of the gauge group and the framing of the spin-networks has interesting physical ramifications that we
will explore shortly.

\chapter{Generalizing the Kodama State: Physical Interpretation\label{KState2}}

We now turn to the physical interpretation of the generalized Kodama state (or states). From an initial data
perspective, we
will see that among the
generalized states, the zero curvature state has the natural interpretation as the flat slicing of quantum de Sitter
space. This state has the special property that the loop transform of the state is well known as the Kauffman bracket 
of framed and q-deformed spin networks. This will yield evidence of both quantization at the Planck scale and
the existence of cosmological horizons in a background independent context.
We will also show that the full set of states naturally fall into the category of WKB state corresponding to de Sitter
space. This will yield evidence that the multiplicity of states may be related by a sector of the Lorentz group that is
cut off by our gauge fixing procedure. The full set of states are invariant under CPT, and they are expected to be
delta-function normalizable.

\section{Physical interpretation: canonical analysis}
First we will discuss the
physical interpretation from the context of initial data on a time slice in the 3+1 formulation of general relativity.
We would like to show that the classical counterparts of the quantum operator equations satisfied by the generalized
Kodama state are in fact the equations for the initial data of de Sitter space. The calculation will be
complicated somewhat due to the Immirzi terms, which require careful treatment.

We start by recalling that the generalized Kodama state satisfies the quantum version of the classical condition
\beq
(1+{\ts\frac{1}{\beta^{2}}})\,R^{ij}-{\ts\frac{1}{\beta^{2}}}
F^{ij}-{\ts\frac{\lambda}{3}}\Sigma^{ij}=0\,. \label{condition1}
\eeq
Since the particular combination of variables occurs explicitly in the Hamiltonian constraint itself, any solution to
the above trivially satisfies this constraint. We still have the Gauss and diffeomorphism constraints to deal with.
As in the quantum case, the Gauss constraint is
\beq
\int_{\Sigma}\lambda_{ij}\,D_{A}(\Sigma^{ij})\approx 0\,.
\eeq
Using the above condition this reduces to
\beq
\int_{\Sigma}\lambda_{ij}\,D_{A}R^{ij}\approx 0
\eeq
where we have used the Bianchi identity $D_{A}F=0$ to simplify. The diffeomorphism constraint is 
\beq
\int_{\Sigma}\mathcal{L}_{\bar{N}}A_{ij}\w \Sigma^{ij} \approx 0\,.
\eeq
Using our condition (\ref{condition1}), this reduces to
\beq
\int_{\Sigma} \mathcal{L}_{\bar{N}}A_{ij}\w R^{ij} \approx 0
\eeq
where we have used the identity
\beq
\int_{\Sigma}-2\mathcal{L}_{\bar{N}}A_{ij}\w F^{ij}
=\int_{\Sigma}\mathcal{L}_{\bar{N}} Y_{CS}[A],
\eeq
which vanishes whenever $\Sigma$ has no boundary. Thus, we clearly see that these two remaining constraints,
can be solved, for example, by the additional condition $R^{ij}=0$, though this is not a unique solution. For example,
we can also solve all of the constraints by the additional condition $R^{ij}=F^{ij}$. This solution to the constraints corresponds
to the initial data on a different slicing of de Sitter space.

Let us now assume that we have found a set of fields $(A_{0}, \Sigma_{0})$ defined on the whole of the Cauchy
surface $\Sigma$
that solve the Hamiltonian constraint by virtue of condition (\ref{condition1}), and also solve the Gauss and
diffeomorphism constraints. Thus, $(A_{0}, \Sigma_{0})$ form a good set of initial data. We could equivalently
characterize the initial data by the variables $(K_{0}, E_{0})$ by the definitions $\Sigma^{ij}=E^{i}\w E^{j}$ and
$K^{i}=\frac{1}{2\beta}{\epsilon^{i}}_{jk}(\Gamma^{jk}-A^{jk})$. Choosing particular functions $N$, $\bar{N}$, and
$\lambda^{ij}$ for the Lagrange multipliers in the constraints, the Hamiltonian evolution generated by the constraints
pushes the data forward in time so that at any time $t$, $(K_{t}, E_{t})$ still solves the constraints. We can
identify the Lagrange multipliers with physical fields as follows (in the time gauge):
\beqa
N&=&e^{0}_{0}\nn\\
N^{a}&=& E^{a}_{i}\,e^{i}_{0}\nn\\
\lambda^{ij}&=&\omega^{ij}(N\bar{n})\nn\\
{\omega^{i}}_{0}(\bar{\eta})&=&0\,.
\eeqa
With these identifications, we can reconstruct the four-dimensional fields $\omega^{IJ}_{\mu}(t)$ and
$e^{K}_{\nu}(t)$. Since the constraints plus the evolution equations reproduce the Holst form of the Einstein-Cartan
equations, the four-dimensional fields will be a solution to the latter. With this hindsight, we can use our knowledge
about the four-dimensional field solutions to gain insight into the data on any time slice. In particular, we recall
that although the Holst form of the Einstein-Cartan equations contains the Immirzi parameter explicitly, since it
occurs in front of a term that depends explicitly on the torsion, and the torsion vanishes when the full set of
equations is solved, any solution $\omega^{IJ}_{\mu}$ and $e^{K}_{\nu}$ that solves these equations must be
independent of the Immirzi parameter. Choosing a time slice of the spacetime, we can construct the initial data
$(K_{0}, E_{0})$ by pulling back the fields to the slice and the Lagrange multipliers $(N, \bar{N}, \lambda)$ by the
above identifications. Since the four-dimensional data is independent of $\beta$, so will be this initial data. Now,
let us suppose we have chosen a solution that satisfies the initial data equation (\ref{condition1}), which we now
write in terms of $K$ and $E$:
\beq
R^{ij}+K^{i}\wedge K^{j}+{\ts\frac{1}{\beta}}\,{\epsilon^{ij}}_{k}D_{\Gamma}K^{k}
-{\ts\frac{\lambda}{3}}E^{i}\w E^{j}=0\,.
\eeq
Now, since the initial data is independent of the Immirzi parameter, we can take the derivative of the above equation
with respect $\beta$:
\beqa
\frac{\partial}{\partial \beta}\left(R^{ij}
+K^{i}\wedge K^{j}+{\ts\frac{1}{\beta}}\,{\epsilon^{ij}}_{k}D_{\Gamma}K^{k}
-{\ts\frac{\lambda}{3}}E^{i}\w E^{j}\right)=0 \nn\\
\longrightarrow D_{\Gamma}K^{i}=0\,.
\eeqa
Putting these two conditions together, we have
\beqa
R^{ij}+K^{i}\wedge K^{j}&=&{\ts\frac{\lambda}{3}}E^{i}\w E^{j}\\
D_{\Gamma}K^{k}&=&0\,.
\eeqa
Once again, we recognize the above as precisely the initial data formulation, (\ref{dScondition0.1}), of de Sitter space!
Indeed, one can
verify that any set of data that satisfies the above solves all the constraints, and that the first equations are the
de Sitter condition (in the time gauge) pulled back to the three-manifold:
\beqa
^{(4)}R^{ij}&=&^{(3)}R^{ij}+K^{i}\w K^{j}={\ts\frac{\lambda}{3}}E^{i}\w E^{j} \nn\\
^{(4)}{R^{i}}_{0} &=& DK^{i}=0
\eeqa
Thus, the classical analogues of the operator equations satisfied by the generalized Kodama state define the initial
data for de Sitter space. 

In deriving the above, we have in essence used the fact that the transformation from the canonical position variable
$K$ to the Ashtekar-Barbero connection, $A$, is a canonical transformation in the classical vacuum theory---that is,
the transformation changes the definition of the canonical variables but does not affect the equations, which remain
independent of $\beta$ (at least in vacuum). It is well known that this canonical transformation cannot be implemented
without anomalies in the quantum theory, and that the Planck scale discreteness is fine tuned by the Immirzi
parameter. Thus, we should expect that although the classical analogue of the generalized Kodama state is
$\beta$-independent,
the quantum theory will almost certainly have $\beta$-dependent properties. This we will verify in upcoming sections. 

\section{Physical interpretation: WKB analysis}
In constructing the generalized Kodama states we exploited many of the properties that the states share in common with
the ordinary single particle momentum eigenstates, properties which we exploited in
the construction of the states. In this section, we will show that the analogy can be extended to the physical
interpretation
as well---in common with the momentum eigenstates, the generalized Kodama states are WKB states (see e.g.
\cite{Rovelli:book}) in
addition to
being exact quantum states. We will suggest that, in contrast to the original Kodama state, the set of generalized states
includes not only a WKB analogue of de-Sitter space, but also first order vacuum perturbations to de Sitter space.
Although one might expect that this could explain the multiplicity of states, but the interpretation of these extra
states will remain somewhat unclear until the next chapter. 

Let us first briefly review the WKB construction of the momentum eigenstates. We recall that in the WKB approximation,
the wave function is split into amplitude and phase, $\Psi=\rho e^{i \Theta /\hbar}$. This splits the Schr\"{o}dinger
equation into two pieces---one piece expresses the conservation of the probability current density, and the other piece
is the Hamilton-Jacobi equation for $\Theta$ plus a small correction proportional to $\hbar$ that is interpreted as a
quantum potential. The WKB approximation consists of solving the Schr\"{o}dinger equation in successive powers of
$\hbar$. Since the quantum potential is proportional to $\hbar$, to lowest order in $\hbar$, the wave function can be
approximated by $\Psi \simeq e^{i S_{0}/\hbar}$ where $S_{0}$ is a solution to the Hamilton-Jacobi equation. A preferred
solution to the Hamilton-Jacobi equation is obtained by evaluating the action on the fixed points of a variation as a
function of the endpoints of the variation. The momentum eigenstates can be
constructed in exactly this way. To see this, consider the non-relativistic free particle action
\beq
S=\int \frac{1}{2}m \dot{x}^{2}\ dt=\int
\left(p \cdot \dot{x}
-\frac{p^{2}}{2m}\right)\, dt.
\eeq
Hamilton's equations of motion obtained by finding the fixed points of the action are
\beqa
\dot{x}=p/m & & \dot{p}=0
\eeqa
whose general solution is $p=p_{0}=constant $, $x=x_{0}+p_{0}/m\  t$. Inserting this into the action we obtain
\beqa
S_{0}&=&\int^{t}_{t=0}\left(p_{0}\cdot\frac{dx}{dt}-\frac{p_{0}^{2}}{2m}\right) dt\nn\\
&=& \int^{x}_{x=0} p_{0} \cdot dx-\int^{t}_{t=0}\frac{p_{0}^{2}}{2m}dt \nn\\
&=& p_{0} \cdot x-\frac{p_{0}^{2}}{2m}t
\eeqa
The corresponding zeroeth order WKB state is clearly the momentum eigenstate corresponding to $p_{0}$:
\beq
\Psi^{WKB}_{p_{0}}(x)=e^{i S_{0}}=e^{i(p_{0}\cdot x-E\,t)}
\eeq
From the perspective of the WKB approximation, at this level the state is only an approximate state, however, 
inserting it into the Schr\"{o}dinger equation shows that it is an exact
solution. Thus, in a sense it is an exact quantum state that is as close to classical as a quantum state can be.
Because of this, it inherits many familiar properties of the classical solution. We note that the simplicity of the
above derivation followed from the fact that the action evaluated on this class of solutions to Hamilton's equations
was a total derivative, and therefore independent of the path connecting the endpoints---one only needs to specify the
boundary data to evaluate the integral. We will see that the action for vacuum quantum gravity has the same property
which allows for exact WKB states. 

To illustrate the WKB nature of the Kodama states, we begin with the Holst action 
\beq
S_{H}=\frac{1}{k}\int \star e\w e\w R +\frac{1}{\beta}e\w e\w R -\frac{\lambda}{6}\star e\w e\w e\w e 
\eeq
and recall that the general solution to the equations of motion takes the form
\beqa
& R=\frac{\lambda}{3}e\w e +C &\nn\\
&T=0. \label{GeneralSolution}&
\eeqa
We note that for $C=0$ the solution to the above is de Sitter space. As
in the non-relativistic case, we expect that to lowest order in $\hbar$, the wave function has the form
$\Psi=\mathcal{N}e^{i S_{0}/\hbar}$ where $S_{0}$ is the action evaluated on a particular solution to the equations of
motion.
Thus, we need to evaluate the action by inserting (\ref{GeneralSolution}) back into the Holst action and choosing a set
of boundary data which restricts one to a particular solution. In the following we will assume that the Weyl tensor is
small, keeping only first order terms in $C$, so we are dealing with small vacuum perturbations to
de Sitter space such as linearized gravitational waves propagating through an expanding universe. Setting the torsion
to zero in the Holst action annihilates the term involving the Immirzi parameter since $ e\w e \w R\sim e\w
DT$. One can view the Immirzi term as a control on the width of fluctuations of torsion in the path
integral\cite{Starodubtsev:MMgravity}. 
This term contains valuable information in the quantum theory so we will keep it by finding an
action which is equivalent to the Holst action at the fixed points. Inserting the equations of motion back into the
Holst action and dropping all $C\w C$ terms, we find
that on shell, the Holst action is equivalent to the topological action:
\beq
S_{0} \simeq \frac{3}{2k\lambda}\int \star R\w R +\frac{1}{\beta}R\w R.
\eeq
The first term is the Euler class and the second term is the second Chern class. Thus, as in the nonrelativistic case,
the action evaluated on the equations of motion is a total derivative. Since the above topological terms are tailored
to detect topological changes in the evolving manifold, it may be necessary to sum over past topological histories as
well as field configurations in the sum over histories---if the topological history of the manifold is trivial the above
terms are zero. Thus, for definiteness one might, for example, choose Hartle and Hawking's ``no boundaries" model and
sum over all closed topologies with a future spacelike boundary $\Sigma$ of a given spatial topology. With such a
choice, the action becomes
\beq
S_{0}\simeq \frac{3}{2k\lambda}\int_{\Sigma}\left(\star+\frac{1}{\beta}\right)\omega\w d\omega +\frac{2}{3}\omega\w
\omega
\w\omega.
\eeq
To make contact with the canonical theory, we partially fix the gauge to the time gauge, thereby reducing the gauge
group to $SU(2)$. In addition, recalling that the vanishing of the three-torsion emerged as a second class constraint
which was solved prior to canonical quantization, we set the three torsion to zero on the spacelike boundary $\Sigma$. We
proceed to rewrite the action in terms of the Ashtekar-Barbero connection, $A$, and a particular spatial triad
configuration whose Levi-Civita curvature, $R=R[E]$, we have seen will serve as a parameterization of a class of
states. The result is
\begin{equation}
S_{0}\simeq\frac{-3}{4k\lambda \beta^{3}}\int_{\Sigma}Y_{CS}[A]-(1+\beta^{2})Y_{CS}[\Gamma] 
+2\beta(1+\beta^{2})Tr(K\wedge R_{\Gamma}), 
\end{equation}
which is precisely the argument of the generalized Kodama state $\Psi_{R}$ for a particular configuration of $R$. As
before, to get a true state one must, in addition, impose the Gauss and diffeomorphism constraints. This is equivalent
to requiring that the bulk symmetries of the action---local $SO(3,1)$ and 4 dimensional diffeomorphism symmetries---are
imposed on the boundary action as well. On the boundary, the diffeomorphism symmetry reduces to $Diff_{3}$, and the
local Lorentz symmetry is reduced to the rotation subgroup due to the gauge fixing on the boundary. These symmetries
are implemented on the boundary action, which can now be viewed as explicitly a functional of $A$ parameterized by a
triad configuration yielding $R[E]$. The constraints that implement these symmetries are precisely the Gauss and
diffeomorphism constraints.

Thus, we have seen that the WKB states corresponding to first order perturbations to de Sitter space are nominally the
generalized Kodama states $\Psi_{R}[A]$. These states are still subject to the Gauss and diffeomorphism constraints,
but we have seen that there exists at least one state in this class that satisfies both of these additional
constraints: the flat curvature solution $\Psi_{R=0}[A]$. Since the analysis we presented is valid not only for the
pure de Sitter solution, but also for first order perturbations therein, it remains a possibility that the different
curvature states are states corresponding to small perturbations to quantum de Sitter space, and this was the first
interpretation the author offered for these additional states\cite{Randono:GKII}. However, it is also possible that these
small
perturbations are simply quantum fluctuations that are absorbed into the ground state, which is identically quantum de
Sitter space. If we adopt this view, the different curvature states may all be different manifestations of the same 
state, which are related by a sector of the local Lorentz group that is cut off by our gauge fixing procedure. Evidence
for this comes from the fact that the functional
\beq
\Psi[\omega]=\exp\left[ \frac{3i}{2k\lambda}\int_{\Sigma}\left(\star
+\frac{1}{\beta}\right)\omega\w d\omega +\frac{2}{3}\omega\w
\omega \w\omega \right]
\eeq
acts as an umbrella state that contains the whole sector of states, $\Psi_{R}[A]$, which are distinguished only by the
information that is held fixed for each state. Since the above functional is invariant under (small) local Lorentz
transformations, it is plausible that all of these disparate states are unified into a single state when the full Lorentz
group is retained. We will
find strong evidence in sections to follow when we attempt to do canonical quantum gravity without fixing to the time
gauge. This will allow us to retain the full local Lorentz group, and we will find strong evidence that the umbrella
state given above is a unique state corresponding to quantum de Sitter space alone.

\section{The flat space de Sitter state}
In the previous section, we saw that the generalized states can be interpreted as WKB states corresponding to de Sitter
space, but that the interpretation of the multitude of states is somewhat unclear. The states can be understood
as quantum states corresponding
to the initial data of de Sitter space or possibly first order pertubations to de Sitter space. Since the specific
form of the initial data depends on the slicing we choose, to get a more clear physical picture of the semi-classical
correspondence it would be valuable to match a state to the initial data formulation of de Sitter space.
We recall that de Sitter space has multiple standard slicings in which the spatial topology is one of $\mathbb{R}^{3}$,
$\mathbb{S}^{3}$, or $\mathbb{H}^{3}$. We will focus on the $\mathbb{R}^{3}$ slicing where the metric takes the following
form:
\beq
ds^{2}=-dt^{2}+e^{2t/r_{0}}(dx^{2}+dy^{2}+dz^{2}).
\eeq
We recall that these coordinates do not cover the whole of de Sitter spacetime due to the presence of a cosmological
horizon
in the expanding universe. It is of particular interest for our problem that in these coordinates, the 3-curvature is
identically zero, $^{(3)}R^{ij}_{ab}=0$ so the 3-space is flat Euclidean space. Thus, to pick out the de Sitter state,
we restrict the spatial topology to be $\mathbb{R}^{3}$ and turn the flat space condition into a quantum operator
equation
\beq
\int_{\mathbb{R}^{3}}\alpha\w\hat{R}\ |\Psi\rangle=0
\eeq
for all values of the test function, $\alpha$. Thus, with the topology restricted to $\mathbb{R}^{3}$, we see that we can
identify the $\Psi_{R=0}$ state with the flat space slicing of de Sitter space
\beqa
&|\Psi_{dS}\rangle =|\Psi_{R=0}\rangle &\nn\\
&\langle A|\Psi_{R=0}\rangle = 
\mathcal{P}\exp\left[-\frac{3i}{4k\lambda\beta^{3}}\int_{\mathbb{R}^{3}} Y_{CS}[A]\right].&
\eeqa
We have seen that this state is distinguished in being identically gauge and diffeomorphism invariant, and it has the
remarkable property that its functional form is known in both the connection and the framed spin-network
representations. 

\subsection{Evidence of Planck scale discreteness and cosmological horizons}
The spin network representation of the $R=0$ state gives us some insight into the quantum geometry of de Sitter space.
In particular, we see evidence of both discretness at the Planck scale, and the existence of a cosmological horizon. 
We recall that in Loop Quantum Gravity, the area operator
has eigenvalues $A=8\pi G \beta\sqrt{j(j+1)}$ where $j$ is the representation of the edge piercing the operator valued
2-surface. A q-deformed spin network has edges labelled by representations of the q-deformed group. The deformation
parameter is given by $q=e^{\frac{2\pi i}{\kappa+2}}$ where $\kappa=\frac{3}{4 G \lambda \beta^{3}}$ is the level of
the
Chern Simons theory in point. If the deformation parameter is a root of unity the representations terminate at a maximum
spin $j_{max}$ where $j_{max}=\frac{\kappa}{2}$. At
large values of $\kappa$, and therefore $j_{max}$, we have $A\simeq 8\pi G \beta j$. This yields a maximum value for
the area of an indivisible surface:
\beq
A_{max}\simeq 4\pi \left( \frac{r_{0}}{\beta}\right)^{2}.
\eeq
Since the Immirzi parameter is believed to be of order unity, this area on the order of area of the de Sitter horizon
$A_{dS}=4\pi r_{0}^{2}$. We interpret this result as evidence of
the existence of a cosmological horizon in the quantum theory. It is of interest to note that in this simple model, we
see a new quantum mechanical feature---the Immirzi parameter, which has no classical effect in vacuum, is not only
significant at very small length scales in the quantum theory where it determines the scale on which Planck scale
discreteness occurs,
\beq
A_{small}\sim \beta\, l_{PL}^{2}\ ,
\eeq
 but it also appears to play a significant role at extremely large, cosmological distances where
it modulates the de Sitter radius,
\beq
A_{large}\sim \left(\frac{r_{0}}{\beta}\right)^{2}\ .
\eeq
It should be stressed that a quantum cosmological horizon is not necessarily composed of
minimally divisible surfaces, and it is not known if this property holds in a more robust treatment. However, since
very large scale, cosmological effects of quantum gravity appear to be the most promising routes to quantum gravity
phenomenology\cite{Glikman:QGPhenomenology}, this tentative result is enticing.

We also note that the identification of cosmological horizons with the representation cut-off of a quantum group
potentially solves a major paradox in identifying horizons in a background independent context. The paradox is that
the identification of cosmological horizons (and one might attempt to argue {\it all} horizons in general relativity)
is fundamentally observer dependent. In de Sitter space, one can show that all geodesic observers will see both
particle and event horizons---however, the location of these horizons depends on where the observer is located in the
spacetime. This problem is especially poignant in a background independent context since different observers will not
agree on the location of a horizon and there is no global reference to which one can retreat as common ground. The
q-deformation scenario resolves this since every observer, no matter where the observer is located will see a cut-off
in the area spectrum on the order of the expected cosmological horizon, and this is a background independent property
of the quantum geometry.

\section{CPT Invariance}
In this section we will discuss the action of the discrete $C$, $P$, and $T$ operations on the generalized Kodama
states, showing that the wave functions violate $CP$ and $T$, while preserving $CPT$ symmetry. Each of the symmetries 
will have action both in the fiber and the base manifold. It will be useful to work in the Clifford algebra
representation to demonstrate the action of the symmetries in the fiber. We recall in the Clifford algebra, there are
two natural inner products. If $A$ and $B$ are arbitrary elements of the Clifford algebra, we have the standard inner
product given by $\langle A, B\rangle=Tr(AB)$, and the metric or $\star$-inner product given by $\langle
A,B\rangle_{\star}=Tr(\star AB)$. The generalized Kodama states utilize both inner products as all of the states are
contained in the umbrella state, which we will make use of in this section:
\beq
\Psi^{\beta}_{R}[A]=\exp \left[\frac{3i}{2k\Lambda}\int_{\Sigma}Tr(\star Y[\omega]+\beta^{-1} Y[\omega])\right] 
\label{CPTKodama}
\eeq
where $Y[\omega]=\omega \wedge d\omega +\frac{2}{3}\omega \wedge \omega \wedge \omega$ (no trace here). In the above it
is understood that to regain a specific state, the three torsion is set to zero on $\Sigma$ and a frame field
is fixed such that $R=R[E]$. We will consider each discrete symmetry separately. 

\subsection{Parity reversal}
On the base manifold, the action of parity simply inverts volume forms on the three-space. Since $Tr(Y)$ and
$Tr(\star Y)$ are ordinary three forms, they are inverted by parity. In addition, parity has action in the fiber which
can be deduced from the ordinary Dirac equation: $(i\gamma^{\mu}\partial_{\mu}-m)\psi[\vec{x},t]=0$. Under parity,
$x=(t,\vec{x})\rightarrow x'=(t,-\vec{x})$. A simple calculation shows that $\psi'(x')= P\psi(x)$, where $P=\eta
i\gamma^{0}$ and $\eta$ is an arbitrary phase factor, 
satisfies the space inverted Dirac equation $(i\gamma^{\mu}\partial '_{\mu}-m)\psi'(x')=0$. The Clifford algebra must
then transform in the adjoint representation so that
\beqa
&\mathcal{P}(\gamma^{\mu})=P\gamma^{\mu}P^{-1} & \nn\\
& (\gamma^{0},\gamma^{i})\rightarrow (\gamma^{0}, -\gamma^{i}) &
\eeqa
as expected. Under this transformation parity preserves the ordinary inner product on the Clifford algebra,
\beqa
& Tr(P AB P^{-1})=Tr(AB) & \nn\\
& \mathcal{P}(\langle A, B\rangle) =\langle A, B\rangle &
\eeqa
but inverts the metric inner product,
\beqa
& Tr(\star P AB P^{-1})=Tr(P^{-1}\star PAB)=-\langle A,B\rangle_{\star} &\nn\\
&\mathcal{P}(\langle A,B\rangle_{\star})=-\langle A,B\rangle_{\star}. &
\eeqa
The net effect on the wave functions \ref{CPTKodama} is an inversion of the Immirzi parameter:
\beq
\Psi^{\beta}_{R}\rightarrow \mathcal{P}(\Psi^{\beta}_{R}) =\Psi^{-\beta}_{R}.
\eeq
This is consistent with the general maxim with a growing body of evidence\cite{Soo:CPT, Rovelli:Torsion,
Freidel:Torsion, Randono:Torsion, Mercuri:Torsion},
\begin{quote}
\textit{The Immirzi parameter is a measure of parity violation built into the framework of quantum gravity.}
\end{quote}

\subsection{Time reversal}
Time reversal in a diffeomorphism invariant theory is somewhat subtle due to the effective dissapearance of time in
the canonical formalism. However, one should expect from general arguments that the quantum mechanical time reversal
operator should be anti-unitary. To see this, let us suppose we have an inner product which annihilates the
Hamiltonian constraint (the matrix elements of the scalar constraints are all zero) but preserves a causal ordering of
the ``in" and ``out" states. Consider then the inner product of $\Psi$ at $t_{2}$ and $\Phi$ at $t_{1}$ where
$t_{2}> t_{1}$ given by
\beq
\langle \Psi, t_{2}|\Phi, t_{1}\rangle .
\eeq
Since the inner product annihilates the Hamiltonian constraint, which generates time reparametrization, the above
inner product cannot depend on the particular values $t_{1}$ and $t_{2}$, though it does depend on their causal
ordering. Therefore, time reversal is equivalent to interchanging $t_{1}\leftrightarrow t_{2}$ (ignoring internal
degrees of freedom) so
the inner product
becomes $\langle \Phi, t_{2}|\Psi, t_{1}\rangle$. Again using the fact that the inner product only depends on the
causal ordering we conclude:
\beq
\mathcal{T}(\langle \Psi |\Phi \rangle)=\langle \Phi |\Psi \rangle =\langle \Psi |\Phi \rangle^{*}.
\eeq
Using the such an inner product to construct the connection representation, $\Psi[A]=\langle A| \Psi\rangle$, we
deduce the (partial) action of the time reversal operator is antiunitary:
\beq
\mathcal{T}(\Psi[A])=U \Psi^{*}[A]
\eeq
where $U$ is a unitary operator which represents the action of time reversal on any remaining internal degrees of
freedom.

To see the net effect of time reversal, we again appeal to the ordinary Dirac equation which we write in Hamiltonian
form: $i\partial_{t}\psi=H\psi$, where $H=i\gamma^{0}\gamma^{i}\partial_{i}-\gamma^{0}m$. The time reversed Dirac
equation is then $Ti\partial_{t}T^{-1}\psi'=THT^{-1}\psi'$, where $\psi'(\vec{x}, -t)=T\psi(\vec{x},t)$.
Time reversal must commute with the Hamiltonian so we must have
\beqa
&T\gamma^{0}T^{-1}=\gamma^{0}& \\\
& T\gamma^{i}T^{-1}=-\gamma^{i}&
\eeqa
and it must reverse the direction of time in the Dirac equation so 
\beq
TiT^{-1}=-i
\eeq
as expected from the previous arguments. On the inner products, we then have
\beqa
\langle A,B\rangle &\rightarrow& \langle A,B\rangle \\
 \langle A,B\rangle_{\star} &\rightarrow&  -\langle A,B\rangle_{\star}.
\eeqa
Thus, in total, the net effect on the generalized Kodama states is
\beq
\Psi^{\beta}_{R}\longrightarrow \mathcal{T}(\Psi^{\beta}_{R})=\Psi^{-\beta}_{R}.
\eeq
Evidently, time reversal undoes the action of parity inversion.

\subsection{Charge Conjugation}
On the surface the action of charge conjugation is simple: gravitons are their own antiparticles so charge conjugation
should not effect the wave-function of pure gravity. However, although one may be able to make this statement precise
in a perturbative context, uncovering the graviton from a non-perturbative framework is a difficult ordeal, and,
regardless, there may be subtle non-perturbative effects which determine the true action of the charge conjugation
operator. 

Like the time reversal operator, the charge conjugation operator is also anti-unitary. Specifically, on a Dirac
spinor, the charge conjugation operator take the form $\mathcal{C}(\psi)=\mathbb{C}\psi^{*}$ where
\beq
\mathbb{C}^{-1}\gamma^{I}\mathbb{C}=-\gamma^{I*}\ .
\eeq 
We expect the the charge conjugation operator to have a similar action on a quantum gravity wave function. However,
since the Kodama states are pure phase, they are not invariant under complex conjugation, and it can be easily checked
that the remaining action of $\mathbb{C}$ on the fiber indices has no effect. Thus, there appears to be a problem. The
resolution comes from the identification of $\int_{\Sigma}Y$ and $\int_{\Sigma} \star Y$ as topological charges. To
justify this we appeal to the gravitational conformal anomaly which states
\beqa
& d\ast J_{5} \sim R\wedge R &\nn\\
& \Delta Q_{5} \sim \Delta\int_{\Sigma}Y &
\eeqa
 where $J_{5}^{\mu}=\bar{\psi}\gamma_{5}\gamma^{\mu}\psi$ is the axial current and $Q_{5}$ is the axial charge which
is inverted by the charge conjugation operator.
Since the above equations are separately $P$ and $T$ invariant, in order for the them to be $CPT$ invariant, it must
be that the right hand side inverts under charge conjugation. Thus, if we identify it with a topological charge, the
charge conjugation must appropriately modify the topological structure on the bundle over $M$ bounded by $\Sigma_{1}\cup
\Sigma_{2}$ 
to invert $\int_{\Sigma}Y$. With this identification the charge operator acts as the identity on the generalized
Kodama states:
\beq
\mathcal{C}(\Psi_{R})=\Psi_{R}.
\eeq

\subsection{Net Effect}
In total we have shown that the states are CPT invariant:
\beqa
& \begin{array}{ccccccc}
  & \mathcal{C} &   & \mathcal{P} &  & \mathcal{T} & \nn\\
\Psi^{\beta}_{R}&\longrightarrow & \Psi^{\beta}_{R} &\longrightarrow &\Psi^{-\beta}_{R} &
\longrightarrow & \Psi^{\beta}_{R} 
\end{array} & \nn\\
& & \nn\\
&\mathcal{CPT}(\Psi^{\beta}_{R}) \longrightarrow \Psi^{\beta}_{R}\ .&
\eeqa
Finally, we note that although it is not known whether the states represent positive semi-definite energy states, one
cannot appeal to the standard argument from the analagous Chern-Simons state in Yang Mills theory to argue for negative
energies. The standard argument states that if there were a positive energy sector, the $CPT$ inverted sector, which
is also in the kernel of the constraints, must be a negative energy sector. Since $CPT$ does not invert the
generalized Kodama states, this argument no longer applies.

\section{Normalizability}
In the previous chapter we showed that the generalized Kodama states form a Hilbert space $\mathfrak{h}$
that is a subset of the space of solutions to the Hamiltonian constraint $\mathcal{H}_{H}$.
With respect to the natural inner product on the function space, the states are delta-function normalizable on
$\mathfrak{h}$. We have seen that there is at least one element, $\Psi_{R=0}$, in $\mathfrak{h}$ that also solves the
remaining
constraints. It follows that this state is also delta-function normalizable with respect to the inner product on
$\mathfrak{h}$. Although this does not imply that the state is normalizable with respect to the physical inner
product (which as of yet has not been constructed), it does give indirect evidence that the state may pass this test.
The proper behavior of the state under CPT is further indirect evidence that the state may be normalizable since the
failure of this invariance in the analogous Yang-Mills state was due to the exchange of normalizable, positive-energy
modes with non-normalizable, negative-energy modes under the action of the CPT operator. We can also appeal to the
results of \cite{Smolin:linearkodama} where it was shown that the linearization of the Euclidean state is
delta-function normalizable under a linearized inner product. As discussed there, the result is a consequence of the
pure phase nature of the Euclidean state. Since our state, $\Psi_{R=0}$, is identical in its functional form to the
Euclidean Kodama state, we fully expect the same result to carry over to this state. In the next section we outline a
path integral inspired method of computing the physical inner product, analogous to the spin foam construction. We
will see that there is an interesting connection between the inner product and the Macdowell-Mansouri partition
function. Thus, the problem of determining the normalizability of the generalized Kodama states may reduce to the problem
of computing the Macdowell-Mansouri partition function.

\section{The physical inner product and the Macdowell-Mansouri formulation of gravity}
In this section we discuss an interesting connection between the true, physical inner product defined by path integral
methods and the Macdowell-Mansouri formulation of gravity\cite{MMoriginal}. The kinematical inner product between two
states
of different 3-curvature given by, $\langle R'| R\rangle \sim \delta(\mathcal{R'}-\mathcal{R})$, is unlikely to be the
proper physical inner product, which is generally defined by a sum over histories as in the Hawking path integral and spin
foam methods. To this end we can formally write the true inner product as a sum over histories connecting two 3-curvature
states on the spacelike boundaries. That is, we take the boundary of our 4-dimensional manifold, M, to be two
3-dimensional spacelike hypersurfaces, $\Sigma_{2}$ and $\Sigma_{1}$, on which the the states $\Psi_{R'}$ and $\Psi_{R}$
are respectively defined. The expected physical inner product is then:
\beq
\langle \Psi_{R'}|\Psi_{R}\rangle_{phys}=\langle \Psi_{R'}|\int^{E_{2}}_{E_{1}}\mathcal{D}\omega 
\mathcal{D}e\ e^{iS_{EC+\beta}}|\Psi_{R}\rangle_{kin}
\eeq
where in the sum over histories we have fixed the spatial triad configurations $E_{2}$ and $E_{1}$ whose Levi-Civita
curvatures are $R'$ and $R$ respectively. In computing the path integral it will be useful to work in the connection
representation so that the total inner product takes the form
\beqa
\langle \Psi_{R'}|\Psi_{R}\rangle_{phys} &=& \int \mathcal{D}A'\mathcal{D}A\ \Psi^{*}_{R'}[A'] \Psi_{R}[A]
\int^{E_{2}}_{E_{1}}\mathcal{D}\omega \mathcal{D}e \ e^{iS_{H+\lambda}}\nn \\
&=& \int^{E_{2}}_{E_{1}}\mathcal{D}\omega \mathcal{D}e\  e^{-i\frac{3}{2k\lambda\beta}
\int_{M}\star R\wedge R +\frac{1}{\beta} R\wedge R}
e^{iS_{EC+\beta}}
\eeqa
where in the last line we have used the relation
\beq
e^{-i\frac{3}{2k\lambda\beta}
\int_{M}\star R\wedge R +\frac{1}{\beta} R\wedge R}\,\Big|^{E_{2}}_{E_{1}}
=\Psi^{*}_{R'}[A']\Psi_{R}[A]\,,
\label{Psi*Psi}
\eeq
and it is understood that on the left hand side we are fixing the triad configurations $E_{1}$ and $E_{2}$ on the
two spacelike boundary endcaps of $M$.
We now claim that the topological terms that enter into the physical inner product are precisely the topological
terms which enter into the Macdowell Mansouri action. We recall that in our introduction we to the Macdowell-Mansouri
action we generalized the action to include the Immirzi parameter. The result was that the action reduced to the
following
\beq
S_{MM+\beta}=S_{topo}+S_{H+\Lambda}
\eeq
where $S_{H+\Lambda}$ is the Holst action with a positive cosmological constant and
\beq
S_{topo}=-\frac{3}{2k\lambda}\int_{M} \star R\wedge R+\frac{1}{\beta}R\wedge R.
\eeq
But this is precisely the negative of the argument of the generalized Kodama states in four-dimensional form!
Furthermore, this is exactly the term which enters into the true inner product as seen by equation (\ref{Psi*Psi}):
\beqa
\langle \Psi_{R'}|\Psi_{R}\rangle_{phys} &=& \int \mathcal{D}A'\mathcal{D}A\ \Psi^{*}_{R'}[A'] \Psi_{R}[A]
\int^{E_{2}}_{E_{1}}\mathcal{D}\omega \mathcal{D}e \ e^{iS_{H+\lambda}} \nn\\
&=& \int^{E_{2}}_{E_{1}}\mathcal{D}\Lambda \ e^{iS_{MM+\beta}}\,.
\eeqa

We conclude that the difference between the Macdowell-Mansouri formulation of gravity and the Einstein-Cartan
formulation is that the former already has the generalized Kodama states built into the theory as ground states. This
is similar to the two formulations of the $\theta$-ambiguity of Yang-Mills theory (see e.g. \cite{Ashtekar:book}). There
one finds that different
sectors of the phase space are connected via large gauge transformations. This ambiguity is reflected in the states,
which are not invariant under large gauge transformations but transform by a phase factor: $\Psi\rightarrow e^{i n
\theta}\Psi$. In an
attempt to salvage gauge invariance, one might normalize all the states by multiplying all states by the phase factor 
$e^{-i\frac{\theta}{8\pi^{2}}\int Y_{CS}}$ so that the $\theta$-ambiguity
is cancelled. However, the ambiguity simply reemerges in the inner product as the measure transforms to include a
factor of the second Chern class $e^{i\frac{\theta}{8\pi^{2}} \int F\wedge F}$. A similar phenomenon appears to 
be happening in the present situation.

\chapter{Canonical Classical and Quantum Gravity Without Gauge-Fixing\label{GaugeFree}}

In the previous two chapters we saw that generalizing the Kodama state to real values of the Immirzi parameter opened
up a large Hilbert space of states. Among these states there is at least one state that is in the physical Hilbert
space, $\mathcal{H}_{phys}=\mathcal{H}_{\{DGH\}}$, and it can be identified with the quantum version of the flat space
slicing of de Sitter space. The
interpretation of the other states we left deliberately open ended. It is natural to assume that the multiplicity of
states come from perturbations to de Sitter space. However, the possibility remains that the disparity of these states
is truly a reflection of the gauge fixing
process and that they are all different representations of a single state, which are connected by a portion of the full
Lorentz group that is cut-off by the gauge fixing procedure. We have already seen evidence of this from the WKB
interpretation of the state. There the action was evaluated on the submanifold spanned by solutions to the Einstein
equations, and it was found that when the Weyl tensor vanishes, the action reduces to a boundary term. The exponent of
this boundary term gives precisely the full set of generalized Kodama states, which are obtained by gauge fixing to
the time gauge and fixing the metric information on the boundary to give the Levi-Civita curvature that parametrizes
the class of states. The WKB state prior to gauge fixing plays the role of an umbrella state that contains all the
information of the whole sector of gauge-fixed states. Here we will obtain qualitative evidence that the same state is
an exact solution to all the quantum constraints in a gauge-free formalism. The evidence is qualitative since a full
demonstration would require solving a set of primary constraints. We will then suggest a route to dealing
with the primary constraints---not by the standard methods of gauge fixing or introducing Dirac brackets, but instead by
avoiding them altogether. In the process it will becoming increasingly clear that the true reason for the existence of
the Kodama state is an underlying local de Sitter symmetry which is necessarily broken on the kinematical phase space
as a whole, but is still retained by a large portion of the phase space including de Sitter space itself.
This underlying de Sitter symmetry manifests itself in the constraint algebra, which we will show is a modification of the
de Sitter Lie algebra.

Thus, in this section we will attempt to reconstruct Hamiltonian General Relativity without partially gauge fixing
to the time gauge. In the context of the Kodama state, the main goal is to provide strong evidence that the state
can be constructed as a state whose underlying gauge group is the full Lorentz
group and not simply the rotation subgroup. In addition, this we
will show that this formalism unifies the infinite class of generalized Kodama states into a single state (in fact, the
umbrella state discussed previously)
which is quantum de Sitter space.

\section{The Kodama state and local de Sitter symmetry}
\subsection{The Macdowell-Mansouri action and local de Sitter symmetry}
Our first indication that the de Sitter group may be the true group underlying the gravitational action comes from the
Macdowell-Mansouri formulation of gravity. Recall from previous sections that the
Macdowell-Mansouri action begins with the curvature
of a de Sitter connection, $\Lambda=\omega+\frac{i}{r_{0}}e$. The curvature is given by
\beq
F=d\Lambda+\Lambda\wedge \Lambda=R-\ts{\frac{1}{{r_0}^2}}e\wedge e +\ts{\frac{i}{r_0}}T.
\eeq
de Sitter space can then be written in compact notation as $F=0$. The Macdowell Mansouri action,
\beq
S_{MM}=-\frac{2}{3k\lambda}\int_M \star F\wedge F,
\eeq
is equivalent to the Einstein-Cartan action up to boundary terms that, as discussed previously, are directly related
to the generalized Kodama states. Thus, it appears that under the Macdowell-Mansouri construction gravity is an
ordinary gauge theory whose gauge group is the de Sitter group. However, we recall that there are several important
differences. First, the ``dual" given by the operator $\star$ is not an external dual but an internal dual---that
is, it is not a dual on the base manifold, but a dual on the fiber itself. Second, the dual itself breaks full de
Sitter invariance (while retaining Lorentz invariance). To see this, consider the action of a small de Sitter
transformation generated by the local Lorentz generator $\alpha=\alpha_{[IJ]}\ts{\frac{1}{4}}\gamma^I\gamma^J$ and the
pseudo-translation generator $\eta=\ts{\frac{1}{2}}\eta_I \gamma^I$. Under an infinitesimal de Sitter transformation,
the connection transforms as:
\beq
\Lambda\rightarrow \Lambda-D_{\Lambda}(\alpha+\ts{\frac{i}{r_{0}}}\eta)
\eeq
and the curvature transforms as
\beq
F\rightarrow F+[\alpha+\ts{\frac{i}{r_{0}}}\eta, F]
\eeq
The action then transforms by $S_{MM}\rightarrow S_{MM}+\delta S_{MM}$ where
\beq
\delta S_{MM}=-\frac{2}{3k\lambda}\int_M [\star, \alpha+\ts{\frac{i}{r_{0}}}\eta]F\wedge F.
\eeq
Since $\alpha$ commutes with $\star$ and the trace of any odd number of gamma-matrices is zero, we have
\beq
\delta S_{MM}=\frac{2}{k}\int_{M}\star \, \eta \, T \wedge(R-\ts{\frac{1}{r_{0}^{2}}}e\wedge e)
\eeq
Thus, we see first of all that the Macdowell-Mansouri is not invariant under the full de Sitter group---it {\it is} invariant
under local Lorentz transformations, but it is {\it not} invariant the pseudo-translations of the de Sitter group. On the
other hand, the action is invariant whenever the integrand above is zero. In particular, it is invariant whenever the
relatively weak condition of vanishing torsion holds. Thus, local de Sitter symmetry of
the action is retained by a relatively large sector of the phase space (much larger, in fact, than the on-shell phase
space). Thus, we might expect that the local evolution of gravitational degrees of freedom might reflect this de
Sitter symmetry. This we will confirm.

\subsection{The na\"{i}ve canonical constraints reflect de Sitter symmetry}
In this section we will construct a model where the underlying de Sitter symmetry of the theory is explicit in 
the canonical theory itself. In particular, we will give an action for vacuum gravity whose canonical phase space
consists of the tetrad and the spin connection as position variables whose conjugate momenta are subject to certain primary
constraints. Prior to implementation of these primary constraints, the na\"{i}ve Poisson algebra of the constraints is
isomorphic to the de Sitter Lie algebra.

Instead of the usual Holst modifidication of the Einstein Cartan action, we will begin with a slightly different
modification that is equivalent to the Holst action up to topological terms. In particular, we begin with the action
\beq
S=S_{EC+\lambda}-\frac{1}{2k \beta}\int_{M}T\wedge T.
\eeq
The additional terms is equal to the Holst modifier $\frac{1}{k\beta}\int_{M} e\wedge e\wedge R$ up to a topological
term proportional to the Nieh-Yan class $\int_{\partial M} (e\wedge T)$. Thus, ignoring boundary terms, the equations
of motion are identical to those of the Holst action, which in turn are identical to those of the Einstein-Cartan
action. Our reason for this choice of action is that, as is typical with topological terms, the term does change the
definition of the kinematical phase space. Performing the standard Legendre transform of the action, we have
\beq
S=\int_{\mathbb{R}}\tilde{\eta}\ \frac{1}{k}\int_{\Sigma}\star e\wedge e\wedge \mathcal{L}_{\bar{t}}\, \omega 
-\ts{\frac{1}{\beta}}\ T\wedge \mathcal{L}_{\bar{t}}\, e -(Constraints)
\eeq
Thus, we see that dynamical variables are the spin connection $\omega$ whose conjugate momentum is proportional to the
area two-form, $\Pi_{\omega}=\frac{1}{k}\Sigma$, and the tetrad whose conjugate momentum is proportional to the
torsion, $\Pi_{e}=-\frac{1}{k\beta}T$. These variables are subject to the primary constraints:
\beqa
\Sigma&=&\star e\wedge e \\
T&=& D_{\omega}e\,.
\eeqa
As usual, all dynamical variables are pulled back to the spatial slices.
We note that as opposed to other formalisms, the primary constraints simply tell us that the momenta are simple 
functions of the position variables. Thus, it may be easier to implement the primary constraints in this approach.

Since we are refraining from gauge fixing to the time gauge, the constraints of this theory have some peculiar
properties. Following the standard ADM decomposition, we have decomposed time into a component normal to the spatial
slice, and a component parallel to it: $\bar{t}=\bar{\eta}+\bar{N}$. It will not be necessary to further decompose
$\bar{\eta}$ into a lapse times a unit normal. The diffeomorphism constraint that follows from the Legendre
transformation contains no surprises:
\beq
C_{D}(\bar{N})=\frac{1}{k}\int_{\Sigma}\mathcal{L}_{\bar{N}}\, \omega \wedge \star e\wedge e 
-\ts{\frac{1}{\beta}}\mathcal{L}_{\bar{N}}\,e \wedge De
\eeq
This constraint simply generates diffeomorphisms on the phase space consisting of $\omega$, $e$ and their conjugate
momenta. Since the diffeomorphisms are purely spatial, this constraint reduces the total Hamiltonian degrees of freedom
by three. The $SO(3,1)$-gauge, or Gauss constraint is given by
\beq
C_{G}(\lambda)=\frac{1}{k}\int_{\Sigma}-D\lambda \wedge \star e\wedge e 
-\ts{\frac{1}{\beta}}[\lambda, e] \wedge De .
\eeq
In performing the Legendre transformation, we have made the usual identification, $-\omega(\bar{\eta})\equiv\lambda$,
which
serves as the generator of the gauge group, and as expected this constraint simply generates Lorentz transformations.
Without gauge fixing, the boosts and rotations
of the Lorentz group are treated on equal footing and the constraint reduces the total degrees of freedom by six. 

Finally we come to the Hamiltonian constraint. The Hamiltonian constraint has the peculiar property in this formalism that
is gains extra degrees of freedom. In particular, the Hamiltonian constraint is not a scalar constraint at all, rather,
it is {\it vectorial}. Explicitly we have
\beq
C_{H}(\eta)=\frac{1}{k}\int_{\Sigma} -[\eta, e]\wedge\left(\star R -\ts{\frac{\lambda}{3}}\star e\wedge e\right)-
\ts{\frac{1}{\beta}}D\eta \wedge De\ .
\eeq
In defining this constraint from the Legendre transformation we have made the identification $\eta \equiv
e(\bar{\eta})$. That is, the generator, $\eta$, is simply the normal component of the time vector projected into the
fiber. In most standard treatments, $\bar{\eta}$ is written $\bar{\eta}=N\bar{n}$ where $N$ is the lapse and $\bar{n}$
is the unit normal. When the normal is projected into the fibre it becomes the unit four vector $n^{I}\equiv
e^{I}(\bar{n})$. The time gauge is achieved by once and for all fixing the direction of this vector in the fiber:
$n^{I}=(1,0,0,0)$. All other indices are decomposed accordingly, and the local gauge freedom is restricted to
those transformations that preserve this direction, thereby reducing the gauge group to spatial rotations. Since we
are not working in the time gauge, the Hamiltonian constraint retains its vectorial character. As we will see, the
generators, $\eta$, are directly related to the generators of the pseudo-translations of the de Sitter group. 

Further justification for the vectorial nature of the Hamiltonian constraint comes from a simple counting argument. It
is well known that total Hamiltonian degree of freedom of vacuum general relativity is equal to two. In the
simplest perturbative framework, these degrees of freedom are associated with the two degrees of freedom of a massless
spin-two graviton. In a background independent, non-perturbative context, the physical degrees of freedom are generally
much more difficult to pinpoint. Nevertheless, the Hamiltonian degrees of freedom found by counting the total degrees
of freedom of the phase space minus the constraints is independent of the formalism used. Typically one counts these
degrees of
freedom by counting the total number of degrees of freedom of the position variables and then subtracting the total
degrees
of freedom of the constraints. For example, in the time gauge, the dynamical position variable is the Ashtekar-Barbero
connection, $A^{i}_{a}$ which has a total of $3\times 3=9$ degrees of freedom. Modding out by the three-dimensional
local rotation group and three-dimensional diffeomorphisms, we have a total of $9-(3+3)=3$ degrees of freedom remaining.
The Hamiltonian constraint is a scalar in the time gauge so we are left with exactly two Hamiltonian degrees of freedom.
Alternatively we could simply take the total degrees of freedom of the phase space consisting of $A^{i}_{a}$ and its
conjugate momenta $(E^{[i}\wedge E^{j]})_{[ab]}$ (which also has nine degrees of freedom), divide by two and subtract
the constraints. This is more suitable for a phase space where it is difficult to distinguish position from momentum
variables, or when the momentum variables are not independent of the position variables as in our case. For our case,
the position variables are the spin connection ${\omega^{[IJ]}}_{a}$, and the tetrad, $e^{I}_{a}$, both pulled back to
the three manifold and their conjugate momenta ${\Sigma^{[IJ]}}_{[ab]}$ and $T^{I}_{[ab]}$ which add no extra degrees
of freedom due to the primary constraint that they are simply functions of the position variables. The total degrees
of freedom in the phase space are thus, $3\times 6+3\times 4=30$. Dividing this by two and subtracting the three
diffeomorphism degrees of freedom and the six local Lorentz transformations, we are left with $\frac{30}{2} - (3+6)=6$.
If the
Hamiltonian were now a scalar constraint, there would be a total of five local degrees of freedom and the theory could
not describe general relativity. But since the Hamiltonian constraint itself has four degrees of freedom we are left
with $\frac{30}{2} -(3+6+4)=2$. Thus, although we have not yet demonstrated that this is general relativity, the theory
does 
have the appropriate number of local degrees of freedom, and the Hamiltonian must be vectorial so long as there are no
second class constraints that crop up in determining the closure of the constraint algebra.

We now wish to compute the algebra of the constraints under the Poisson bracket. Prior to the implementation of the
primary constraints, the symplectic structure yields the canonical Poisson bracket:
\beq
\left\{A,B\right\}=k\int_{\Sigma}\frac{\delta A}{\delta \omega}\wedge \frac{\delta B}{\delta \Sigma}
-\beta \frac{\delta A}{\delta e}\wedge \frac{\delta B}{\delta T} \ -\ (A \leftrightarrow B) \label{PB}
\eeq
Here $A$ and $B$ are assumed to be integral functionals on the three manifold. In the above treatment, we expressed
all the constraints as functions of the position variables alone. Since the canonical Poisson bracket of any two
functionals that are purely functionals of the position variables is zero (prior to implementing the primary
constraints), we need to re-express the constraints in terms of position an momentum variables. Since the momentum and
position variables are not truly independent, there
is considerable ambiguity in which variables we call momentum and which variables we call position variables. Once the
primary constraints are implemented this ambiguity should be completely
resolved---when the primary constraints
are implemented properly by, for example by the implementation of a non-canonical Poisson bracket modified to take
these constraints into account, the algebra will be independent of what we call position and momentum variables. In
the following section we will develop one approach to solving this problem. For the present, since we are using this
example for the purpose of illustration, we will be content with defining the constraints in such a way that the
resulting algebra retains some of the key features of the true algebra. To this end we define the constraints in the
very natural way:
\beqa
C_{D}(\bar{N})&=&\frac{1}{k}\int_{\Sigma}\mathcal{L}_{\bar{N}}\, \omega \wedge \Sigma-
\ts{\frac{1}{\beta}}\mathcal{L}_{\bar{N}}\,e \wedge T \nn\\
C_{G}(\lambda) &=& \frac{1}{k}\int_{\Sigma}-D\lambda \wedge \Sigma 
-\ts{\frac{1}{\beta}}[\lambda, e] \wedge T \nn\\
C_{H}(\eta)&=&\frac{1}{k}\int_{\Sigma} -[\eta, e]\wedge\left(\star R -\ts{\frac{\lambda}{3}}\Sigma\right)-
\ts{\frac{1}{\beta}}D\eta \wedge T\ . \label{C2}
\eeqa
The na\"{i}ve constraint algebra is then computed by taking the commutators of the above constraints under the canonical
Poisson bracket given above. The calculations are tedious but straightforward, and we will simply state the result:
\beqa
\{ C_{D}(\bar{N}_{1}), C_{D}(\bar{N}_{2})\}&=&C_{D}([\bar{N}_{1},\bar{N}_{2}])\nn\\
\{C_{D}(\bar{N}), C_{G}(\lambda)\}&=&C_{G}(\mathcal{L}_{\bar{N}}\lambda)\nn\\
\{C_{D}(\bar{N}), C_{H}(\eta)\}&=& C_{H}(\mathcal{L}_{\bar{N}}\eta) \nn\\
\{C_{G}(\lambda_{1}), C_{G}(\lambda_{2})\} &=& C_{G}([\lambda_{1}, \lambda_{2}]) \nn\\
\{C_{G}(\lambda), C_{H}(\eta)\} &=& C_{H}([\lambda, \eta]) \nn\\
\{C_{H}(\eta_{1}), C_{H}(\eta_{2})\} &=& -\ts{\frac{\lambda}{3}}\,C_{G}([\eta_{1}, \eta_{2}]) . \label{Alg}
\eeqa
Remarkably, the algebra closes. Let us analyze this algebra in pieces. The first commutator simply tells us that the
commutator of two infinitesimal diffeomorphisms is a just another infinitesimal diffeomorphism generated by the Lie
commutator of the two vectors: $[\bar{N}_{1}, \bar{N}_{2}]^{\nu}\equiv N^{\mu}_{1}\partial_{\mu}N^{\nu}_{2}
-N^{\mu}_{2}\partial_{\mu}N^{\nu}_{1}$. This is a standard result, which simply indicates that the diffeomorphisms act
freely on the phase space and there are no anomalies in the Poisson algebra which produces these transformations.
Similarly, the next two commutators simply tell us the the algebra is the semi-direct product of the diffeomorphism
group with another closed algebra. The commutator of two Gauss constraints is equal to another Gauss constraint
generated by the Lie bracket of the two elements of the $so(3,1)$ Lie algebra. This simply means that the full Lorentz
group acts freely on the phase space and itself constitutes a closed subalgebra. The remaning two commutators are new
to this approach. In fact, the last three commutators combined are isomorphic to a familiar Lie algebra: the ten
dimensional de Sitter Lie algebra discussed in section (\ref{deSitterChapter})! This is the main feature of this approach
that we wanted to
illustrate since it will give us valuable insight into the nature of the Kodama state. In total then we see that this
na\"{i}ve canonical algebra of constraints forms a representation of the Lie algebra of $dS_{4}\rtimes Diff_{3}$.
Naturally, this cannot be general relativity since the constraints themselves define the full evolution of the
gravitational field which has local degrees of freedom---thus, the algebra cannot be isomorphic to a Lie
algebra. It appears that the constraints defined above with the given canonical Poisson bracket pick out a topological
sector of the theory with an exact de Sitter symmetry. We should expect that the unique (up to topology) solution to the
full set of constraints is de Sitter space itself. 

Let us now consider the quantum theory defined loosely from the above na\"{i}ve canonical algebra. It may seem premature
to define a quantum theory before we have even implemented the primary constraints which define the full classical
theory. Nevertheless, we will proceed. Our justification is twofold. First, the above theory appears to retain enough
of the full classical theory to give us non-trivial results. In addition, it isolates just the aspects of the
full classical that we want: it picks out a classical sector with an exact de Sitter symmetry. We will see that this
alone will give us considerable insight into the true nature of the Kodama state. Second, it is not unprecedented to
construct a quantum theory before the full set of constraints is implemented, and history has shown that there may be
significant advantages in proceeding this way. The BF formulation\cite{Baez:SpinFoam} of spin-foam
dynamics is a classic example. There one begins with a topological theory which can be constrained to give classical
general relativity. One then constructs spin-foam amplitudes from the topological theory prior to the implementation
of the constraints. The constraints are then implemented in the full quantum theory to yield the local degrees of
freedom of general relativity. Our approach has much of the same flavor as the BF spin foam models. The unconstrained
algebra is topological, and the local degrees of freedom must emerge from the implementation of the primary
constraints.

The task of defining a quantum theory based on the given canonical Poisson bracket and the constraint algebra is
to choose a set of fundamental Poisson brackets that will carry over to operator commutators without modification.
Then one uses this set of fundamental operators to define the constraints in such a way that operator ordering
anomalies are minimized. Here, the natural set of fundamental commutators carry over to operator commutators is:
\beqa
\{\omega^{IJ} |_{P}, \Sigma_{KL} |_{Q}\}= k\,\delta^{IJ}_{KL}\,\delta(P,Q) 
&\rightarrow& 
\left[\hat{\omega}^{IJ} |_{P}, \hat{\Sigma}_{KL} |_{Q}\right]= i\,k\,\delta^{IJ}_{KL}\,\delta(P,Q)
\nn\\
\{e^{I} |_{P}, T_{J} |_{Q}\}= -k\beta\,\delta^{I}_{K}\,\delta(P,Q) 
&\rightarrow &
\left[\hat{e}^{I} |_{P}, \hat{T}_{J} |_{Q}\right]= -i\,k\beta\,\delta^{I}_{J}\, \delta(P,Q) \nn \\
\{\omega^{IJ} |_{P}, T_{K} |_{Q}\}=0 &\rightarrow& \left[\hat{\omega}^{IJ} |_{P}, \hat{T}_{K} |_{Q}\right]=0\nn\\
\{e^{I} |_{P}, \Sigma_{KL} |_{Q}\}=0 &\rightarrow& \left[\hat{e}^{I} |_{P}, \hat{\Sigma}_{KL} |_{Q}\right] =0\nn \\
\{\omega^{IJ} |_{P}, e_{K} |_{Q}\}=0 &\rightarrow &\left[\hat{\omega}^{IJ} |_{P}, \hat{e}_{K} |_{Q}\right]=0\ .
\eeqa
The most natural definition of operators is the ``position" representation,
\beqa
\hat{\omega}=\omega &\ \ \  & \hat{\Sigma}= -ik \, \frac{\delta}{\delta \omega} \nn\\
\hat{e}=e &\ \ \  & \hat{T} = ik\beta \, \frac{\delta}{\delta e}\ .
\eeqa
We choose the angular momentum operator ordering where position variables occur to the left of momentum
variables---this is identical to the ordering given in the constraints (\ref{C2}).
Fortunately, this choice works, and the resulting constraint algebra under the operator commutator is identical to the
canonical Poisson algebra (\ref{Alg}).

\subsection{The Kodama state in this formalism}
We now set out to find solutions to the constraints in this formalism. As mentioned previously, the philosophy we will
adopt is similar to that of BF spin foam models---the procedure is to quantize the topological theory first and then
later attempt to impose the constraints which yield the local degrees of freedom of vacuum general relativity.
Although we won't proceed to the second step in this model, we will address the issue in another formalism in
proceeding sections. 

Since the constraint algebra is isomorphic to the de Sitter algebra, it is natural to guess that the states that
annihilate the constraints may be simple de Sitter invariant functionals of $\omega$ and $e$. This is partially true.
However, the constraints are a peculiar representation of the de Sitter algebra and do not act on functionals of
$\omega$ and $e$ in the standard representation $\omega\rightarrow -\frac{1}{r_{0}}[\eta, e]$ and 
$e\rightarrow -r_{0}\,D_{\omega}\eta$. To see this, let us analyze the action of the constraints explicitly. The Gauss
and diffeomorphism constraints contain no surprises and they simply tell us that the states must be diffeomorphism
invariant and local Lorentz invariant functionals under the standard action of the groups on $\omega$ and $e$. Now
consider the Hamiltonian constraint:
\beq
C_{H}(\eta)=\frac{1}{k}\int_{\Sigma}-[\eta, e]\wedge\star R+\frac{i}{r_{0}}
\int_{\Sigma}-{\ts \frac{1}{r_{0}}}\,[\eta,e]
\wedge\frac{\delta}{\delta \omega}-r_{0}\,D_{\omega}\eta\wedge\frac{\delta}{\delta e}.
\eeq
We see that the second two terms simply generate de Sitter pseudo-translations on functionals of $\omega$ and $e$ in
the standard way, however, the first term is peculiar to this representation. Any functional of de Sitter connection 
$\Lambda=\omega +\frac{i}{r_{0}}e$ that is invariant under the standard action of the de Sitter group will annihilate
the second two terms of the constraint. Thus, we look for functionals of the form:
\beq
\Psi[\omega, e]=\mathcal{N}\,\psi[\omega]\,\chi[\Lambda]
\eeq
where $\chi[\Lambda]$ is invariant under local de Sitter transformations. The action of the Hamiltonian constraint
then reduces to
\beq
\widehat{C_{H}}(\eta)\ \Psi[\omega, e]=\mathcal{N}\left(\frac{1}{k}\int_{\Sigma}
-[\eta, e]\wedge\left(\star R+{\ts i\frac{k\lambda}{3}}
\frac{\delta}{\delta \omega}\right)\psi[\omega]\right)\, \times\, \chi[\Lambda].
\eeq
Thus, the choice $\psi[\omega]=\exp\left(i\frac{3}{2k\lambda}\int \star Y\right)$ works, and the corresponding state
$\Psi[\omega,e]$ solves all of the constraints. There is considerable freedom in defining the functional
$\chi[\Lambda]$. For example, $\chi[\Lambda]$ could be any cylindrical function defined on the space of spin networks
with edges labeled by representations of the de Sitter group and nodes labeled with de Sitter intertwiners. Consider
the choice:
\beq
\chi[\Lambda]=e^{i \frac{3}{2k\lambda\beta} \int_{\Sigma} Y[\Lambda]}.
\eeq
Using the identity 
\beq
\int Y[\Lambda]=\int Y[\omega]-\frac{1}{{r_{0}}^{2}}Y_{NY}[\omega, e]
\eeq
where $\int Y_{NY}=\int e \wedge D_{\omega} e$ is the Nieh-Yan invariant we have
\beq
\Psi[\omega, e]=\mathcal{N}\, \exp\left[
i \frac{3}{2k\lambda} \int_{\Sigma}\star Y[\omega]+{\ts \frac{1}{\beta}}\,Y[\omega] \right]
\times 
\exp \left[\frac{-i}{2k\beta} \int_{\Sigma} Y_{NY}[\omega, e]\right].
\eeq
From the preceding arguments, it is clear that this state solves all of the constraints. We recognize the first term in
the above as the umbrella state containing the full sector of generalized Kodama states.
The second term is to be expected since the action we began with differs from the Holst action by precisely the term
$-\frac{1}{2k\beta}\int_{\partial M}e\wedge De$. It follows that the state can still be interpreted as a WKB state
corresponding to de Sitter space. Let us now consider the primary constraints. The primary constraint on the torsion
is, in fact, already solved by this state, since:
\beq
\hat{T}\, \Psi[\omega, e]=ik\beta \,\frac{\delta \Psi}{\delta e}=De \, \Psi[\omega, e].
\eeq
Now consider the action of $\hat{\Sigma}$ on the state:
\beq
\hat{\Sigma}\, \Psi[\omega, e]=\left({\ts\frac{3}{\lambda}}\left(\star R+{\ts \frac{1}{\beta}}\,R\right)
-{\ts\frac{1}{\beta}}\,e\wedge e\right)\,\Psi[\omega, e].
\eeq
Rearranging this we have
\beq
P_{\star}\,R\ \Psi[\omega, e]={\ts \frac{\lambda}{3}}\left(\hat{\Sigma}+{\ts \frac{1}{\beta}}e\wedge e\right)
\, \Psi[\omega,e].
\eeq 
We recall $P_{\star}=\star+\frac{1}{\beta}$ is invertible whenever $\beta\neq \pm i$. Thus, using this property we
see that although
the primary constraint $\widehat{\Sigma}=\star\widehat{ e\wedge e}$ is not automatically satisfied, whenever it is
satisfied, the state satsifies the operator version of de Sitter space:
\beq
\widehat{R}\ \Psi = {\ts \frac{\lambda}{3}}\,\widehat{e\wedge e}\ \Psi
\eeq

\section{The non-canonical Poisson evolution}
The procedure above for evolving the the gravitational degrees of freedom with a vector-valued Hamiltonian constraint
is incomplete without a proper treatment of the primary constraints. Indeed, the dynamics that results from it does not
appear to have local degrees of freedom. In this section we present one method of dealing with primary constraints.
However, rather than explicitly imposing the primary constraints, we will avoid them altogether. To do this we will
exploit the
coordinate invariant (by coordinates we here mean coordinates on the infinite dimensional phase space) formulation of
Hamiltonian dynamics. This will allow us to avoid the Legendre transformation altogether so that we will never have
need to define the ``momentum" variables. In many respects, this is extremely natural in the context of Palatini
general relativity since the theory is already first order. One of the main advantage of the Legendre transform is that
it turns a system of second order differential equations into a coupled set of first order differential equations.
Since Palatini general relativity is already first order, the Legendre transform is somewhat unnatural, and, indeed,
we can define a symplectic structure and a non-canonical Poisson algebra without it.

The approach presented here is motivated by the need to solve the problems associated with the model described
in the previous section\footnote{After several fruitless attempts at imposing the primary constraints explicitly, I
concluded
that the easiest way to get around this problem is to avoid defining the momentum variables entirely, which, after
all, do not add any extra degrees of freedom to the phase space defined by $\omega$ and $e$ on the three-manifold.}.
Here we outline the important features and conclusions abstracted from the model:

\begin{itemize}
\item It appears to be possible to define the Hamiltonian evolution of the gravitational field without partially gauge
fixing
to the time gauge. It may be that there are no second-class constraints and it is not necessary to introduce the
torsion-free spin connection, which severely complicates the canonical theory.
\item The true dynamical degrees of freedom of the phase space can be coordinatized by the spin connection, $\omega$,
and the tetrad, $e$, both pulled-back to the spacelike hypersurface. The ``momentum" variables do not add extra degrees
of freedom to the phase space and should be viewed as unnecessary artifices of the Legendre transformation. 
\item The Hamiltonian constraint is vectorial, and in the presence of a cosmological constant its generator is closely
related to the pseudo-translation generator of the de Sitter group.
\item The true degrees of freedom (DOF) are counted as follows: 
\beqa
\frac{DOF[e]+DOF[\omega]}{2} &-& \left(DOF[C_{D}]+DOF[C_{G}]+DOF[C_{H}]\right)\nn \\
&=& \frac{4\times 3 + 6\times 3}{2}-\left(3+6+4\right)\nn\\
&=& 2\nn
\eeqa
\item The true constraint algebra algebra will follow from a bracket that falls into the general category of a
non-canonical Poisson bracket. From the perspective of the model described above, this follows from the deformation of
the given canonical Poisson bracket to account for the primary constraints. From the current perspective, the primary
constraints do not exist however the bracket that we will introduce is still non-canonical.
\item The true constraint algebra is likely to be a deformation of the de Sitter Lie algebra. The deformation itself
contains the local degrees of freedom of general relativity. Since the action posesses an exact de Sitter symmetry
on a very large portion of the phase space (anywhere $T=0$), the constraint algebra is likely to reduce to the de Sitter
algebra on a large portion of the constraint manifold. 
\item This intimate connection between the constraint algebra and the de Sitter Lie algebra is the true reason for the
existence of the Kodama state. 
\end{itemize}

\subsection{The coordinate invariant treatment of symplectic dynamics}
We begin with a brief overview of the coordinate-independent formulation of symplectic Hamiltonian dynamics (see
\cite{Carlip:3DGravity} and references therein).
Let us begin with the simple one-dimensional action defined a real manifold with endcaps, $[0,1]\times \mathbb{R}$:
\beq
S=\int^{t_{2}}_{t_{1}}\left({\ts\frac{1}{2}}m \dot{x}^{2}-V(x)\right)dt\,.
\eeq
We assume that the variational principle holds on the endcaps as well as in the bulk. Thus, the variational principle
tells us:
\beqa
\bm{\delta}S &=&\int^{t_{2}}_{t_{1}}\left[\frac{d}{dt}\left(m\dot{x}\, 
\bm{\delta}x\right)-\left(m\ddot{x}+\frac{\partial V}{\partial x}
\right)\bm{\delta} x\right] dt \\
&=& m\dot{x}\, \bm{\delta}x \,\Big|^{t_{2}}_{t_{1}}-\int^{t_{2}}_{t_{1}}
\left(m\ddot{x}+\frac{\partial V}{\partial x}\right)\bm{\delta} x\ dt \ .
\eeqa
The vanishing of the bulk variation yields the equations of motion, and the vanishing of the variation on the endcaps
yields a conserved current, $\bm{J}(t_{2})-\bm{J}(t_{1})=0$, where
\beq
\bm{J}=m\dot{x}\, \bm{\delta}x\
\eeq
is referred to as the symplectic one-form. We can think of the variation $\bm{\delta}$ as the exterior
derivative on the phase space itself. Here we have adopted the notation that objects in \textbf{bold} will represent
forms and vectors in the phase space. The symplectic form is then obtained by taking the exterior derivative of the
symplectic one-form:
\beq
\bm{\Omega}=-\bm{\delta}\bm{J}=\dl m\dot{x}\wedge \dl x=\dl x\wedge\dl p\ .
\eeq
Hamilton's equations can then be written geometrically in a way that is independent of the coordinates on the phase
space:
\beq
\bm{\Omega}(\bm{\bar{t}}, \ )=\dl H \label{Ham}
\eeq
where $\bm{\bar{t}}=\frac{\bm{d}}{\bm{d}t}$ is the time evolution vector field. In a particular set of coordinates $(x,p)$,
the time evolution vector field is $\bm{\bar{t}}=\dot{x}\frac{\bm{\partial}}{\bm{\partial}x}
+\dot{p}\frac{\bm{\partial}}{\bm{\partial}p}$. Then the above form of Hamilton's equations reduces to
\beq
\dot{x}\,\dl p-\dot{p}\,\dl x=\frac{p}{m}\,\dl p+\frac{\partial V}{\partial x}\,\dl x\ .
\eeq
Identifying the $\dl x$ and $\dl p$ components separately in the above equation we have the usual form of Hamilton's
equations:
\beqa
\dot{x}&=&\frac{p}{m}\nn\\
\dot{p}&=&-\frac{\partial V}{\partial x}
\eeqa

\subsection{The coordinate independent approach to Hamiltonian general relativity}
We now wish to apply the above techniques to Hamiltonian general relativity. In order to avoid proliferation of
symbols, we will drop all explicit traces, indices, and wedge products on the base manifold. When a wedge product does
enter
into a formula, it is understood that this is the wedge product on the (infinite-dimensional) phase space. As in the
previous section, one-forms and vectors in the phase space will be written in {\bf bold} font. In this section it will
be useful to distinguish the tetrad on the 4-manifold from the same tetrad pulled back to the spacelike hypersurface.
Thus, we denote the tetrad by $\vep=\frac{1}{2}\gamma_{I}\vep^{I}$, and its pullback $\phi^{*}\vep\equiv e $,
where $\phi$ is the embedding of the spacelike hypersurface in the 4-manifold. As usual we are using a Clifford
algebra representation where $\omega=\frac{1}{4}\gamma_{I}\gamma_{J}\omega^{IJ}$, and $\star=-i\gamma_{5}$.

In this section we will return to the Holst action with a cosmological constant given by
\beq
S_{H}=\frac{1}{k}\int_{M}P_{\star} \ \vep\,  \vep\,  
R-\ts{\frac{\lambda}{6}}\star \vep\, \vep\, \vep \, \vep 
\eeq
with $P_{\star}=\star+\frac{1}{\beta}$. Here we take the boundary of the manifold to be two spacelike hypersurfaces at
$t_{1}$ and $t_{2}$: $\partial M=\Sigma(t_{1})\cup \Sigma(t_{2})$.
Variation with respect to the dynamical variables $\vep$ and $\omega$ yields, upon integration by parts,
\beqa
\dl S_{H}&=&\frac{1}{k}\int_{\partial M}P_{\star} e\, e\, \dl \omega \nn\\
&+&\int_{M}-D(P_{\star}\vep\,\vep)\, \dl\omega +\left(P_{\star}R\, \vep -\vep\,P_{\star}R-
{\ts\frac{2\lambda}{3}}\star \vep\,\vep\,\vep \right)\dl \vep
\eeqa
The bulk variation vanishes whenever the equations of motion hold, which we repeat here:
\beqa
P_{\star}R\,\vep-\vep\,P_{\star}R-{\ts\frac{2\lambda}{3}}\star \vep\,\vep\,\vep&=&0 \label{EOM1}\\
D(P_{\star}\vep\,\vep)&=& 0 \label{EOM2}
\eeqa
The boundary term gives us the symplectic one-form, 
\beq
\bm{J}=\frac{1}{k}\int_{\Sigma}P_{\star}\, e\, e\,\dl\omega ,
\eeq
which the variational principle tells us is conserved:
\beq
\bm{J}(t_{2})-\bm{J}(t_{1})=0.
\eeq
Taking $\dl$ to be the exterior derivative on the infinite dimensional function space with coordinates $e$ and
$\omega$, we define the symplectic two-form $\bm{\Omega}$ to be the negative of the exterior derivative of $\bm{J}$:
\beq
\bm{\Omega}=-\dl \bm{J} = \frac{1}{k}\int_{\Sigma}P_{\star}\, \dl\omega \w \dl (e\, e)  \, .
\eeq

To obtain GR from Hamilton's equations we first need the constraints. These can be found by performing the Legendre
transformation but stopping short of defining the momentum variables. To this end, we fix a foliation of the manifold
and define a ``time" evolution variable. Since we are not gauge fixing to the time gauge, the foliation need not
necessarily be a foliation into spacelike hypersurfaces so long as each slice allows for a well-defined evolution. 
In this respect, the ``time" variable need not necessarily be a timelike vector field. Nevertheless, we will
refer to the dynamics along the one-dimensional integral curves as evolution in time. We split the time evolution vector
field into perpendicular and parallel components: $\bar{t}=\bar{\eta}+\bar{N}$. As in the previous section, we define
the $so(3,1)$ generator, $\lambda\equiv -\omega(\bar{\eta})$,
and the pseudo-translation generator, $\eta\equiv \vep(\bar{\eta})$. The constraints then become
\beqa
C_{D}(\bar{N}) &=& \frac{1}{k}
\int_{\Sigma}\mathcal{L}_{\bar{N}}\omega\, P_{\star}\, e\, e  \ \approx\ 0\nn\\
C_{G}(\lambda)&=&\frac{1}{k}\int_{\Sigma}-D\lambda \, P_{\star}\, e\, e \ \approx\  0\nn\\
C_{H}(\eta)&=& \frac{1}{k}\int_{\Sigma} -[\eta, e]\,
\left(P_{\star} R -\ts{\frac{\lambda}{3}}\star e\, e\right)\ \approx\  0.
\eeqa
Alternatively, we can define the constraints from the equations of motion themselves. First, we notice that the Gauss
and Hamiltonian constraint are none other than the equations of motion (\ref{EOM1}) and (\ref{EOM2}) pulled back to
$\Sigma$ and
appropriately smeared and integrated over the slice. Likewise, the diffeomorphism constraint can be derived from the
equations of motion. To see this, smear (\ref{EOM1}) by $\mathcal{L}_{\bar{N}}\vep$, and (\ref{EOM2}) 
by $\mathcal{L}_{\bar{N}}\omega$. Adding these together, and integrating over the whole manifold we have
\beqa
\frac{1}{k}\int_{M}P_{\star}\mathcal{L}_{\bar{N}}(\vep\, \vep)\, R -\frac{2\lambda}{3}
\star\,(\mathcal{L}_{\bar{N}}\vep)\,\vep\,\vep\,\vep + P_{\star}\,
\mathcal{L}_{\bar{N}}\omega \, D \vep \,\vep \,.
\eeqa
This expression reduces to the variation of the action under a diffemorphism:
\beqa
\delta S_{H}&=& \frac{1}{k}\int_{M}\mathcal{L}_{\bar{N}}\tilde{L} \nn\\
&=& \frac{1}{k}\int_{\partial M} \tilde{L}(\bar{N}) \nn\\
&=& \frac{1}{k}\int_{\partial M} [e(\bar{N}), e]\,
(P_{\star}\,R-{\ts \frac{\lambda}{3}}\star\,e\,e) -P_{\star}\,e\,e\,D \omega({\bar{N}})
+P_{\star}\, e\, e \, \mathcal{L}_{\bar{N}}\omega \nn \\
&\approx & 0 \,.
\eeqa
The first two terms in the second to last equation vanish whenever the Gauss and Hamiltonian constraints vanish. The
only independent constraint, therefore, comes from the third term, and is precisely the constraint
$C_{D}(\bar{N})$. Thus, we see that the constraints themselves are essentially the equations of motion pulled back to
the Cauchy slice. 

We can now give the full set of equations of motion from Hamilton's equation (\ref{Ham}). First we define the ``time"
evolution vector field, $\bm{\bar{t}}$, on the infinite dimensional phase space. Using our intuition that the dynamical
variables, $\omega$ and $e$, are a good set of coordinates on the phase space, in these coordinates the time evolution
vector field can be written
\beq
\bm{\bar{t}}=\frac{\dl}{\dl t}=\int_{\Sigma}\mathcal{L}_{\bar{t}}\,\omega\, \frac{\dl}{\dl \omega}
+\mathcal{L}_{\bar{t}}\,e \, \frac{\dl}{\dl e}.
\eeq
Hamilton's equations are now
\beq
\bm{\Omega}(\bm{\bar{t}}, \ )=\dl C_{D}(\bar{N}) +\dl C_{G}(\lambda)+\dl C_{H}(\eta) \label{Ham2}
\eeq
The left hand side of the above is
\beq
\bm{\Omega}(\bm{\bar{t}}, \ )=\int_{\Sigma}(e\, P_{\star} \mathcal{L}_{\bar{t}}\,\omega 
+P_{\star} \mathcal{L}_{\bar{t}}\,\omega \, e)\, 
\dl e - P_{\star}\left(\mathcal{L}_{\bar{t}}\,e\, e +e\, \mathcal{L}_{\bar{t}}\,e\right)\, \dl \omega
\eeq
We notice from the above a peculiarity of the gravitational equations of motion: the time evolution vector field
$\bm{\bar{t}}$ is not uniquely determined by the symplectic evolution, rather, only the particular combination of
variables given above is determined. Since the fields $e$ are not invertible maps, the components 
$\mathcal{L}_{\bar{t}}\, e $ and $\mathcal{L}_{\bar{t}}\,\omega $ are not uniquely determined from Hamilton's
equations alone. The resolution to this paradox comes from the constraints themselves---when the full set of
constraints are solved, and the symplectic evolution is computed, only then can one uniquely determine the time
evolution of the dynamical fields $\omega$ and $e$. 

To compute the right hand side of Hamilton's equations (\ref{Ham2}) we need to compute the 
exterior derivative of the constraints:
\beqa
\dl C_{D}(\bar{N})&=&\frac{1}{k}\int_{\Sigma}-P_{\star}\,\mathcal{L}_{\bar{N}}(e\,e)\,\dl\omega
+\left(e\, P_{\star}\,\mathcal{L}_{\bar{N}}\omega+P_{\star}\,\mathcal{L}_{\bar{N}}\omega\right)\, \dl e\nn\\
\dl C_{G}(\lambda)&=&\frac{1}{k}\int_{\Sigma}-[\lambda, P_{\star}\,e\,e]\,\dl\omega
-\left(e P_{\star}\, D\lambda+P_{\star}\,D\lambda\,e\right)\, \dl e \nn\\
\dl C_{H}(\eta)&=& \frac{1}{k}\int_{\Sigma}-P_{\star}\,D[\eta, e]\,\dl\omega \nn\\
&\ & +\left([\eta, P_{\star}\,R]+\frac{2\lambda}{3}\star(\eta\,e\,e-e\,\eta\,e+e\,e\,\eta)\right)\,\dl e\, .
\eeqa
Putting the two sides of Hamilton's equations together and identifying components independently for arbitrary
variation, $\dl \omega$ and $\dl e$, we have
\beqa
P_{\star}(\mathcal{L}_{\bar{t}}e\, e +e\, \mathcal{L}_{\bar{t}}e) &=& P_{\star}\,\mathcal{L}_{\bar{N}}(e\,e)+[\lambda,
P_{\star}\,e\,e] +P_{\star}\,D[\eta, e] \label{EE2a}\\
e\, P_{\star} \mathcal{L}_{\bar{t}}\,\omega +P_{\star} \mathcal{L}_{\bar{t}}\,\omega \, e
&=& \left(e\, P_{\star}\,\mathcal{L}_{\bar{N}}\omega+P_{\star}\,\mathcal{L}_{\bar{N}}\omega\,e\right)
-\left(e \,P_{\star}\, D\lambda+P_{\star}\,D\lambda\,e\right)\nn \\
&\ & +[\eta, P_{\star}\,R]+\ts{\frac{2\lambda}{3}}\star(\eta\,e\,e-e\,\eta\,e+e\,e\,\eta)\,. \label{EE2b}
\eeqa
These complicated looking expressions can, in fact, be deciphered rather easily. The first equation, (\ref{EE2a}), is
precisely
the time component of the equation of motion found from varying the action with respect to $\omega$, (\ref{EOM2}),
\beq
i_{\bar{\eta}}\left(D(P_{\star}\, \vep\,\vep)\right)=0
\eeq
and the second set, (\ref{EE2b}), is the time component of the equation of motion found from varying the action with
respect to $\vep$, (\ref{EOM1}),
\beq
i_{\bar{\eta}}\left(P_{\star}R\, \vep -\vep\,P_{\star}R-\ts{\frac{2\lambda}{3}}\star \vep\,\vep\,\vep
\right) = 0 \,.
\eeq
Thus, with the given set of constraints, Hamilton's equations give us precisely the time components of the Einstein-Cartan
equations. Recalling that, aside from the diffeomorphism constraint, the constraints themselves are the equations of
motion pulled back to the Cauchy slice, the remaining components of the Einstein Cartan equations are the vanishing of
the constraints themselves. 
\subsection{The constraint algebra}
In the previous section, we have given a Hamiltonian formulation of gravity that gives precisely the
Einstein equations of the Einstein-Cartan formulation. We have not had need to gauge fix, and there are no primary
constraints in our formalism. We now need to compute the Poisson algebra of the constraints in order to check that the
constraint algebra closes. In order to do this, we will exploit the coordinate invariant definition
of the Poisson bracket, which we briefly review below. 

Given any integral functional $f$ over the Cauchy slice, one
can (partially) define a canonical vector field $\bm{\bar{X}}_{f}$ associated with $f$ by
\beq
\bm{\Omega}(\bm{\bar{X}}_{f},\ )=\dl f\,. \label{CanonicalV}
\eeq
Hamilton's equations then simply tell us that the time evolution
vector field, $\bm{\bar{t}}$, is the canonical vector field associated with the total Hamiltonian, $H=C_{D}+C_{G}+C_{H}$:
\beq
\bm{\bar{t}}=\bm{\bar{X}}_{H}\, .
\eeq
Given two functionals $f$ and $g$ and their associated canonical vector fields $\bm{\bar{X}}_{f}$ and 
$\bm{\bar{X}}_{g}$, the coordinate invariant definition of Poisson bracket is\cite{Jose&Saletan}
\beq
\{f,g\}\equiv \bm{\Omega}(\bm{\bar{X}}_{g},\bm{\bar{X}}_{f})\,.
\eeq
This should allow us to compute the Poisson bracket without resorting to explicit expressions for the bracket such as
(\ref{PB}), which generally come from the Legendre transform resulting in a canonical Poisson bracket. The only
problem we will face is that the components of the vector field $\bm{\bar{X}}_{f}$ are not uniquely determined by
(\ref{CanonicalV}). We saw this before in when computing the components of the time evolution vector field. Writing
the vector field in component notation 
$\bm{\bar{X}}_{f}=\int_{\Sigma}\delta_{f} \omega\, \frac{\dl}{\dl \omega}+\delta_{f} e\, \frac{\dl}{\dl e}$, 
the definition (\ref{CanonicalV}) becomes
\beqa
(e\, P_{\star} \delta_{f}\omega 
+P_{\star} \delta_{f}\omega \, e) &=& k\,\frac{\delta f}{\delta e} \\
P_{\star}(\delta_{f}e\, e +e\, \delta_{f} e) &=& -k\,\frac{\delta f}{\delta \omega} \label{CanVect}
\eeqa
and we see that the components $\delta_{f}\omega$ and $\delta_{f}e$ are only determined up to the particular
combination of variables given above.
This, in turn, means
that the Poisson bracket is a non-canonical bracket. Nevertheless, the commutator satisfies out intuitive notion of
evolving one function along the canonical vector field of the other. To see this, consider the commutator of
any integral functional $f=f(\omega,e)$ with another functional $g=g(\omega,e)$. The components of the canonical vector
field associated with $f$ are given above in (\ref{CanVect}), and similarly for $g$. The commutator is
\beqa
\{g,f\}&=&\bm{\Omega}(\bm{\bar{X}}_{f}, \bm{\bar{X}}_{g})\nn\\
&=&\frac{1}{k}\int_{\Sigma}P_{\star}\delta_{f}\omega\,\delta_{g}(e\,e)-\delta_{g}\omega 
\, P_{\star}\delta_{f}(e\,e)\nn\\
&=&\int_{\Sigma} \frac{\delta f}{\delta e}\,\delta_{g}e
+\frac{\delta f}{\delta \omega}\,\delta_{g}\omega\nn\\
&=&\bm{\mathcal{L}}_{\bm{\bar{X}}_{g}}f\,.
\eeqa
Thus, even though the bracket is non-canonical, it
still serves the ordinary purpose of defining the evolution of one function along the canonical vector field of the
other. Most importantly, since the canonical vector field associated with the total Hamiltonian is the time evolution
vector field, we still have
\beqa
\{H,f\}&=&\int_{\Sigma} \frac{\delta f}{\delta e}\,\mathcal{L}_{\bar{t}}e
+\frac{\delta f}{\delta \omega}\,\mathcal{L}_{\bar{t}}\omega\nn\\
&=& i_{\bm{\bar{t}}}\,\bm{\delta}f \nn\\
&=&\bm{\mathcal{L}_{\bar{t}}}f\ .
\eeqa

The fact that the components of the canonical vector fields are not determined uniquely presents problems in 
computing the commutator of two arbitrary functionals. Nevertheless, we will show that the canonical vector fields
associated with the constraints are sufficiently well defined to compute all of the commutators of the constraints
relatively straightforwardly, with the exception of the commutator of two Hamiltonian constraints. This commutator
will require a bit more work, but it also can be computed using this method. Aside from the commutator of two
Hamiltonian constraints, the constraint algebra is:
\beqa
\{ C_{D}(\bar{N}_{1}), C_{D}(\bar{N}_{2})\}&=&C_{D}([\bar{N}_{1},\bar{N}_{2}])\nn\\
\{C_{D}(\bar{N}), C_{G}(\lambda)\}&=&C_{G}(\mathcal{L}_{\bar{N}}\lambda)\nn\\
\{C_{D}(\bar{N}), C_{H}(\eta)\}&=& C_{H}(\mathcal{L}_{\bar{N}}\eta) \nn\\
\{C_{G}(\lambda_{1}), C_{G}(\lambda_{2})\} &=& C_{G}([\lambda_{1}, \lambda_{2}])\nn \\
\{C_{G}(\lambda), C_{H}(\eta)\} &=& C_{H}([\lambda, \eta]) \nn\\
\{C_{H}(\eta_{1}),C_{H}(\eta_{2})\}&=& ??
\eeqa
Thus, with respect to all but the last commutator, our na\"{i}ve canonical Poisson bracket (\ref{PB}) of the previous model
gave
identical results, (\ref{Alg}), for the commutators. Let us now consider the commutator of two Hamiltonian constraints.
It will be useful to split the constraint into two separate pieces $C_{H}=C_{H_{0}}+C_{H_{\lambda}}$ where
\beqa
C_{H_{0}}(\eta)&=&\frac{1}{k}\int_{\Sigma}-[\eta, e]\,P_{\star}\,R \\
C_{H_{\lambda}}(\eta)&=&\frac{1}{k}\int_{\Sigma}\ts{\frac{\lambda}{3}}\,[\eta, e]\,\star e\, e\,.
\eeqa
The commutator we wish to evaluate now becomes,
\beqa
\{C_{H}(\eta_{1}),C_{H}(\eta_{2})\}&=&\{C_{H_{0}}(\eta_{1}),C_{H_{0}}(\eta_{2})\}
+\{C_{H_{\lambda}}(\eta_{1}),C_{H_{\lambda}}(\eta_{2})\}\nn\\
& & +\{C_{H_{0}}(\eta_{1}),C_{H_{\lambda}}(\eta_{2})\}
+\{C_{H_{\lambda}}(\eta_{1}),C_{H_{0}}(\eta_{2})\}\,.\nn\\
& &
\eeqa
Since $C_{H_{\lambda}}$ does not contain $\omega$, clearly we have
\beq
\{C_{H_{\lambda}}(\eta_{1}), C_{H_{\lambda}}(\eta_{2})\}=0\,.
\eeq
Computing cross-terms we have
\beqa
\{C_{H_{0}}(\eta_{1}), C_{H_{\lambda}}(\eta_{2})\}
&+&\{C_{H_{\lambda}}(\eta_{1}), C_{H_{0}}(\eta_{2})\}\nn\\
&=& \frac{1}{k}\int_{\Sigma}{\ts\frac{\lambda}{2}}\star [\eta_{2}, e]\,D[\eta_{1}, e]
-\frac{1}{k}\int_{\Sigma}{\ts\frac{\lambda}{2}}\star [\eta_{1}, e]\,D[\eta_{2}, e] \nn\\
&=& 0\, .
\eeqa
Thus, we see that the commutator reduces to
\beq
\{C_{H}(\eta_{1}), C_{H}(\eta_{2})\}=\{C_{H_{0}}(\eta_{1}), C_{H_{0}}(\eta_{2})\}\,.
\eeq
Proceeding, we take the gradient of $C_{H_{0}}$ which yields, upon identification of components:
\beqa
(e\, P_{\star} \delta_{\eta_{1}}\omega 
+P_{\star} \delta_{\eta_{1}}\omega \, e)&=& [\eta_{1}, P_{\star}R] \nn\\
P_{\star}(\delta_{\eta_{1}}e\, e +e\, \delta_{\eta_{1}} e) &=& P_{\star}\,D[\eta_{1},e]\,.
\eeqa
The symplectic form contracted onto the canonical vector fields takes the general form
\beq
\bm{\Omega}(\bm{\bar{X}}_{C_{H_{0}}(\eta_{2})},\bm{\bar{X}}_{C_{H_{0}}(\eta_{1})})=
\frac{1}{k}\int P_{\star}\delta_{\eta_{2}}\omega\,\delta_{\eta_{1}}e\,e 
-P_{\star}\delta_{\eta_{1}}\omega\,\delta_{\eta_{2}}e\,e\,.
\eeq
Inserting the the $\delta \omega $ components first we have
\beq
\frac{1}{k}\int_{\Sigma} [\eta_{2},P_{\star}R]\,\delta_{\eta_{1}}e
-[\eta_{1},P_{\star}R]\,\delta_{\eta_{2}}e\,. \label{HalfCom}
\eeq
At this point we are stuck. Only the particular combination $P_{\star}(\delta e\, e +e\,\delta e)$ of the canonical
vector field associated with $C_{H_{0}}$ are determined from the symplectic form, yet we simply need the components
$\delta e$ to evaluate the above. We could attempt to re-evaluate the expression by inserting the $\delta e$ components
first and we arrive at 
\beqa
\{C_{H_{0}(\eta_{1})},C_{H_{0}(\eta_{1})}\}&=&\frac{1}{k}
\int_{\Sigma}P_{\star}\delta_{\eta_{2}}\omega\,D[\eta_{1},e]
-P_{\star}\delta_{\eta_{1}}\omega\,D[\eta_{2},e] \nn\\
&=& \frac{1}{k}\int_{\Sigma}D[\eta_{2},\eta_{1}]\, P_{\star} R 
+P_{\star}\delta_{\eta_{2}}\omega\,[\eta_{1},T]
-P_{\star}\delta_{\eta_{1}}\omega\,[\eta_{2},T] \nn\\
&=& \frac{1}{k}\int_{\Sigma}P_{\star}\delta_{\eta_{2}}\omega\,[\eta_{1},T]
-P_{\star}\delta_{\eta_{1}}\omega\,[\eta_{2},T] \,.
\eeqa
We see we are stuck with the same problem---we need the components of the canonical vector field
$\delta_{\eta}\omega$, but only the combination, $e\, P_{\star} \delta_{\eta}\omega 
+P_{\star} \delta_{\eta}\omega \, e$, is given by the symplectic form. The root of the problem is that the constraint,
$C_{H_{0}}$, contains only one factor of $e$, whereas the symplectic structure is quadratic in $e$. We were able to
evaluate the other commutators because at least one of the constraints in the commutator was quadratic or more in $e$.
In the next section we will show that the commutator can be evaluated by employing the Ricci decomposition of the
Riemann tensor. 

\subsection{Resolving the commutator $\{C_{H(\eta_{1})},C_{H(\eta_{2})}\}$}
In the previous section we reached an impasse in evaluating the commutator of two Hamiltonian constraints. The problem
essentially boiled down to the constraint being linear as opposed to quadratic in the tetrad. Here we will show that
the commutator can in fact be evaluated by use of the Ricci decomposition of the Riemann tensor. This decomposition
effectively introduces enough $e$'s into the calculation so that the commutator can be evaluated. Recall that the Ricci
decomposition of Riemann tensor splits the tensor into three pieces
\beq
R^{IJ}=C^{IJ}+E^{IJ}+S^{IJ}\,.
\eeq
Here $S^{IJ}$ is the scalar part which contains only information about the Ricci scalar, 
$R=R^{IJ}(\bar{\vep}_{I},\bar{\vep}_{J})$. Specifically, it is given by
\beq
S^{IJ}={\ts \frac{1}{12}}\, \vep^{I}\wedge \vep^{J}\,R\,.
\eeq
The tensor $E^{IJ}$ is the semi-traceless tensor defined such that 
$E^{IJ}(\bar{\vep}_{I},\ )=R^{I}-{\ts\frac{1}{4}}\, \vep^{J}\,R $ where 
$R^{J}\equiv R^{IJ}(\bar{\vep}_{J},\ )$ is the Ricci tensor. This implies 
$E^{IJ}(\bar{\vep}_{I}, \bar{\vep}_{J})=0$, hence, it is semi-traceless. Specifically, it is given by
\beq
E^{IJ}={\ts \frac{1}{2}}\,(\vep^{I}\wedge R^{J}-\vep^{J}\wedge R^{I})-{\ts\frac{1}{4}}
\,\vep^{I}\wedge \vep^{J}R\,.
\eeq
The remaining piece is the Weyl tensor and it is defined to be completely trace free: $C^{IJ}(\bar{\vep_{I}},\ )=0$.
In total then, we have (we will not have need to distinguish $E^{IJ}$ from $S^{IJ}$ so we will lump these together)
\beq
R^{IJ}={\ts \frac{1}{2}}\,(\vep^{I}\wedge R^{J}-\vep^{J}\wedge R^{I})-{\ts\frac{1}{6}}
\,\vep^{I}\wedge \vep^{J}R +C^{IJ}.
\eeq
In the index free, Clifford notation, we distinguish the Ricci tensor and scalar as follows
\beqa
\stackrel{\bullet}{R}&\equiv& R^{IJ}(\bar{\vep}_{I},\bar{\vep}_{J})\nn\\
\stackrel{\circ}{R}&\equiv & {\ts \frac{1}{2}}\gamma_{J}\, R^{IJ}(\bar{\vep}_{I},\ )\,.
\eeqa
With these definitions, the Riemann tensor, which we will express pulled back to the three-manifold, is given by
(dropping the explicit wedge product):
\beq
R={\ts \frac{1}{2}}\,\left(e\cR +\cR e \right)-{\ts\frac{1}{6}}\,e\,e\bR +C\,.
\eeq
At first glance it appears that the above expression simply defers the problem---although the first few terms contain
extra factors of $e$, which we need, we are left with the bare Weyl tensor. More explicitly, substituting the above
expression for $R$ we have
\beq
\frac{1}{k}\int_{\Sigma} \left[\,\eta_{2}\,,\,P_{\star}\left({\ts \frac{1}{2}}\,\left(e\cR +\cR e \right)
-{\ts\frac{1}{6}}\,e\,e\bR\right)\,\right]\,\delta_{\eta_{1}}e + [\eta_{2},P_{\star}C]\,\delta_{\eta_{1}}e
-(1\leftrightarrow 2)\,. 
\eeq
The first couple terms in the above can be evaluated since they are at least quadratic in $e$, however, we still
appear to be stuck with the terms involving the Weyl tensor. In fact, this term can be evaluated as well. The result
is easier to interpret when the Immirzi terms are not present, so we will first present the case where $\beta
\rightarrow
\infty $, and later generalize to an arbitrary Immirzi parameter. The term in consideration is,
\beqa
\frac{1}{k}\int_{\Sigma}[\eta_{2},\star C]\,\delta_{\eta_{1}}e
-(1\leftrightarrow 2) =\frac{1}{4k}\int_{\Sigma}-\epsilon_{IJKL}\,\eta^{I}_{2}
\,C^{JK}\,\delta_{\eta_{1}}e^{L}-(1\leftrightarrow 2)\,. \label{HamCom2}
\eeqa
Since the Weyl tensor is defined to be trace free, it can be shown that the following identity holds on the
four-manifold:
\beq
\epsilon_{IJKL}\,\vep^{J}\wedge C^{KL}=0\,.
\eeq
Contracting this expression onto the normal and pulling back to $\Sigma$ we have
\beq
\epsilon_{IJKL}\,\eta^{I}\wedge C^{KL}=\epsilon_{IJKL}\,e^{I}\wedge C^{KL}(\bar{\eta})\,.
\eeq
But, the left hand side is precisely the term that occurs in (\ref{HamCom2}). Thus, in total we have:
\beqa
\frac{1}{k}\int_{\Sigma} \left[\,\eta_{2}\,,\,P_{\star}\left({\ts \frac{1}{2}}\,\left(e\cR +\cR e \right)
-{\ts\frac{1}{6}}\,e\,e\bR\right)\,\right]\,\delta_{\eta_{1}}e 
&+&\star
C(\bar{\eta}_{2}) \,\delta_{\eta_{1}}(e\,e)\nn\\
&-&(1\leftrightarrow 2)\,. 
\eeqa
Since all terms are now at least quadratic in $e$, we can evaluate the above using the identity 
$\star \delta_{\eta}(e\, e) = \star\,D[\eta , e]$. The final result for the commutator is
\beqa
\{C_{H}(\eta_{1}),C_{H}(\eta_{2})\}&=& \frac{1}{k}\int_{\Sigma}\star [\eta_{1},\eta_{2}]\,[T, \cR]
-{\ts\frac{1}{6}}\bR\star[\eta_{1},\eta_{2}]\,[T,e] \nn\\
&\
&+2\star(\eta_{1}\,C(\bar{\eta}_{2})-\eta_{2}\,C(\bar{\eta}_{1}))\,T \,.
\eeqa
We note that even though all of the expressions are pulled back to the three manifold, the right-hand side does
contain first order time derivatives. We justify this by first recalling that
the commutators of the constraints essentially determine the second order time evolution of the phase space. In
evaluating the second order time evolution, we must first assume that the first order evolution is available to us.

There are several interesting properties that we can derive from the above. First we note that all terms depend
explicitly on the torsion. Since the full set of equations of motion tell us that torsion must be zero and the
constraints are essentially the equations of motion pulled back to $\Sigma$,
no set of initial data with non-zero torsion can solve the constraints. Thus, the torsion must vanish on the
constraint submanifold. This in turn implies that the above commutator is weakly vanishing. Since all of the other
commutators are also weakly vanishing, we have a closed first class algebra! There are no second class constraints. In
retrospect this was a foregone conclusion. After all, the canonical variables are simply the pull-back of the
dynamical Lagrangian variables to $\Sigma$, and the constraints are themselves simply the Einstein equations
pulled-back to $\Sigma$ together with the diffeomorphism constraint. The symplectic evolution of the system simply
gives us the remaining components of Einstein's equations in the four-manifold. Thus, the question of whether the
constraint algebra closes weakly is equivalent to the question: {\it are Einstein's equations self-consistent?} The
answer is, of course, {\it yes!} Phrased another way, suppose we have a set of initial data 
$\omega_{t_{0}}$ and $e_{t_{0}}$ on the initial Cauchy surface $\Sigma_{t_{0}}$. If the data set is a good data set, it
will solve the constraint equations---in other words it will solve Einstein's equations pulled-back to
$\Sigma_{t_{0}}$. The symplectic evolution simply enforces the remaining equations of motion, but for our purposes it
also serves to evolve $\omega_{t_{0}}$ and $e_{t_{0}}$ on $\Sigma_{t_{0}}$ to $\omega_{t_{0}+\Delta t}$ 
and $e_{t_{0}+\Delta t}$ on the new Cauchy surface $\Sigma_{t_{0}+\Delta t}$. Now the questions is {\it does the new
data satisfy Einstein's equations pulled back to the new Cauchy slice?} If it does then it will satisfy the
constraints on $\Sigma_{t_{0}+\Delta t}$. If it doesn't then this will be reflected in the non-closure of the
constraints, which would indicate that the evolution generated by the given constraints and symplectic
structure pulls the initial data off the constraint submanifold. This, in turn, would indicate a need for more
constraints. But, since two of the constraints plus the evolution equation are {\it precisely} the full set of Einstein's
equations, and Einstein's equations are self-consistent (barring the emergence of singularities
where various physical quantities become singular), this cannot happen
so the constraint algebra {\it must} close. The diffeomorphism constraint does not change this argument---it simply says
that we are free to choose new
coordinates related by an infinitesimal one-parameter diffeomorphism in evolving from $\Sigma_{t_{0}}$ to 
$\Sigma_{t_{0}+\Delta t}$.

Another interesting property arises from the constraint algebra. Our intuition from the previous model considered
suggested that the true algebra was likely to
be a deformation of the de Sitter algebra that reduces exactly to the de Sitter algebra on a large portion of the
phase space including de Sitter space itself.
Consider the above commutator evaluated on the equation of motion
solved by
$R=\frac{\lambda}{3}\,\vep\,\vep +C$. This implies that the Ricci scalar and tensor are
respectively, $\bR=4\lambda$ and 
$\cR=\lambda\,e$. Using these substitutions, (but keeping the torsion terms) the commutator becomes
\beq
\{C_{H}(\eta_{1}), C_{H}(\eta_{2})\}\approx -{\ts \frac{\lambda}{3}}\,C_{G}([\eta_{1},\eta_{2}])
-C_{G}(C(\bar{\eta}_{1}, \bar{\eta}_{2}))\,.
\eeq
As expected, the algebra is in fact a deformation of the de Sitter algebra, and when the Weyl
tensor is zero, as it is for de Sitter space, the algebra reduces to the de Sitter algebra exactly. Furthermore, we
see the local degrees of freedom of general relativity emerging out of the algebra itself via the Weyl terms in the
commutator.

In the presence of the Immirzi parameter, the same trick works in evaluating the commutator. The only difference is
that the curvature is modified by an Immirzi parameter dependent term as follows:
\beq
R\longrightarrow \mathcal{R}=(1-{\ts \frac{1}{\beta}}\star)R\,.
\eeq
Making the obvious substitutions,
\beqa
\stackrel{\bullet}{R} &\longrightarrow & \stackrel{\bullet}{\mathcal{R}}
=\mathcal{R}^{IJ}(\bar{\vep}_{I},\bar{\vep}_{J}) \nn\\
\stackrel{\circ}{R}&\longrightarrow& \stackrel{\circ}{\mathcal{R}}
={\ts\frac{1}{2}}\gamma_{J}\mathcal{R}^{IJ}(\bar{\vep}_{I},\ )
\eeqa
the Ricci decomposition takes the form
\beq
\mathcal{R}={\ts \frac{1}{2}}\,\left(e\stackrel{\circ}{\mathcal{R}} 
+\stackrel{\circ}{\mathcal{R}} e \right)
-{\ts\frac{1}{6}}\,e\,e\stackrel{\bullet}{\mathcal{R}} +\mathcal{C}\,.
\eeq
where $\mathcal{C}$ is the completely trace-free part of $\mathcal{R}$.
The commutator then becomes
\beqa
\{C_{H}(\eta_{1}),C_{H}(\eta_{2})\}&=& 
\frac{1}{k}\int_{\Sigma}\star [\eta_{1},\eta_{2}]\,[T, \stackrel{\circ}{\mathcal{R}}]
-{\ts\frac{1}{6}}\stackrel{\bullet}{\mathcal{R}}\star[\eta_{1},\eta_{2}]\,[T,e] \nn\\
&\
&+2\star(\eta_{1}\,\mathcal{C}(\bar{\eta}_{2})-\eta_{2}\,\mathcal{C}(\bar{\eta}_{1}))\,T \,.
\eeqa

\section{The quantum theory in this formalism}
In this section we will hint at possible routes to constructing a quantum theory based on the formalism
developed in the previous section. The basic idea is to define a set of fundamental commutators that will carry over
to operator commutators without modification under the generic substitution 
$\{\,\cdot\,,\,\cdot\,\}\rightarrow -i[\,\cdot\,,\,\cdot\,]$. Defining the smeared operators 
\beqa
\omega_{\alpha}&=&\int_{\Sigma} \alpha \,\omega \nn\\
\Sigma^{(M)}_{\beta}&=& \int_{\Sigma} \beta\,M\,e\,e 
\eeqa
where $\alpha$ and $\beta$ are appropriate smearing functions, and $M$ is any matrix that commutes with bivector
elements of the Clifford algebra,
the natural fundamental commutators to choose are
\beqa
\{\omega_{\alpha_{1}},\omega_{\alpha_{2}}\}=0 &\rightarrow & 
[\hat{\omega}_{\alpha_{1}},\hat{\omega}_{\alpha_{2}}] =0\\
\{\Sigma^{(P\star)}_{\beta_{1}},\Sigma^{(P\star)}_{\beta_{2}}\}=0 &\rightarrow& 
[\hat{\Sigma}^{(P\star)}_{\beta_{1}},\hat{\Sigma}^{(P\star)}_{\beta_{2}}]=0\\
\{\omega_{\alpha},\Sigma^{(P\star)}_{\beta}\}=-\int_{\Sigma}\alpha\,\beta &\rightarrow & 
[\hat{\omega}_{\alpha},\hat{\Sigma}^{(P\star)}_{\beta}]=i\int_{\Sigma}\alpha\,\beta \,.
\eeqa
The first commutator is particularly significant since it tells us that the Lorentz connection is fully commutative.
This is in direct contrast to the standard approach to covariant canonical gravity\cite{Alexandrov:Covariant,
Livine:Covariant} where the rotation
and boost parts of connection do not commute. The reason for the non-commutativity in these approaches comes down to
the primary simplicity constraint that induces an additional second-class constraint. These constraints must be
either solved, or implemented via the Dirac bracket. The constraint effectively
ensures the rotational part of the torsion vanishes. Thus, solving the constraint replaces the rotational part of the
connection with the Levi-Civita connection: ${\omega^{ij}}_{a}\rightarrow {\Gamma^{ij}}_{a}[E]$ where $E$ is the
rotational part of the triad. With this substitution, the components of the connection no longer commute, and the
Hamiltonian constraint is complicated significantly. Since there are no second class constraints in our approach, this
is not necessary. We note, however, that the dynamical variables, $\omega$ and $e$, will not commute in our approach.
Thus, if one were to introduce a de Sitter connection $\Lambda=\omega +\frac{i}{r_{0}}e$, as is common in
Macdowell-Mansouri like approaches, the Lorentz and translation components of this connection will almost certainly not
commute. 

For defining the Hamiltonian constraint, we appeal to a trick first introduced by Chopin Soo in the context of
Ashtekar gravity\cite{Soo:HConstraint}. The trick is to express part of the Hamiltonian constraint as a commutator of the
volume operator and the Chern-Simons functional. In our context, consider the functional
\beq
C_{Y}=-\frac{1}{\lambda k}\int_{\Sigma} \star P_{\star}^{2} \,Y
\eeq
where $Y=d\omega +\frac{2}{3}\omega\,\omega\,\omega$.
The commutator of the volume piece of the Hamiltonian constraint with this functional is
\beq
\{C_{H_{\lambda}}(\eta), C_{Y}\}=C_{H_{0}}(\eta)\,.
\eeq
This suggests that we define the Hamiltonian constraint by
\beq
\hat{C}_{H}=-i[\hat{C}_{H_{\lambda}}(\eta),\hat{C}_{Y}]+\hat{C}_{H_{\lambda}}(\eta)\,.
\eeq
The advantage of this is twofold. First, the operators $\hat{C}_{H_{\lambda}}$ and $\hat{C}_{Y}$ may be easy to define
in the quantum theory since each are built out of operators that commute. Second, there is no operator ordering
ambiguity in the above expression for the Hamiltonian constraint.

Finally, we discuss the Kodama state in this context. Consider the functionals,
\beqa
\Sigma_{\alpha}&=&\int \alpha\,e\,e \nn\\
I &=&\int_{\Sigma}P_{\star}Y\,.
\eeqa
The Poisson bracket of these two functionals is
\beq
\{\Sigma_{\alpha}, I\}=-2k\int_{\Sigma}\alpha\, R\,.
\eeq
Assuming that we can define a set of fundamental commutators under which the above commutator carries into the quantum
theory without modification, we have
\beq
\left[\,{\ts\frac{\lambda}{3}}\hat{\Sigma}_{\alpha}\,,\,{\ts\frac{3i}{2k\lambda}}\hat{I}\,\right]
=\int_{\Sigma}\alpha\,\hat{R}\,. 
\eeq
Since the commutator itself commutes with $\hat{I}$, taking the exponent and using the Campbell-Baker-Hausdorff identity
we have 
\beq
\left[\,{\ts \frac{\lambda}{3}}\,\hat{\Sigma}_{\alpha}\,,\,e^{\frac{3i}{2k\lambda}\hat{I}}\right]
=\int_{\Sigma}\alpha\,\hat{R}\,\times\, e^{\frac{3i}{2k\lambda}\hat{I}}\,.
\eeq
Let us assume that there exists a representation of the fundamental commutators where $\hat{\omega}$ is
multiplicative. We will interpret the operator above as the wave function itself in this representation: 
$\Psi[\omega]=\exp^{\frac{3i}{2k\lambda}\int P_{\star}Y[\omega]}$, which is the umbrella state containing the full
sector of generalized Kodama states.
From the commutator we then have 
\beq
{\ts\frac{\lambda}{3}}\hat{\Sigma}_{\alpha}\,\Psi[\omega]
=\int_{\Sigma}\alpha\,R\ \Psi[\omega] -\Psi[\omega]\,{\ts\frac{\lambda}{3}}\hat{\Sigma}_{\alpha}\,.
\eeq
Since $\Psi[\omega]$ is the wave function, the last term on the right vanishes since $\hat{\Sigma}_{\alpha}$ has
nothing to act on\footnote{If this seems strange, recall that the same relation holds for the momentum eigenstates.
There the commutator $[\hat{x}, \hat{p}]=i$, yields the commutator $\left[\hat{p},e^{i p_{0}\cdot
\hat{x}}\right]=p_{0}e^{ip_{0}\cdot\hat{x}}$. When $e^{ip_{0}\cdot x}$ is viewed as a wavefunction in the position
representation, we have $\hat{p}\,e^{ip_{0}\cdot x}=p_{0}\,e^{ip_{0}\cdot x}+e^{ip_{0}\cdot x}\hat{p_{0}}$. This is
the ordinary momentum eigenstate condition since the term on the far right has nothing to act on, so it vanishes.}.
In an arbitrary representation, we then have (dropping the smearing function):
\beq
\widehat{R}\,|\Psi\rangle ={\ts\frac{\lambda}{3}}\,\widehat{e\,e}\,|\Psi\rangle\,.
\eeq
Thus, we see once again that $|\Psi\rangle $ can be interpreted as the quantum version of de Sitter space.

\chapter{Concluding Remarks and Open Problems\label{Conclusions}}
The search for quantum de Sitter space is an important first step in the construction of a theory of quantum gravity.
Not only does the spacetime display many of the subtleties and complexities of full general relativity, it is also
physically relevant since it appears that we are living in an increasingly lambda dominated universe that is
asymptotically de Sitter in the future and possibly in the past as well. de Sitter space is simple enough that we
should expect exact quantum analogues, based on which one might hope to make exact quantum gravitational
predictions. The Kodama state is one such state that bears the promise of a physical state corresponding to de Sitter
space. It is an exact solution to all the constraints of quantum gravity. It has a well-defined semi-classical
interpretation as the initial data state of de Sitter space. And, it fits (reasonably well) within the framework of
the modern theory of quantum gravity. It would be a shame if one could not make physical sense out of this state.

Nevertheless, any quantum state must pass some simple requirements that we expect a physical state to possess. In this
respect, there is strong evidence that original Kodama state fails. However, we have shown in this dissertation that one can
generalize the in order to address the root of this failure. As we have shown, the generalized states resolve, or are
expected to resolve many of these problems.

In addition, the generalization of the Kodama state opens up some intriguing paradoxes, possibly indicating the
limitations of the standard approach to Loop Quantum Gravity. The canonical construction of the
state in the time gauge opened up a large Hilbert space of states, all of which satisfy the Hamiltonian constraint. We
showed that there is at least one state in this set that also satisfies the remaining constraints and can be
interpreted as the quantum analogue of the initial data formulation of de Sitter space in a particular slicing. The
remaining states may simply be a relic of the process we used for the construction of this state, or they may all be
related in some fundamental way. To address this problem, we developed a new approach to covariant classical and
quantum gravity where the role of the de Sitter group in the constraint evolution was emphasized. We found that this
formalism allowed one to return to the question of the existence of a Kodama-like state in a covariant context where
the full Lorentz group is retained, and the relic of de Sitter symmetry persisted in the background. We found that the
quantum theory likely does contain a de Sitter-like solution, it satisfies the quantum operator conditions of quantum de
Sitter space, and it contains the full set of generalized Kodama states in the time gauge. 
In this respect, the covariant framework appears to be a much richer theory where disparate states are unified through
the larger gauge symmetry. There is much work left to be done in this formalism, but we have seen good evidence that
the framework is structurally sound.

Returning to the $R=0$ state in the time gauge, this state provides the promise of quantum gravity phenomenology in de
Sitter space. The state is naturally identified with the quantum analogue of de Sitter space in the co-moving,
inflationary slicing. Since the state is pure phase, perturbations to the state are expected to be normalizable.
By considering perturbations to the connection around the flat space de Sitter solution, one may be able to construct
a theory of linearized gravitons in a familiar slicing of de Sitter space. Alternatively, one could attempt to couple
matter to the theory and derive the free-field propagators. One might expect that the propagator should resemble the
ordinary propagators of perturbative quantum field theory in de Sitter space with quantum mechanical corrections. We
recall that the hallmark achievement of Loop Quantum Gravity
is the discretization of space at the Planck scale. The Kodama state retains this prediction beautifully since its
functional form is known in the spin network representation, which diagonalize the area and volume operators. Thus,
one should expect corrections to the free-field propagator due to Planck scale discreteness. This has been predicted
many times in the past where it usually goes by the name of Doubly Special Relativity or $\kappa$-Poincar\'{e}
algebra. Generally, they describe a set of spacetime symmetries that preserve both a fundamental velocity (the speed
of light) and a fundamental length scale (the Planck length). There are tantalizing clues that the
semi-classical limit of Loop Quantum gravity on a Minkowski or de Sitter background may yield symmetries of this
type\cite{Freidel&Livine:k-Poincare},
but there is no clear consensus of the exact quantum prediction from Loop Quantum Gravity. The generalized Kodama state
may shine in this respect. Already there have been significant advances in the linearization of the Kodama
state\cite{Smolin:linearkodama} and the coupling of the state to matter\cite{EyoEyo:Kodama1}. It
remains to be seen whether this framework can yield a well defined perturbation theory of fields on a fully non-perturbative
quantum gravitational background. Since the large-scale, cosmological implications of quantum gravity may be the most
likely avenue towards quantum gravity phenomenology, making sense of the Kodama state is an important first step
towards a full theory of quantum gravity. We hope to have made some progress in this direction.

%\addcontentsline{toc}{chapter}{Bibliography} 
                                     %% Force Bibliography to appear in contents

%\begin{thebibliography}{..}          %% Start your bibliography here; you can
\bibliography{ThesisArXiv}                  %% also use the \bibliography command

\providecommand{\href}[2]{#2}\begingroup\raggedright\begin{thebibliography}{10}

\bibitem{Kodama:original}
H.~Kodama {\em Prog. Theor. Phys.} {\bf 80} (1988) 1024.

\bibitem{Kodama:original2}
H.~Kodama {\em Phys. Rev. D} {\bf 42} (1990) 2548.

\bibitem{Smolin:kodamareview}
L.~Smolin, ``Quantum gravity with a positive cosmological constant,''
  \href{http://www.arXiv.org/abs/arXiv:hep-th/0209079}{{\tt
  arXiv:hep-th/0209079}}.

\bibitem{Witten:knots}
E.~Witten, ``Quantum field theory and the \uppercase{J}ones polynomial,'' {\em
  Commun. Math. Phys.} {\bf 121} (1989) 351--399.

\bibitem{Pullin:Book}
R.~Gambini and J.~Pullin, {\em Loops, Knots, Gauge Theories, and Quantum
  Gravity}.
\newblock Cambridge University Press, 2000.

\bibitem{Kauffman:TemperleyLieb}
L.~H. Kauffman and S.~L. Lins, {\em Temperley-\uppercase{L}ieb Recoupling
  Theory and Invariants of 3-Manifolds}.
\newblock Princeton University Press, Princeton, New Jersey, 1994.

\bibitem{Glikman:QGPhenomenology}
G.~Amelino-Camelia and J.~Kowalski-Glikman, eds., {\em Planck Scale Effects in
  Astrophysics and Cosmology}.
\newblock Lecture Notes in Physics. Springer, Netherlands, 2005.

\bibitem{Witten:note}
E.~Witten, ``A note on the \uppercase{C}hern-\uppercase{S}imons and
  \uppercase{K}odama wavefunctions,''
  \href{http://www.arXiv.org/abs/arXiv:gr-qc/0306083}{{\tt
  arXiv:gr-qc/0306083}}.

\bibitem{Smolin:linearkodama}
L.~Freidel and L.~Smolin, ``The linearization of the \uppercase{K}odama
  state,'' {\em Class. Quant. Grav.} {\bf 21} (2004) 3831--3844,
  \href{http://www.arXiv.org/abs/arXiv:hep-th/0310224}{{\tt
  arXiv:hep-th/0310224}}.

\bibitem{GockelerSchucker}
M.~G{\"o}ckeler and T.~Sch{\"u}cker, {\em Differential Geometry, Gauge
  Theories, and Gravity}.
\newblock Cambridge University Press, Cambridge, 1987.

\bibitem{Ortin:book}
T.~Ort{\'{i}}n, {\em Gravity and Strings}.
\newblock Cambridge University Press, Cambridge, 2004.

\bibitem{Holst}
S.~Holst, ``Babero's \uppercase{H}amiltonian derived from a generalized
  \uppercase{H}ilbert-\uppercase{P}alatini action,'' {\em Phys.Rev. D} {\bf 53}
  (1996) 5966--5969, \href{http://www.arXiv.org/abs/arXiv:gr-qc/9511026}{{\tt
  arXiv:gr-qc/9511026}}.

\bibitem{Starodubtsev:MMgravity}
A.~Starodubtsev and L.~Freidel, ``Quantum gravity in terms of topological
  observables,'' \href{http://www.arXiv.org/abs/arXiv:hep-th/0501191}{{\tt
  arXiv:hep-th/0501191}}.

\bibitem{Ashtekar:book}
A.~Ashtekar, {\em Lectures on Non-perturbative Canonical Gravity}.
\newblock World Scientific, 1991.

\bibitem{Rovelli:Torsion}
A.~Perez and C.~Rovelli, ``Physical effects of the \uppercase{I}mmirzi
  parameter,'' \href{http://www.arXiv.org/abs/arXiv:gr-qc/0505081 v2}{{\tt
  arXiv:gr-qc/0505081 v2}}.

\bibitem{Freidel:Torsion}
L.~Freidel, D.~Minic, and T.~Takeuchi, ``Quantum gravity, torsion, parity
  violation and all that,'' \href{http://www.arXiv.org/abs/arXiv:hep-th/0507253
  v2}{{\tt arXiv:hep-th/0507253 v2}}.

\bibitem{Randono:Torsion}
A.~Randono, ``A note on parity violation and the \uppercase{I}mmirzi
  parameter,'' \href{http://www.arXiv.org/abs/arXiv:hep-th/0510001}{{\tt
  arXiv:hep-th/0510001}}.

\bibitem{Mercuri:Torsion}
S.~Mercuri, ``Fermions in \uppercase{A}shtekar-\uppercase{B}arbero connections
  formalism for arbitrary values of the \uppercase{I}mmirzi parameter,'' {\em
  Phys. Rev. D} {\bf 73} (2006) 084016,
  \href{http://www.arXiv.org/abs/arXiv:gr-qc/0601013}{{\tt
  arXiv:gr-qc/0601013}}.

\bibitem{Hawking:LSS}
S.~Hawking and G.~Ellis, {\em The large scale structure of space-time}.
\newblock Cambridge University Press, Cambridge, 1973.

\bibitem{Moschella:deSitter}
U.~Moschella, ``The de \uppercase{S}itter and anti-de \uppercase{S}itter
  sightseeing tour,'' in {\em Einstein 1905--2005}, vol.~47 of {\em Progress in
  Mathematical Physics}, pp.~120--133.
\newblock Birkhauser Basel, 2006.

\bibitem{MMoriginal}
S.~Macdowell and F.~Mansouri, ``Unified geometric theory of gravity and
  supergravity,'' {\em Physical Review Letters} {\bf 38} (1977), no.~14,.

\bibitem{Randono:dSgroup}
A.~Randono, ``Think positive.'' Received {\it Honorable Mention} in GRF2007
  essay competition and is pending publication in Dec. 2007 Special Issue of
  IJMPD.

\bibitem{Zee:QFT}
A.~Zee, {\em Quantum Field Theory in a Nutshell}.
\newblock Princeton University Press, 2003.

\bibitem{Ramond}
P.~Ramond, {\em Journeys Beyond the Standard Model}.
\newblock Frontiers in Physics. Perseus Books, Cambridge, MA, 1999.

\bibitem{Mohapatra:left-right}
G.~Senjanovic and R.~N. Mohapatra, ``Exact left-right symmetry and spontaneous
  symmetry breaking,'' {\em Phys. Rev. D} {\bf 12} (1975), no.~5, 1502--1505.

\bibitem{Wise:MMgravity}
D.~K. Wise, ``Macdowell-mansouri gravity and cartan geometry,''
  \href{http://www.arXiv.org/abs/arXiv:gr-qc/0611154}{{\tt
  arXiv:gr-qc/0611154}}.

\bibitem{Baez:Book}
J.~Baez and J.~P. Munian, {\em Gauge Fields Knots and Gravity}, vol.~4 of {\em
  Series on Knots and Everything}.
\newblock World Scientific, Singapore, 1994.

\bibitem{Smolin:MMaction}
L.~Smolin and A.~Starodubtsev, ``General relativity with a topological phase:
  an action principle,''
  \href{http://www.arXiv.org/abs/arXiv:hep-th/0311163}{{\tt
  arXiv:hep-th/0311163}}.

\bibitem{Randono:GKII}
A.~Randono, ``Generalizing the \uppercase{K}odama state \uppercase{II}:
  Properties and physical interpretation,''
  \href{http://www.arXiv.org/abs/arXiv:gr-qc/0611074}{{\tt
  arXiv:gr-qc/0611074}}.

\bibitem{Jacobson:QuadraticSpinor}
R.~S. Tung and T.~Jacobson, ``Spinor one-forms as gravitational potentials,''
  \href{http://www.arXiv.org/abs/arXiv:gr-qc/9502037}{{\tt
  arXiv:gr-qc/9502037}}.

\bibitem{Tung:Immirzi}
C.-H. Chou, R.-S. Tung, and H.-L. Yu, ``Origin of the immirzi parameter,''
  \href{http://www.arXiv.org/abs/arXiv:gr-qc/0509028}{{\tt
  arXiv:gr-qc/0509028}}.

\bibitem{ADM}
R.~Arnowitt, S.~Deser, and C.~Misner, ``The dynamics of general relativity,''
  in {\em Gravitation: an introduction to current research}, ch.~7,
  pp.~227--264.
\newblock Wiley, New York, 1962.
\newblock \href{http://www.arXiv.org/abs/arXiv:gr-qc/0405109}{{\tt
  arXiv:gr-qc/0405109}}.
\newblock Reprint of the original publication now out of print.

\bibitem{Livine:Covariant}
E.~R. Livine, ``Towards a covariant loop quantum gravity,''
  \href{http://www.arXiv.org/abs/arXiv:gr-qc/0608135}{{\tt
  arXiv:gr-qc/0608135}}. Draft Chapter for the book in preparation {\it
  \uppercase{A}pproaches to Quantum Gravity} by Danielle Oriti.

\bibitem{Alexandrov:Covariant}
S.~Alexandrov, ``$\uppercase{SO}(4,\mathbb{C})$--covariant
  \uppercase{A}shtekar-\uppercase{B}arbero gravity and the \uppercase{I}mmirzi
  parameter,'' {\em Classical and Quantum Gravity} {\bf 17} (2000) 4255--4268,
  \href{http://www.arXiv.org/abs/arXiv:gr-qc/9511026}{{\tt
  arXiv:gr-qc/9511026}}.

\bibitem{Thiemann:Book}
T.~Thiemann, {\em Modern Canonical Quantum General Relativity}.
\newblock Cambridge Monographs on Mathematical Physics. Cambridge University
  Press, 1~ed., 2007.

\bibitem{Barbero}
F.~Barbero, ``Real \uppercase{A}shtekar variables for \uppercase{L}orentzian
  signature space-times,'' {\em Phys. Rev. D} {\bf 51} (1995), no.~10,
  5507--5510, \href{http://www.arXiv.org/abs/arXiv:gr-qc/9410014}{{\tt
  arXiv:gr-qc/9410014}}.

\bibitem{Samuel:Connection}
J.~Samuel, ``Is \uppercase{B}arbero's \uppercase{H}amiltonian formulation a
  gauge theory of \uppercase{L}orentzian gravity?,'' {\em Classical and Quantum
  Gravity} {\bf 17} (2000) L141--L148,
  \href{http://www.arXiv.org/abs/arXiv:gr-qc/0005095}{{\tt
  arXiv:gr-qc/0005095}}.

\bibitem{Ashtekar:LQGReview1}
A.~Ashtekar and J.~Lewandowski, ``Background independent quantum gravity: a
  status report,'' \href{http://www.arXiv.org/abs/arXiv:gr-qc/0404018}{{\tt
  arXiv:gr-qc/0404018}}.

\bibitem{Smolin:LQGReview1}
L.~Smolin, ``An invitation to loop quantum gravity,''
  \href{http://www.arXiv.org/abs/arXiv:hep-th/0408048}{{\tt
  arXiv:hep-th/0408048}}.

\bibitem{Rovelli:book}
C.~Rovelli, {\em Quantum Gravity}.
\newblock Cambridge University Press, 2004.

\bibitem{Penrose:SpinNetworks}
R.~Penrose, ``Angular momentum: an approach to combinatorial space-time,'' in
  {\em Quantum Theory and Beyond}, T.~Bastin, ed.
\newblock Cambridge University Press, Cambridge, 1971.

\bibitem{Rovelli&Smolin:SpinNetworks}
C.~Rovelli and L.~Smolin, ``Discreteness of are and volume in quantum
  gravity,'' {\em Nuclear Physics B} {\bf 442} (1995) 593--622,
  \href{http://www.arXiv.org/abs/arXiv:gr-qc/9411005}{{\tt
  arXiv:gr-qc/9411005}}.

\bibitem{Ashtekar:entropy}
A.~Ashtekar, J.~C. Baez, and K.~Krasnov, ``Quantum geometry of isolated
  horizons and black hole entropy,'' {\em Adv. Theor. Math. Phys.} {\bf 4}
  (2000) 1--94, \href{http://www.arXiv.org/abs/arXiv:gr-qc/0005126}{{\tt
  arXiv:gr-qc/0005126}}.

\bibitem{Loll:Volume}
R.~Loll, ``The volume operator in discretized quantum gravity,''
  \href{http://www.arXiv.org/abs/arXiv:gr-qc/9506014}{{\tt
  arXiv:gr-qc/9506014}}.

\bibitem{Soo:thermalkodama}
L.~Smolin and C.~Soo, ``The \uppercase{C}hern-\uppercase{S}imons invariant as
  the natural time variable for classical and quantum cosmology,'' {\em
  Nucl.Phys.} {\bf B449} (1995) 289--316,
  \href{http://www.arXiv.org/abs/arXiv:gr-qc/9405015}{{\tt
  arXiv:gr-qc/9405015}}.

\bibitem{Randono:GKI}
A.~Randono, ``Generalizing the \uppercase{K}odama state \uppercase{I}:
  Construction,'' \href{http://www.arXiv.org/abs/arXiv:gr-qc/0611073}{{\tt
  arXiv:gr-qc/0611073}}.

\bibitem{Randono:GK}
A.~Randono, ``A generalization of the \uppercase{K}odama state for arbitrary
  values of the \uppercase{I}mmirzi parameter,''
  \href{http://www.arXiv.org/abs/arXiv:gr-qc/0504010}{{\tt
  arXiv:gr-qc/0504010}}.

\bibitem{Kauffman:q-deformation}
L.~H. Kauffman, ``Spin networks and the bracket polynomial,'' {\em Knot Theory}
  {\bf 42} (1998) 187--204.

\bibitem{Soo:CPT}
C.~Soo, ``Self-dual variables, positive semi-definite action, and discrete
  transformations in four-dimensional quantum gravity,'' {\em Phys. Rev. D}
  {\bf 52} (1995) 3484--3493,
  \href{http://www.arXiv.org/abs/arXiv:gr-qc/9504042}{{\tt
  arXiv:gr-qc/9504042}}.

\bibitem{Baez:SpinFoam}
J.~C. Baez, ``An introduction to spin foam models of quantum gravity and
  \uppercase{BF} theory,'' {\em Lect. Notes Phys.} {\bf 243} (2000) 25--94,
  \href{http://www.arXiv.org/abs/arXiv:gr-qc/9905087}{{\tt
  arXiv:gr-qc/9905087}}.

\bibitem{Carlip:3DGravity}
S.~Carlip, {\em Quantum Gravity in $2+1$ Dimensions}.
\newblock Cambridge University Press, Cambridge, 1998.

\bibitem{Jose&Saletan}
J.~V. Jos{\'e} and E.~J. Saletan, {\em Classical Dynamics: a Contemporary
  Approach}.
\newblock Cambridge University Press, Cambridge, 1998.

\bibitem{Soo:HConstraint}
C.~Soo, ``Further simplification of the constraints of four-dimensional
  gravity,'' in {\em Proceedings of the 7th Asia Pacific International
  Conference on Gravitation and Astrophysics (ICGA7)}, J.~Nester, C.~Chen, and
  J.~Hsu, eds., pp.~278--283.
\newblock World Scientific, Nat. Central U., Taiwan, 2007.
\newblock \href{http://www.arXiv.org/abs/arXiv:gr-qc/0512025}{{\tt
  arXiv:gr-qc/0512025}}.

\bibitem{Freidel&Livine:k-Poincare}
L.~Freidel and E.~Livine, ``3d quantum gravity and effective non-commutative
  quantum field theory,'' {\em Physical Review Letters} {\bf 96} (2006)
  \href{http://www.arXiv.org/abs/arXiv:hep-th/0512113}{{\tt
  arXiv:hep-th/0512113}}.

\bibitem{EyoEyo:Kodama1}
E.~E. Ita~III, ``Existence of generalized semiclassical kodama states {I}: the
  ashtekar--klein--gordon model,''
  \href{http://www.arXiv.org/abs/arXiv:gr-qc/0703052}{{\tt
  arXiv:gr-qc/0703052}}.

\end{thebibliography}\endgroup
%\end{thebibliography}                %% to generate your bibliography.

\begin{thesisauthorvita}             %% Write your vita here; it can be
Andy Randono grew up in the foothills of the Rocky Mountains in Colorado Springs. Through high-school he was primarily
interested in art, and decided to attend Tufts University to pursue studio art without sacrificing
 academics through their dual degree program with the School of the Museum of Fine Arts in Boston. The five year 
program begins with the first two and a half years primarily at Tufts while taking a few classes at the Museum school,
and the next two and a half years primarily at Museum school while taking some classes at Tufts. Had it been the other
way around, he may have been in art instead of physics now. While at Tufts he became interested in physics and the
philosophy of science, and started a multi-disciplinary program of study between the physics, math, and philosophy
departments. Increasingly interested in physics, after graduating from Tufts in 2001 he spent a summer at CERN, and the
next year at Los Alamos National Laboratory where he researched quantum optics and cryptography. He joined the
graduate program at the University of Texas at Austin in the fall of 2002 and eventually became interested in
canonical quantum gravity. Since nobody at UT does active research in the field, he joined the relativity 
group there and started a long-distance advising situation with Lee Smolin at the Perimeter Institute. 

Though he loves Austin, the mountains are in his blood. When he has the opportunity, he enjoys skiing, mountaineering,
biking, and outdoor sports in general.After leaving UT this fall, he will be joining the quantum gravity group at Penn
State for a postdoc.
                                %% anything in LaTeX2e par-mode.
\end{thesisauthorvita}               %%

\end{document}